%% file: greedy_steiner_forest_submission.tex
\documentclass[11pt,letterpaper]{article}

\usepackage[utf8]{inputenc}
\usepackage[T1]{fontenc}
\usepackage{amsmath,amssymb,amsfonts,amsthm,mathtools}
\IfFileExists{libertine.sty}{\usepackage[tt=false,type1=true]{libertine}}{\usepackage{lmodern}}
\IfFileExists{zi4.sty}{\usepackage[varqu]{zi4}}{}
\IfFileExists{newtxmath.sty}{\usepackage[libertine]{newtxmath}}{}
\usepackage[margin=1in]{geometry}
\usepackage{microtype,parskip,needspace,placeins}
\usepackage{enumitem,booktabs,array,float}
\usepackage[dvipsnames]{xcolor}
\usepackage{tikz}
\usetikzlibrary{arrows.meta,shapes.geometric,positioning,fit,backgrounds,calc,decorations.pathreplacing}
\definecolor{ForestGreen}{rgb}{0.1333,0.5451,0.1333}
\definecolor{DarkRed}{rgb}{0.65,0,0}
\usepackage{hyperref}
\hypersetup{colorlinks=true,linkcolor=DarkRed,urlcolor=cyan,citecolor=ForestGreen,
 pdftitle={The Tight Competitive Ratio of Greedy for Online Steiner Forest},
 pdfauthor={Iman Gholami, Ilya Hajiaghayi, MohammadTaghi Hajiaghayi, Peyman Jabbarzade}}
\renewcommand{\eqref}[1]{\hyperref[#1]{(\ref*{#1})}}
\usepackage[numbers]{natbib}
\usepackage[nameinlink]{cleveref}
\input{macros}

\title{The Tight Competitive Ratio of Greedy for Online Steiner Forest}
\author{Iman Gholami\thanks{University of Maryland. Email: \texttt{igholami@umd.edu}.}
\and Ilya Hajiaghayi\thanks{Takoma Park Middle School. Email: \texttt{ihajiaghayi@gmail.com}.}
\and MohammadTaghi Hajiaghayi\thanks{University of Maryland. Email: \texttt{hajiagha@umd.edu}.}
\and Peyman Jabbarzade\thanks{University of Maryland. Email: \texttt{peymanj@umd.edu}.}}
\date{}

\begin{document}
\hypersetup{pageanchor=false}
\begin{titlepage}
\maketitle
\input{abstract}
\end{titlepage}
\hypersetup{pageanchor=true}

\input{introduction}

\input{overview}

\input{preliminaries}

\input{upper_bound}

\input{lower_bound}

\input{methodology}

\bibliographystyle{alphaurl}
\bibliography{references}


\end{document}

%% file: macros.tex
\newtheorem{theorem}{Theorem}
\newtheorem{corollary}{Corollary}[section]
\newtheorem{lemma}[corollary]{Lemma}

\theoremstyle{definition}
\newtheorem{definition}[corollary]{Definition}
\theoremstyle{remark}

\newcommand{\OPT}{\operatorname{OPT}}
\DeclareRobustCommand{\FullRef}[2]{\hyperref[#2]{#1}}
\newcommand{\ALG}{\operatorname{ALG}}
\newcommand{\dist}{\operatorname{dist}}
\newcommand{\cost}{c}

\newcommand{\E}{\mathbb E}
\newcommand{\Prb}{\mathbb P}
\newcommand{\e}{\mathrm e}
\newcommand{\dd}{\,\mathrm d}

\newcommand{\ScaleSym}{\rho}

\newcounter{algocf}
\newcounter{algoline}[algocf]
\newcommand{\AlgoLine}{\refstepcounter{algoline}\par\noindent\hbox to 1.8em{\hfil\footnotesize\arabic{algoline}:\hspace{0.4em}}\hangindent=1.8em\hangafter=1\ignorespaces}
\newcommand{\AlgoProc}[2]{\AlgoLine\textbf{procedure} \textsc{#1}$(#2)$}
\newcommand{\AlgoInd}[1]{\unskip\hangindent=\dimexpr1.8em+#1\dimexpr12pt\relax\relax\hangafter=1\hspace*{\numexpr#1*12\relax pt}\ignorespaces}
\newcommand{\AlgoComment}[1]{\hfill$\triangleright$\ \emph{#1}}
\newcommand{\KwIn}[1]{\par\noindent\hangindent=1.5em\hangafter=1\hspace*{1.5em}\textbf{Input:} #1\par}
\newcommand{\KwOut}[1]{\par\noindent\hangindent=1.5em\hangafter=1\hspace*{1.5em}\textbf{Output:} #1\par\vspace{2pt}}
\newenvironment{algobox}[2][htbp]{%
  \refstepcounter{algocf}%
  \setcounter{algoline}{0}%
  \begin{figure}[#1]%
  \setlength{\parskip}{2pt}%
  \vspace{2pt}\hrule height 0.8pt\vspace{3pt}%
  \noindent\textbf{Algorithm \thealgocf} #2\par\vspace{2pt}\hrule height 0.4pt\vspace{3pt}%
  \small
}{%
  \par\vspace{3pt}\hrule height 0.8pt\vspace{2pt}%
  \end{figure}%
}
\newcommand{\NumPairs}{k}
\newcommand{\Bought}[1]{P_{#1}}
\newcommand{\BoughtThrough}[1]{P_{\le #1}}
\newcommand{\Pay}[1]{\cost(\Bought{#1})}
\newcommand{\Residual}[1]{\dist_{#1}}
\newcommand{\Fopt}{F^*}
\newcommand{\Fref}{F^\dagger}

\newcommand{\costref}{\cost(\Fref)}
\newcommand{\ForestSaving}{\Delta}

\newcommand{\BatchSaving}[1]{\Delta_{#1}}
\newcommand{\Batch}[1]{B_{#1}}
\newcommand{\BatchFamily}{\mathcal B}
\newcommand{\CurrentBatch}{B}
\newcommand{\Scale}{\ScaleSym}

\newcommand{\ParallelCap}{\mu}
\newcommand{\CoreDegree}{d_{\mathrm{core}}}

\newcommand{\CycleFactor}{\mathcal L}
\newcommand{\LogPairs}{L}
\newcommand{\ClassCount}{h}

\newcommand{\LifeTree}[1]{T_{#1}}
\newcommand{\OldTree}[1]{A_{#1}}
\newcommand{\PairLabel}[1]{\ell_{#1}}
\newcommand{\Far}[1]{u_{#1}}

\newcommand{\ProxyLength}[1]{\delta_{#1}}
\newcommand{\Zone}{Z}
\newcommand{\ZoneTree}[1]{Z_{#1}}
\newcommand{\ZoneCopy}[1]{\widehat Z_{#1}}
\newcommand{\ChargingGraph}{W}
\newcommand{\ZoneGraph}{Q}
\newcommand{\CoreGraph}{Q_{\mathrm{core}}}
\newcommand{\Marks}{\mathcal M}
\newcommand{\CountAux}{q}
\newcommand{\CountMarks}{m}
\newcommand{\CountCore}{N_{\mathrm{core}}}
\newcommand{\CountCoreEdges}{q_{\mathrm{core}}}
\newcommand{\CountCoreComp}{c_{\mathrm{core}}}
\newcommand{\CostCore}{S_{\mathrm{core}}}
\newcommand{\Sep}{\tau}
\newcommand{\LargePairs}{B_{\mathrm{large}}}
\newcommand{\SmallPairs}{B_{\mathrm{small}}}
\newcommand{\FirstPairs}{\mathcal I^{\mathrm{first}}}
\newcommand{\LongPairs}{\mathcal I^{\mathrm{long}}}
\newcommand{\Retained}{\mathcal I^{\mathrm{keep}}}

\newcommand{\InputGraph}{G}
\newcommand{\InputVertices}{V}
\newcommand{\InputEdges}{E}
\newcommand{\EdgeSet}{T}
\newcommand{\PositiveReals}{\mathbb R_{>0}}
\newcommand{\Source}[1]{s_{#1}}
\newcommand{\Target}[1]{t_{#1}}
\newcommand{\TerminalPair}[1]{(\Source{#1},\Target{#1})}
\newcommand{\UpperConstant}{C_*}
\newcommand{\LowerConstant}{c}
\newcommand{\RestrictionSize}{k}
\newcommand{\BatchScale}[1]{\ScaleSym_{#1}}
\newcommand{\FutureMarks}[2]{\Marks_{#1,#2}}
\newcommand{\AuxLeft}[1]{x_{#1}}
\newcommand{\AuxRight}[1]{z_{#1}}
\newcommand{\AuxEdge}[1]{e_{#1}}
\newcommand{\AuxPath}[1]{\pi_{#1}}
\newcommand{\ZoneVertex}[2]{#1^{(#2)}}

\newcommand{\TestCost}{S}
\newcommand{\CycleRank}{r}
\newcommand{\KernelGraph}{H}
\newcommand{\KernelCost}{S_H}
\newcommand{\KernelVertices}{n_v}
\newcommand{\KernelEdges}{n_e}
\newcommand{\KernelRank}{r_H}
\newcommand{\WalkSteps}{t}
\newcommand{\SelectedCount}{q}
\newcommand{\ThresholdIndices}{I}
\newcommand{\PartitionTree}{T}
\newcommand{\ParallelCount}{s}
\newcommand{\ParallelCutoff}{s_0}
\newcommand{\ZoneLog}{\eta}
\newcommand{\Centers}{\mathcal{Z}_{\mathrm{cen}}}

\newcommand{\RequiredCount}{t}
\newcommand{\BranchCount}{n_{\mathrm{br}}}
\newcommand{\OutsideMarks}{m_{\mathrm{out}}}
\newcommand{\CoreRank}{r_{\mathrm{core}}}

\newcommand{\SmallTree}{T_{\mathrm{small}}}
\newcommand{\PositivePairs}{k_+}
\newcommand{\PaymentWeight}[1]{w_{#1}}
\newcommand{\MetricTree}{T}
\newcommand{\MetricCost}{S}
\newcommand{\PointCount}{q}
\newcommand{\InsertionDistance}[1]{d_{#1}}
\newcommand{\FutureCount}{q_{\mathrm{next}}}

\newcommand{\LowerLevel}{h}
\newcommand{\LowerLogSize}{L}
\newcommand{\LowerVertices}{N}
\newcommand{\LowerMatchingSize}{K}
\newcommand{\LowerPrefixFraction}{\alpha_0}
\newcommand{\LowerFirstRank}{i_0}
\newcommand{\LowerPrice}[1]{p_{#1}}
\newcommand{\LowerMatchingEdge}[1]{e_{#1}}
\newcommand{\LowerPath}{F}
\newcommand{\LowerMatching}{M}
\newcommand{\LowerAuxiliaryGraph}{H}
\newcommand{\LowerTestGraph}[1]{H_{#1}}
\newcommand{\LowerPartner}[1]{M(#1)}
\newcommand{\LowerEndpointA}[1]{a_{#1}}
\newcommand{\LowerEndpointB}[1]{b_{#1}}
\newcommand{\LowerSubdivision}[1]{z_{#1}}
\newcommand{\LowerTotalPrice}{S_0}
\newcommand{\LowerBadPrice}{B}
\newcommand{\LowerTestCost}[1]{\omega_{#1}}
\newcommand{\LowerRankFraction}{\alpha}
\newcommand{\LowerRadius}{R}
\newcommand{\LowerTilt}{\theta}
\newcommand{\LowerMeanWeight}{\beta}
\newcommand{\LowerRunSum}{a}
\newcommand{\LowerPathSteps}{\ell}
\newcommand{\LowerShortcutCount}{r}
\newcommand{\LowerMaxPrice}{p_{\max}}
\newcommand{\LowerRequestCount}[1]{k_{#1}}
\newcommand{\LowerVerticesAt}[1]{N_{#1}}
\newcommand{\LowerLogSizeAt}[1]{L_{#1}}
\newcommand{\LowerDeletedEdge}{e}
\newcommand{\ElementarySymmetric}[1]{e_{#1}}
\newcommand{\PolyConstant}{a_0}
\newcommand{\PolyLinearCoeff}{b_0}
\newcommand{\PolyProductCoeff}{c_0}

\newcommand{\SamplePopulation}{m}
\newcommand{\SampleSize}{r}
\newcommand{\SampleVector}{x}
\newcommand{\SampleValue}[1]{x_{#1}}
\newcommand{\SampleIndex}[1]{J_{#1}}
\newcommand{\MetricPoint}[1]{v_{#1}}

\tikzset{
  triangle/.style args={#1}{
    insert path={
        ++(90:1.2*#1)
      -- ++($(90:-1.2*#1)+(210:1.2*#1)$)
      -- ++($(210:-1.2*#1)+(330:1.2*#1)$)
      -- cycle
    }
  },
  pentagon/.style args={#1}{
    insert path={
      ++(90:1.05*#1)
      -- ++($(90:-1.05*#1)+(162:1.05*#1)$)
      -- ++($(162:-1.05*#1)+(234:1.05*#1)$)
      -- ++($(234:-1.05*#1)+(306:1.05*#1)$)
      -- ++($(306:-1.05*#1)+(18:1.05*#1)$)
      -- cycle
    }
  },
  square/.style args={#1}{
    insert path={
        ++(45:1.1*#1)
      -- ++($(45:-1.1*#1)+(135:1.1*#1)$)
      -- ++($(135:-1.1*#1)+(225:1.1*#1)$)
      -- ++($(225:-1.1*#1)+(315:1.1*#1)$)
      -- cycle
    }
  },
  star/.style args={#1}{
    insert path={
      ++(90:1.1*#1)
      \foreach \a in {90,162,234,306,378} {
        -- ++($(\a:-1.1*#1) + (\a+36:0.5*#1)$)
        -- ++($(\a+36:-0.5*#1) + (\a+72:1.1*#1)$)
      }
      -- cycle
    }
  },
  center arc/.style args={#1:#2:#3}{
    insert path={+ (#1:#3) arc (#1:#1+#2:#3)}
  }
}

%% file: abstract.tex
\begin{abstract}
\normalsize
The online Steiner Forest problem is a fundamental network design problem
in which terminal pairs arrive one at a time and an algorithm must
irrevocably purchase edges connecting each pair. The natural greedy
algorithm connects each terminal pair by a shortest path after making
all previously purchased edges free. Awerbuch, Azar, and Bartal
(SODA 1996) proved that this algorithm is $O(\log^2 \NumPairs)$-competitive
and conjectured an $O(\log \NumPairs)$ bound, where $\NumPairs$ is the number of terminal pairs.
Bamas, Drygala, and Maggiori (SODA 2022) obtained an
$O(\log \NumPairs\log\log \NumPairs)$ bound for greedy on a restricted class of
inputs, leaving the general case open.

In this paper, we resolve the competitive ratio of the greedy algorithm for
online Steiner Forest, proving that it is tightly
$\Theta(\log \NumPairs\log\log \NumPairs)$. Specifically, we establish an
$O(\log \NumPairs\log\log \NumPairs)$ upper bound for arbitrary instances and
construct a matching $\Omega(\log \NumPairs\log\log \NumPairs)$ lower bound.
Our results refute the $O(\log \NumPairs)$ conjecture of Awerbuch, Azar, and
Bartal and establish a tight $\Theta(\log\log \NumPairs)$ separation between
the greedy algorithm and the best possible online algorithm for Steiner Forest.
\end{abstract}

%% file: introduction.tex
\section{Introduction}
\label{sec:intro}

The Steiner Forest problem, also known as the Generalized Steiner Tree
problem, is a fundamental optimization problem on edge-weighted graphs.
Given an undirected graph $\InputGraph=(\InputVertices,\InputEdges)$ with strictly positive edge costs and a
collection of terminal pairs, the objective is to find a minimum-cost
subgraph connecting the two endpoints of every pair. The problem
generalizes Steiner Tree, in which all terminals must belong to one
connected component. In the online setting, the graph is known in
advance, but the terminal pairs arrive one at a time. The algorithm must
satisfy each terminal pair before seeing the next one, and purchased edges
cannot be removed.

A natural algorithm satisfies each terminal pair as cheaply as possible.
To connect the $i$th terminal pair $\TerminalPair{i}$, it makes previously purchased edges free
and buys the unpurchased edges of a shortest $\Source{i}$--$\Target{i}$ path in the resulting metric. We refer to
this algorithm as \emph{greedy} (the \emph{Min-Cost} algorithm of Awerbuch, Azar, and Bartal~\cite{AAB04} and \emph{Rule~1} of Bamas, Drygala, and Maggiori~\cite{BDM22}). Let $\ALG$ denote the total cost of
edges purchased by greedy, and let $\OPT$ denote the minimum cost of
an offline solution satisfying the entire sequence of terminal pairs.
The rule of greedy is simple, but its competitive ratio has remained
difficult to determine. Awerbuch, Azar, and
Bartal~\cite{AAB04} proved an $O(\log^2 \NumPairs)$ bound in work first presented
at SODA 1996, and explicitly conjectured in the journal version that greedy is $O(\log \NumPairs)$-competitive.
Here $\NumPairs$ denotes the number of terminal pairs. This conjecture would
place greedy on the same asymptotic footing as the best online
algorithms for Steiner Forest.

The corresponding question for online Steiner Tree was resolved much
earlier. Imase and Waxman~\cite{IW91} showed that connecting each new
terminal to the existing tree gives an $O(\log \NumPairs)$-competitive
algorithm, matching a logarithmic lower bound. However, this analysis
does not extend to arbitrary pairs. Greedy can maintain many connected
components of purchased edges, even when a single inexpensive tree connects all terminals in the
input pairs. Consequently, a common reference tree does not reduce the
online Steiner Forest execution to an online Steiner Tree execution.

Westbrook and Yan~\cite{WY95} initiated the analysis of greedy for online
Steiner Forest. After the $O(\log^2 \NumPairs)$ analysis of Awerbuch, Azar, and
Bartal, Berman and Coulston~\cite{BC97} obtained an
$O(\log \NumPairs)$-competitive algorithm by purchasing additional edges
beyond the cheapest route for the current terminal pair. This establishes the
optimal order for the problem, but leaves open whether those additional
purchases are necessary to improve on greedy.

Bamas, Drygala, and Maggiori~\cite{BDM22} proved an
$O(\log \NumPairs\log\log \NumPairs)$ bound for greedy when each pair's
\emph{contraction}, the ratio of its original endpoint distance to its
greedy payment, is polylogarithmically bounded
(their Theorem~1.1, which includes our full purchased-path contraction rule, whereas their unrestricted-contraction results for a single-tree comparator or non-increasing greedy payments or original endpoint distances concern terminal-only contraction, their Rule~3). The competitive ratio of greedy on general instances remained open and was again highlighted as an open problem by Long, Mahabadi, Sarkar, and Tarnawski~\cite{LMST26}.

In this work, we determine the competitive ratio of greedy. We prove an
$O(\log \NumPairs\log\log \NumPairs)$ upper bound for arbitrary instances and a matching
$\Omega(\log \NumPairs\log\log \NumPairs)$ lower bound. Thus the logarithmic conjecture
is false, and the additional $\log\log \NumPairs$ factor is necessary.

Our analysis shares with Bamas, Drygala, and Maggiori~\cite{BDM22} the
grouping of cost scales into residue classes with polylogarithmic
separation. Their proof uses dual balls and balanced dual solutions to
recursively decompose the instance. We track greedy components through
lifetime trees to extract paths of bounded original cost, and use these
paths to replace dense regions of an evolving reference forest. Charging
to the resulting cost savings removes the bounded-contraction assumption.

\subsection{Main Results}

Our first result bounds the total cost of greedy in terms of the
offline optimum.

\begin{theorem}[Upper bound]
\label{thm:main}
Let $\InputGraph$ be a finite undirected graph with strictly positive edge costs, and let
$\TerminalPair{i}_{i=1}^{\NumPairs}$, with $\NumPairs\ge1$, be a sequence of terminal pairs
such that the endpoints of every pair are connected by a path in $\InputGraph$.
Greedy purchases edges of total cost at most
\[
 \UpperConstant\,\OPT\log(2\NumPairs)\log\log(4\NumPairs),
\]
where $\UpperConstant$ is an absolute constant.
\end{theorem}

The bound of \FullRef{Theorem~\ref*{thm:main}}{thm:main} holds for every choice of tie-breaking among shortest paths and depends only on the number of terminal pairs $\NumPairs$, independently of the number of Steiner vertices in $\InputGraph$.

\begin{theorem}[Lower bound]
\label{thm:lower}
There is an infinite family of online Steiner Forest instances with
unbounded numbers $\NumPairs$ of terminal pairs on which
\[
 \ALG\ge \LowerConstant\,\OPT\log \NumPairs\log\log \NumPairs
\]
for an absolute constant $\LowerConstant>0$.
\end{theorem}

In \FullRef{Theorem~\ref*{thm:lower}}{thm:lower}, randomness is used only to prove the existence of a fixed deterministic graph and terminal-pair sequence; each requested pair already has a unique simple shortest path in its residual graph, so the lower bound holds independently of tie-breaking. Together, the two theorems give the following characterization.

\begin{corollary}
\label{cor:tight}
The worst-case competitive ratio of greedy for online Steiner Forest
with $\NumPairs$ terminal pairs is
$\Theta(\log \NumPairs\log\log \NumPairs)$.
\end{corollary}

\FullRef{Table~\ref*{tab:history}}{tab:history} summarizes the distinction between the problem's
optimal competitive ratio and the competitive ratio of greedy.

\begin{table}[ht]
\centering
\begin{tabular}{@{}>{\raggedright\arraybackslash}p{0.30\linewidth}>{\raggedright\arraybackslash}p{0.31\linewidth}>{\raggedright\arraybackslash}p{0.30\linewidth}@{}}
\toprule
Setting & Guarantee & Reference\\
\midrule
Online Steiner Tree, greedy & $\Theta(\log \NumPairs)$ & Imase--Waxman~\cite{IW91}\\
Online Steiner Forest, greedy & $O(\log^2 \NumPairs)$ & Awerbuch--Azar--Bartal~\cite{AAB04}\\
Online Steiner Forest, augmented & $O(\log \NumPairs)$ & Berman--Coulston~\cite{BC97}\\
Greedy, polylogarithmic contraction & $O(\log \NumPairs\log\log \NumPairs)$ & \cite{BDM22}\\
Online Steiner Forest, greedy & $\Theta(\log \NumPairs\log\log \NumPairs)$ & This paper\\
\bottomrule
\end{tabular}
\caption{Competitive guarantees for irrevocable edge purchases.}
\label{tab:history}
\end{table}

\subsection{Related Work}
\label{sec:related}

\paragraph{Steiner Tree.}
Offline Steiner Tree has a long algorithmic history. Dreyfus and
Wagner~\cite{DW71} gave an exact dynamic program parameterized by the
number of terminals. On the approximation side, the shortest-path
heuristic of Takahashi and Matsuyama~\cite{TM80} and the metric
minimum-spanning-tree approach of Kou, Markowsky, and
Berman~\cite{KMB81} give factor-two guarantees. Mehlhorn~\cite{M88}
developed a faster implementation of the latter approach. These
algorithms can choose which terminal or component to connect next,
whereas an online algorithm must serve the current terminal pair.

Zelikovsky~\cite{Z93} broke the factor-two barrier with an
$11/6$-approximation. Subsequent improvements include the algorithms of
Karpinski and Zelikovsky~\cite{KZ97} and Robins and
Zelikovsky~\cite{RZ05}; the latter achieves a ratio approaching
$1+\tfrac12\ln3$. Borchers and Du~\cite{BD97} provided the
$\RestrictionSize$-restricted component decomposition underlying several of these
approaches. Byrka, Grandoni, Rothvo\ss, and
Sanit\`a~\cite{BGRS13} obtained a $\ln4+\varepsilon$ approximation by
iterative randomized rounding. Goemans, Olver, Rothvo\ss, and
Zenklusen~\cite{GORZ12} analyzed hypergraphic relaxations through
matroid methods. 
These results concern offline approximation and do not imply an online
competitive bound.

\paragraph{Steiner Forest.}
Agrawal, Klein, and Ravi~\cite{AKR95}, in work first presented at STOC
1991, gave the classical factor-two approximation for Steiner Forest.
Goemans and Williamson~\cite{GW95} extended the primal-dual approach to
constrained forest problems, and Jain~\cite{J01} obtained a factor-two
algorithm for generalized Steiner networks through iterative rounding.
Other perspectives include the cost-sharing and timed-growth methods
of Gupta, Kumar, P\'al, and Roughgarden~\cite{GKPR03} and K\"onemann,
Leonardi, Sch\"afer, and van Zwam~\cite{KLS08}.

The role of greedy methods is particularly relevant here. Gupta and
Kumar~\cite{GK15} proved a constant approximation for an offline
\emph{gluttonous} algorithm, which connects suitable nearby active
components rather than serving a fixed terminal-pair order. Gro\ss, Gupta,
Kumar, Matuschke, Schmidt, Schmidt, and Verschae~\cite{GGK18} developed
a constant-factor local-search algorithm. Neither algorithm is the
online greedy rule analyzed in this paper.

Ahmadi, Gholami, Hajiaghayi, Jabbarzade, and Mahdavi~\cite{AGH25}
broke the factor-two barrier for offline Steiner Forest with a
$(2-10^{-11})$-approximation, combining moat growing, local search, and
autarkic pairs. Gupta and Traub~\cite{GT26} subsequently simplified and
extended this approach to obtain a $1.994$-approximation. Stronger
relaxations have also been investigated by Byrka, Grandoni, and
Traub~\cite{BGT26}. The static replacement argument used here allows
changes to a reference forest, but is an analysis of the fixed online
greedy execution rather than an offline optimization algorithm.

\paragraph{Online Network Design.}
The online Steiner Tree and Forest results of Imase and
Waxman~\cite{IW91}, Westbrook and Yan~\cite{WY95}, Awerbuch, Azar, and
Bartal~\cite{AAB04}, and Berman and Coulston~\cite{BC97} form the
starting point of our work. Umboh~\cite{U15} gave a framework based on
hierarchical decompositions that recovers logarithmic bounds for online
Steiner Tree and the augmented Steiner Forest algorithm. Chen,
Roughgarden, and Valiant~\cite{CRV10} used greedy Steiner Forest
analysis in network protocol design. Dehghani, Ehsani, Hajiaghayi,
Liaghat, and Seddighin~\cite{DEHLS18} studied greedy algorithms for
online survivable network design, including stochastic input models.

Additional structure can change the online problem. Alon and
Azar~\cite{AA93} studied online Steiner trees in the Euclidean plane;
Matsubayashi~\cite{M21} studied non-greedy algorithms on outerplanar
graphs. Naor, Panigrahi, and Singh~\cite{NPS11} considered online
node-weighted Steiner problems, while Hajiaghayi, Liaghat, and
Panigrahi~\cite{HLP17} developed disk-painting methods for node-weighted
Steiner Forest. Qian, Umboh, and Williamson~\cite{QUW18} studied online
constrained forests and prize-collecting network design. In a terminal-only metric,
Long, Mahabadi, Sarkar, and Tarnawski~\cite{LMST26} obtained a constant-competitive Steiner Forest
algorithm with logarithmic amortized edge recourse.

\paragraph{Prize-Collecting Problems.}
For variants allowing penalties for unsatisfied pairs, Hajiaghayi and
Jain~\cite{HJ06} developed a primal-dual approach; Ahmadi, Gholami,
Hajiaghayi, Jabbarzade, and Mahdavi obtained a factor-two approximation for
Steiner Forest~\cite{AGHpcsf25} and a $1.79$-approximation for Steiner
Tree~\cite{AGHpcst24}.

\paragraph{Girth and Lower Bounds.}
Girth arguments are central to earlier greedy analyses and
examples~\cite{AAB04,BDM22}.
Large-girth cubic graphs, such as those surveyed by
Biggs~\cite{Biggs98}, explain how a single cost scale can already incur
a logarithmic loss. The Moore bound and its extension to irregular
graphs by Alon, Hoory, and Linial~\cite{AHL02} formalize the relation
between density and short cycles. We prove the weighted cycle-rank
estimate needed in our upper bound directly. Our lower bound uses random matching shortcuts to support a harmonic
range of prices. All future shortcuts are included when testing whether
a price can be forced.

\paragraph{Organization.}
\FullRef{Section~\ref*{sec:overview}}{sec:overview} gives an overview of the
upper- and lower-bound proofs and their main ingredients.
\FullRef{Section~\ref*{sec:prelim}}{sec:prelim} fixes notation and conventions.
\FullRef{Section~\ref*{sec:upper}}{sec:upper} proves the upper bound
(\FullRef{Theorem~\ref*{thm:main}}{thm:main}), and
\FullRef{Section~\ref*{sec:lower}}{sec:lower} presents the lower-bound
construction and proves \FullRef{Theorem~\ref*{thm:lower}}{thm:lower}.


%% file: overview.tex
\section{Overview of the Proofs}
\label{sec:overview}

This section gives a self-contained proof sketch of both the upper bound and the matching lower bound. Our goal is to explain the main
conceptual obstacles, motivate each structural definition at the moment it is
needed, and show how these ideas combine to establish the tight
$\Theta(\log \NumPairs\log\log \NumPairs)$ competitive ratio without getting
bogged down in low-level constant calculations.

\subsection{Upper Bound: Evolving the Optimal Forest Across Scales}

The upper bound (\FullRef{Theorem~\ref*{thm:main}}{thm:main}) is a retrospective
charging argument. Its input consists of the underlying graph
$\InputGraph=(\InputVertices,\InputEdges)$ with positive edge costs
$\cost:\InputEdges\to\mathbb{R}_{>0}$, the completed online greedy execution on
$\NumPairs$ terminal pairs $\TerminalPair{1},\ldots,\TerminalPair{\NumPairs}$,
and a fixed optimal offline Steiner forest $\Fopt$ of cost $\OPT=\cost(\Fopt)$.
When terminal pair $\TerminalPair{i}$ arrives at step $i$, we define
$\Bought{i}\subseteq\InputEdges$ to be the set of new (previously unpurchased)
edges bought by greedy to connect $\Source{i}$ and $\Target{i}$, and we define
greedy's step-$i$ payment to be $\Pay{i}=\cost(\Bought{i})$ and its total
expenditure to be $\ALG=\sum_{i=1}^{\NumPairs}\Pay{i}$. Because greedy reuses
already-purchased edges for free, we define the \emph{residual cost} of an edge
$e\in\InputEdges$ before step $i$ to be $0$ if $e\in\bigcup_{j<i}\Bought{j}$ and
$\cost(e)$ otherwise, and we define the \emph{residual distance}
$\Residual{i}(u,v)$ to be the shortest-path distance between $u$ and $v$ in
$\InputGraph$ under these residual costs before step $i$. Thus greedy pays
$\Pay{i}=\Residual{i}(\Source{i},\Target{i})$ at step $i$, and immediately
afterward $\Residual{i+1}(\Source{i},\Target{i})=0$.

For this sketch assume $\NumPairs\ge4$ and set
$\LogPairs=\lceil\log_2(8\NumPairs)\rceil+1$; the formal shifted-logarithm
bounds also cover smaller inputs. The upper-bound construction follows
\[
 \begin{gathered}
 \text{greedy history}\longrightarrow\text{auxiliary pairs}\longrightarrow\text{zone core}\\
 \longrightarrow\text{forest replacement}\longrightarrow\text{stride accounting}.
 \end{gathered}
\]

\paragraph{1. Cost batches, the classical $O(\log^2 \NumPairs)$ girth argument, and how to beat it (\FullRef{Section~\ref*{sec:upper-objects}}{sec:upper-objects}).}
Any terminal pair with $\Pay{i}<\OPT/\NumPairs$ is cheap: even $\NumPairs$ such
pairs pay less than $\NumPairs\cdot(\OPT/\NumPairs)=\OPT$ in total. Since no
single greedy step ever pays more than $\OPT$, all remaining pairs have payments
in $[\OPT/\NumPairs,\OPT]$. We partition them into $O(\log \NumPairs)$ doubling
intervals by defining the \emph{cost batches}
$\Batch{b}=\{i:\BatchScale{b}\le\Pay{i}<2\BatchScale{b}\}$, where we define the
\emph{batch scale} $\Scale=\BatchScale{b}=2^b\OPT/\NumPairs$. Within a single
batch $\CurrentBatch$ of scale $\Scale$, every pair pays $\Theta(\Scale)$.

Why does the classical analysis of greedy~\cite{AAB04} give
$O(\OPT\log^2 \NumPairs)$ across these $O(\log \NumPairs)$ batches? The
classical argument compares every batch $\CurrentBatch$ against the same
\emph{static} optimal forest $\Fopt$. Adding a zero-cost edge between
$\Source{i}$ and $\Target{i}$ for each pair $i\in\CurrentBatch$ to $\Fopt$
creates cycles whose $\Fopt$-edges must have total cost at least $\Scale$: right
before the latest pair $i$ on a cycle arrived, all earlier pairs on the cycle
had residual distance $0$, so the $\Fopt$-edges on the cycle formed a walk
between $\Source{i}$ and $\Target{i}$ of residual cost at least
$\Residual{i}(\Source{i},\Target{i})=\Pay{i}\ge\Scale$. By the classical Moore
bound (or cycle-rank girth bound), a graph of total edge cost $\OPT$ with minimum
cycle length $\Omega(\Scale)$ supports at most
$O\bigl(\frac{\OPT}{\Scale}\log \NumPairs\bigr)$ added edges, bounding a single
batch's payment by $O(\OPT\log \NumPairs)$ and the sum over all $O(\log \NumPairs)$
scales by $O(\OPT\log^2 \NumPairs)$.

To improve $O(\log^2 \NumPairs)$ to $O(\log \NumPairs\log\log \NumPairs)$,
observe where the $O(\log \NumPairs)$ girth factor in a single batch actually
comes from. When $m_0$ zero-cost edges are added to a tree of cost $S$ without
creating cycles shorter than $\Omega(\Scale)$, the girth bound states that
$m_0\Scale=O(S\log(2m_0))$: the logarithmic overhead is $O(\log(2m_0))$ (the
logarithm of the number of edges added to that local piece) rather than
$O(\log \NumPairs)$! This suggests a dichotomy between sparse/small regions and
dense regions of the optimal trees:
\begin{itemize}[leftmargin=*,nosep]
\item \emph{Small or sparsely hit regions of the optimal trees:} If a connected
piece of an optimal tree has cost $S=\operatorname{poly}(\log \NumPairs)\,\Scale$,
then at most $m_0=\operatorname{poly}(\log \NumPairs)$ batch pairs can fall
inside it. Since $\log(2m_0)=O(\log\log \NumPairs)$, those pairs pay at most
$O(\log\log \NumPairs)\,S$ in total, losing only a $\log\log \NumPairs$ factor.
\item \emph{Densely hit regions of the optimal trees:} If an expensive region of
an optimal tree of cost $S\gg\operatorname{poly}(\log \NumPairs)\,\Scale$ is hit
by $m_0$ batch pairs with $\log m_0=\Theta(\log \NumPairs)$, those pairs pay
$O(\log \NumPairs)\,S$, which we cannot afford to charge to $S$ across multiple
scales. However, precisely because greedy has just connected all $m_0$ of those
pairs at scale $\Scale$, the edges greedy purchased form a dense network of
cheap connections across that region! If we use greedy's purchased edges to
\emph{surgically replace} that expensive region of cost $S$ by a cheaper
subgraph of cost at most $S/2$, we permanently reduce the cost of our comparison
forest by $\ForestSaving\ge S/2$ and charge the $O(\log \NumPairs)\,S$ payment
directly to $O(\log \NumPairs)\,\ForestSaving$.
\end{itemize}
Motivated by this dichotomy, instead of comparing every batch against the static
optimal forest $\Fopt$, we define an evolving \emph{reference forest} $\Fref$
(initialized to $\Fopt$) as a forest of original edges of $\InputGraph$ that
continues to connect the remaining terminal pairs while its cost $\costref$
decreases by $\ForestSaving=\cost(\Fref_{\mathrm{old}})-\cost(\Fref_{\mathrm{new}})\ge 0$
whenever a dense cluster of batch connections allows a profitable surgical
replacement. For a batch $\CurrentBatch$ of scale $\Scale$ and current reference
forest $\Fref$, we define a connected component of $\Fref$ to be a \emph{small
tree} if its cost is less than $2^{16}\LogPairs^3\Scale$ and a \emph{large tree}
otherwise, and we define the \emph{small pairs}
$\SmallPairs\subseteq\CurrentBatch$ and \emph{large pairs}
$\LargePairs\subseteq\CurrentBatch$ to be the batch pairs whose endpoints lie in
a small tree or a large tree of $\Fref$, respectively. For $\SmallPairs$, each
small tree already has cost $\operatorname{poly}(\log\NumPairs)\,\Scale$, so the
local girth bound directly charges $\SmallPairs$ to $\costref$ with only an
$O(\log\log\NumPairs)$ overhead. Turning the surgical replacement idea on the
large-tree pairs $\LargePairs$ into a rigorous proof requires overcoming three
obstacles, addressed in the next three subsections of
\FullRef{Section~\ref*{sec:upper}}{sec:upper}.

\paragraph{2. Obstacle 1: Zero-cost reuse in greedy's forest and lifetime trees (\FullRef{Section~\ref*{sec:history-idea}}{sec:history-idea}).}
To use a large pair $i\in\LargePairs$ (which pays $\Pay{i}=\Theta(\Scale)$) to
replace an expensive piece of $\Fref$, we need a path between its endpoints
whose \emph{original} edge cost in $\InputGraph$ is small, say
$O(\Scale\log^3\NumPairs)$. The first obstacle is that $\Pay{i}=\Theta(\Scale)$
only bounds the newly purchased edges $\Bought{i}$: greedy's path between
$\Source{i}$ and $\Target{i}$ may traverse a preexisting connected component of
previously purchased edges (of residual cost $0$ at step $i$) whose original
edge cost in $\InputGraph$ is arbitrarily larger than $\Scale$.

How can we extract short-original-cost paths when greedy reuses a massive
preexisting component? The key insight is that if many pairs in the current
batch touch the same preexisting greedy component, their attachment points must
crowd close to one another along that component! To measure proximity inside
greedy's growing components without double-counting edges across merges, we
track the merger history of the connected components of the forest
$\bigcup_{j\le i}\Bought{j}$ of edges purchased by greedy:
\begin{itemize}[leftmargin=*,nosep]
\item Whenever $\Bought{i}$ creates a brand-new connected component (touching no
previously purchased edges), we assign it a fresh integer label $\ell$.
\item Whenever $\Bought{i}$ merges two or more existing greedy components, the
single heaviest participating component (by current original edge cost) keeps
its integer label $\ell$ and absorbs $\Bought{i}$ together with all lighter
components, while every lighter merged component \emph{retires} its label.
\item For each component label $\ell$, we define its \emph{lifetime tree}
$\LifeTree{\ell}$ to be a spanning tree of all edges ever held by component
$\ell$ from its birth until it retires (or the end of the execution).
\end{itemize}
Whenever a component retires in a merger, the winning component is at least as
heavy, so the total cost of the component containing its edges at least doubles.
A purchase $\Bought{i}$ can therefore belong to at most
$1+\log_2(\ALG/\Pay{i})$ lifetime trees. Weighting this count by $\Pay{i}$ and
using that the normalized positive payments have entropy at most
$\log_2\NumPairs$ gives
$\sum_{\ell}\cost(\LifeTree{\ell})\le\ALG(1+\log_2\NumPairs)$
(\FullRef{Lemma~\ref*{lem:mass}}{lem:mass}). This is a weighted aggregate bound;
an individual edge need not belong to only logarithmically many lifetime trees.

For each large pair $i\in\LargePairs$, we define its \emph{greedy component
label} $\PairLabel{i}$ to be the label of the heaviest greedy component touched
by $\Bought{i}$ before step $i$ (or the newly created label if $\Bought{i}$
touches none), so $\Bought{i}$ and all components it touches at step $i$ belong
to $\LifeTree{\PairLabel{i}}$. Birth pairs always use their own purchased path,
of original cost $\Pay{i}<2\Scale$, as $\AuxPath{i}$, with auxiliary pair
$\TerminalPair{i}$. For every nonbirth pair, define $\Far{i}$ to be its endpoint
farther in residual distance from the old surviving component. We distinguish
two nonbirth cases, as in
\FullRef{Algorithm~\ref*{alg:extract-aux}}{alg:extract-aux}:
\begin{enumerate}[label=\textnormal{\arabic*.},leftmargin=*,nosep]
\item \emph{Cheap lifetime tree ($\cost(\LifeTree{\PairLabel{i}})\le2^6\LogPairs^2\Pay{i}$):}
Here $\Source{i}$ and $\Target{i}$ are already connected in
$\LifeTree{\PairLabel{i}}$ by a path of original cost
$O(\Scale\log^2\NumPairs)\le O(\Scale\log^3\NumPairs)$. We therefore define the
\emph{auxiliary pair} $(\AuxLeft{i},\AuxRight{i})=(\Source{i},\Target{i})$ and
define its \emph{realizing path} $\AuxPath{i}$ to be the unique path between
$\Source{i}$ and $\Target{i}$ in $\LifeTree{\PairLabel{i}}$.
\item \emph{Expensive lifetime tree ($\cost(\LifeTree{\PairLabel{i}})>2^6\LogPairs^2\Pay{i}$):}
When $\LifeTree{\PairLabel{i}}$ is too expensive to route between $\Source{i}$
and $\Target{i}$ directly, observe that our girth and surgery arguments on
$\Fref$ do not actually require the endpoints $\TerminalPair{i}$ themselves: all
we need from step $i$ is \emph{some} pair of vertices $(\AuxLeft{i},\AuxRight{i})$
in the large trees of $\Fref$ that was separated by residual distance
$\ge\Scale/2$ right before step $i$, connected to residual distance $0$ right
after step $i$, and joined in $\InputGraph$ by a cheap original-cost path
$\AuxPath{i}$. To find such a pair, note that since
$\Residual{i}(\Source{i},\Target{i})=\Pay{i}\ge\Scale$ and all vertices already
in component $\PairLabel{i}$ before step $i$ have residual distance $0$ from one
another, $\Source{i}$ and $\Target{i}$ cannot both lie within residual distance
$<\Scale/2$ of component $\PairLabel{i}$ before step $i$. Thus at least one
endpoint in $\{\Source{i},\Target{i}\}$, its \emph{far endpoint} $\Far{i}$,
lies at residual distance at least $\Scale/2$ from the
entire preexisting component $\PairLabel{i}$ before step $i$, and is absorbed
into $\LifeTree{\PairLabel{i}}$ at step $i$. If $i$ is not the first nonbirth large pair
with label $\PairLabel{i}$ in the batch, we pair $\Far{i}$ with the nearest
\emph{earlier} far endpoint $\Far{j}$ of a nonbirth large pair ($j<i$) from the same batch that shares
label $\PairLabel{j}=\PairLabel{i}$ (minimizing tree distance in
$\LifeTree{\PairLabel{i}}$), defining the substitute \emph{auxiliary pair}
$(\AuxLeft{i},\AuxRight{i})=(\Far{i},\Far{j})$ and defining its \emph{realizing
path} $\AuxPath{i}$ to be the unique path between $\Far{i}$ and $\Far{j}$ in
$\LifeTree{\PairLabel{i}}$. Because $\Far{j}$ already belonged to component
$\PairLabel{i}$ by step $j\le i-1$ whereas $\Far{i}$ was at residual distance
$\ge\Scale/2$ from all of component $\PairLabel{i}$ before step $i$, we have
$\Residual{i}(\Far{i},\Far{j})\ge\Scale/2$ before step $i$ and
$\Residual{i+1}(\Far{i},\Far{j})=0$ after step $i$.
\end{enumerate}

By an Euler-tour packing argument on $\LifeTree{\ell}$
(\FullRef{Lemma~\ref*{lem:nn}}{lem:nn}),
the sum of distances from $m_0$ sequential far endpoints in $\LifeTree{\ell}$ to
their nearest predecessors is at most
$2\cost(\LifeTree{\ell})\ln m_0=O(\cost(\LifeTree{\ell})\log \NumPairs)$. In
each batch, we set aside two corner-case subsets of large pairs with expensive
lifetime trees: we define the \emph{first-pair set} $\FirstPairs$ to consist of
the first nonbirth large pair with each label $\ell$ in the batch when its
lifetime tree is expensive (which has no earlier nonbirth partner $j<i$), and
we define the \emph{long-distance set} $\LongPairs$ to consist of the remaining
expensive-lifetime pairs whose nearest earlier nonbirth large partner in
$\LifeTree{\PairLabel{i}}$ is at tree distance greater than
$2^8\LogPairs^3\Scale$. In each batch, the single
pair in $\FirstPairs$ for label $\ell$ pays
$O(\Scale)=O\bigl(\cost(\LifeTree{\ell})/\log^2\NumPairs\bigr)$, and by Markov's
inequality on the $O(\cost(\LifeTree{\ell})\log \NumPairs)$ nearest-predecessor
distance sum, the pairs in $\LongPairs$ with label $\ell$ also pay at most
$O\bigl(\cost(\LifeTree{\ell})/\log^2\NumPairs\bigr)$. Summing over all
$O(\log \NumPairs)$ batches and all labels $\ell$ bounds the total payment of
$\FirstPairs\cup\LongPairs$ across the entire execution by
$O\bigl(\frac{1}{\log \NumPairs}\sum_{\ell}\cost(\LifeTree{\ell})\bigr)=O(\ALG)$,
which with our choice of constants is at most $\ALG/32$
(\FullRef{Lemma~\ref*{lem:excluded-payments}}{lem:excluded-payments}). Every remaining large
pair in the batch, which we define as a \emph{retained pair} in $\Retained$,
supplies an auxiliary pair $(\AuxLeft{i},\AuxRight{i})$ in the large trees of
$\Fref$ with $\Residual{i}(\AuxLeft{i},\AuxRight{i})\ge\Scale/2$,
$\Residual{i+1}(\AuxLeft{i},\AuxRight{i})=0$, and a realizing path $\AuxPath{i}$
in $\InputGraph$ of original cost at most $2^8\LogPairs^3\Scale$.

\paragraph{3. Obstacle 2: Charging small trees and peeling sparse zones at $O(\log\log \NumPairs)$ overhead (\FullRef{Sections~\ref*{sec:construction}}{sec:construction} and~\FullRef{\ref*{sec:replacement}}{sec:replacement}).}
With $\FirstPairs\cup\LongPairs$ absorbed into $\ALG/32$, it remains in each
batch $\CurrentBatch$ of scale $\Scale$ to bound the payments of the small pairs
$\SmallPairs$ and the retained large pairs $\Retained$, where we define
$\CountAux=|\Retained|$ to be the number of retained large pairs. Both bounds
rely on a self-contained weighted cycle-rank bound: in any multigraph for which
we define $\TestCost$ to be the total positive edge cost,
$\CycleRank=|E|-|V|+\text{components}\ge 1$ to be the \emph{cycle rank}, and
$\Sep>0$ to be a cycle-cost threshold such that every cycle has cost at least
$\Sep$, one has $\CycleRank\Sep=O(\TestCost\log(2\CycleRank))$
(\FullRef{Lemma~\ref*{lem:girth}}{lem:girth}). Pruning bridges and suppressing
degree-two vertices preserves cycle rank. Choose a component whose ratio of
cost to cycle rank is at most $\TestCost/\CycleRank$. A pure-cycle component
is handled directly; otherwise the component has minimum degree at least $3$
and at most $2\CycleRank$ vertices. A sufficiently long
$\Theta(\log(2\CycleRank))$-step nonbacktracking random walk has expected cost
$O\bigl(\frac{\TestCost}{\CycleRank}\log(2\CycleRank)\bigr)$ and enough possible
continuations to force two short walks to collide, yielding a cycle of that cost.

Applying this cycle-rank bound to each \emph{small tree} $T_{\mathrm{small}}$ of
$\Fref$ (which has $\cost(T_{\mathrm{small}})=O(\Scale\log^3\NumPairs)$) with
$m_0\ge 1$ small pairs from $\SmallPairs$ added as zero-cost edges gives
$m_0\Scale=O(\cost(T_{\mathrm{small}})\log(2m_0))$. Since
$\cost(T_{\mathrm{small}})/\Scale=O(\log^3\NumPairs)$, this implies
$m_0=O(\log^4\NumPairs)$ and hence $\log(2m_0)=O(\log\log \NumPairs)$, so the
small pairs in $T_{\mathrm{small}}$ pay at most
$O(\log\log \NumPairs)\,\cost(T_{\mathrm{small}})$ and all small pairs together
pay $\sum_{i\in\SmallPairs}\Pay{i}=O(\log\log \NumPairs)\,\costref$.

For the $\CountAux=|\Retained|$ retained large pairs, a large tree of $\Fref$
may have huge total cost while the auxiliary pairs only cluster in a small part
of it. To localize our analysis and surgery to uniform-cost units, we define a
\emph{zone decomposition} by partitioning each large tree bottom-up (allowing
edge subdivisions) into connected subtrees called \emph{zones}, each of cost in
$[2^{16}\LogPairs^3\Scale,3\cdot2^{16}\LogPairs^3\Scale]$. Thus every realizing
path costs at most $2^{-8}$ times the minimum zone cost. Because the zones
are edge-disjoint, the total number of zones across all large trees of $\Fref$
is at most $O\bigl(\frac{\costref}{\Scale\log^3\NumPairs}\bigr)$. We then define
the \emph{charging graph} $\ChargingGraph$ by taking disjoint copies of all
zones (disconnecting shared boundary vertices between zones so that paths can
only move between zones via auxiliary edges) and adding a zero-cost
\emph{auxiliary edge} $\AuxEdge{i}$ between the zone copies of $\AuxLeft{i}$ and
$\AuxRight{i}$ for each $i\in\Retained$. Every cycle $C$ in $\ChargingGraph$ has tree-edge cost
at least $\Scale/2$: if $\AuxEdge{i}$ is the highest-index auxiliary edge on $C$,
every other auxiliary edge $\AuxEdge{j}$ ($j<i$) on $C$ had
$\Residual{i}(\AuxLeft{j},\AuxRight{j})=0$ before step $i$, so projecting
$C\setminus\{\AuxEdge{i}\}$ to $\InputGraph$ gives a walk between $\AuxLeft{i}$
and $\AuxRight{i}$ of residual cost at least
$\Residual{i}(\AuxLeft{i},\AuxRight{i})\ge\Scale/2$.

We next define the \emph{zone multigraph} $\ZoneGraph$ by contracting each zone
copy in $\ChargingGraph$ to a single vertex while keeping the $\CountAux$
auxiliary edges. A single zone with $\ParallelCount\ge1$ self-loops has cycle
rank $\ParallelCount$, while two zones with $\ParallelCount\ge2$ parallel
auxiliary edges have cycle rank $\ParallelCount-1$. In both cases the
tree-edge cost is $O(\Scale\log^3\NumPairs)$, so the cycle-rank bound gives
$\ParallelCount\Scale=O(\Scale\log^3\NumPairs\cdot\log(2\ParallelCount))$.
The one-edge parallel bundle is trivial. These bounds let
us define a \emph{multiplicity cap}
$\ParallelCap=\Theta(\log^3\NumPairs\log\log \NumPairs)$ bounding every
self-loop bundle and parallel edge bundle in $\ZoneGraph$. Replacing zones in
$\Fref$ will only be profitable where zones form a high-degree cluster (so
boundary-reconnection costs and retained zone trees can be amortized over many
distinct neighboring zones). To isolate these dense clusters, we define the
\emph{dense zone core} $\CoreGraph$ by deleting all self-loops from $\ZoneGraph$
and repeatedly peeling away any vertex of degree strictly less than a \emph{core
degree threshold}
$\CoreDegree=\Theta(\ParallelCap)=\Theta(\log^3\NumPairs\log\log \NumPairs)$,
which we define to be a large constant multiple of $\ParallelCap$. Each zone
loses at most $\ParallelCap$ self-loops and fewer than $\CoreDegree$ incident
edges when peeled, so at most
$\ParallelCap+\CoreDegree=O(\log^3\NumPairs\log\log \NumPairs)$ auxiliary edges
are discarded per zone. Multiplying by the number of zones
$O\bigl(\frac{\costref}{\Scale\log^3\NumPairs}\bigr)$ cancels the
$\log^3\NumPairs$ factor, so the total greedy payment of all retained pairs
discarded outside $\CoreGraph$ is at most $O(\log\log \NumPairs)\,\costref$.

\paragraph{4. Obstacle 3: Forest surgery on the dense zone core (\FullRef{Section~\ref*{sec:replacement}}{sec:replacement}).}
For the dense zone core $\CoreGraph$, we define $\CountCore$ to be its number of
core zones, $\CountCoreEdges$ to be its number of core auxiliary edges,
$\CountCoreComp$ to be its number of connected components, and
$\CostCore=\Theta(\CountCore\Scale\log^3\NumPairs)$ to be the total original
tree cost of its $\CountCore$ core zones. Because $\CoreGraph$ has no
self-loops, minimum degree $\CoreDegree$ equal to a large constant multiple of
$\ParallelCap$, and at most $\ParallelCap$ parallel edges between any two
vertices, every core zone has a large constant number $\CoreDegree/\ParallelCap$
of \emph{distinct} neighboring core zones, so $\CountCoreComp$ is at most a
small constant fraction of $\CountCore$. Moreover, since $\CoreGraph$ has
minimum degree $\CoreDegree\ge 4$, we define its cycle rank
$\CoreRank=\CountCoreEdges-\CountCore+\CountCoreComp\ge\CountCoreEdges/2$;
applying the cycle-rank bound to the subgraph of $\ChargingGraph$ induced by the
$\CountCore$ core zones yields
$\CostCore=\Omega(\CountCoreEdges\Scale/\log \NumPairs)$.

We now surgically replace all $\CountCore$ core zones in $\Fref$ by a subgraph
of $\InputGraph$ of cost at most
$\CostCore/2+O(\CountMarks\Scale\log^3\NumPairs)$ while preserving all connections
in $\Fref$ among a given set $\Marks\subseteq\InputVertices$ of
$\CountMarks=|\Marks|$ vertices that we define as \emph{marked vertices}
(representing endpoints of terminal pairs in later batches):
\begin{enumerate}[label=\textnormal{\arabic*.},leftmargin=*,nosep]
\item \emph{Replacement skeleton $S_{\mathrm{skel}}$ across each component of $\CoreGraph$:}
Why can we not simply replace each component of $\CoreGraph$ by a spanning tree
of realizing paths $\AuxPath{i}$? Because each vertex of $\CoreGraph$ is an
entire zone $\Zone$: two realizing paths $\AuxPath{i},\AuxPath{j}$ incident to
the same zone $\Zone$ may land at different vertices inside $\Zone$ and fail to
connect unless we also keep $\Zone$'s tree, yet keeping all $\CountCore$ zone
trees would save nothing! To keep only a tiny fraction of zone trees, we exploit
the fact that every core zone in $\CoreGraph$ has a large constant number of
distinct neighbors: random sampling selects a dominating set
$\mathcal{Z}_{\mathrm{cen}}\subseteq V(\CoreGraph)$ of \emph{center zones} whose
size $|\mathcal{Z}_{\mathrm{cen}}|$ is a small constant fraction of $\CountCore$.
We hook each noncenter core zone to an adjacent center zone via one auxiliary
edge in $\CoreGraph$ (so each noncenter leaf zone has degree $1$ in its star and
touches only a single realizing path!) and link the stars in each component of
$\CoreGraph$ via $|\mathcal{Z}_{\mathrm{cen}}|-\CountCoreComp$ interstar
auxiliary edges. We then define the \emph{replacement skeleton}
$S_{\mathrm{skel}}\subseteq\InputEdges$ to consist of the realizing paths
$\AuxPath{i}$ of these $\CountCore-\CountCoreComp<\CountCore$ star and interstar
auxiliary edges together with the full zone trees of \emph{only} the
$|\mathcal{Z}_{\mathrm{cen}}|$ center zones and the at most
$2(|\mathcal{Z}_{\mathrm{cen}}|-\CountCoreComp)$ endpoint zones of the interstar
edges (at most $3|\mathcal{Z}_{\mathrm{cen}}|$ zone trees in total). Whenever
two selected realizing paths meet at a common core zone, that zone is a center
or interstar endpoint whose full tree is in $S_{\mathrm{skel}}$ and bridges
them, while every noncenter leaf zone is visited by the endpoint of its single
star-assignment realizing path. Since each realizing path costs a small constant
fraction of a zone and only $3|\mathcal{Z}_{\mathrm{cen}}|\ll\CountCore$ full
zone trees are kept, $\cost(S_{\mathrm{skel}})\le\CostCore/4$.
\item \emph{Restoring external connections and marks via the boundary-hub graph $H_{\mathrm{hub}}$ and attachment edge set $P_{\mathrm{att}}$:}
Deleting the internal edges of noncenter core zones could sever paths in $\Fref$
that pass through core zones on their way between external edges of $\Fref$ or
marked vertices in $\Marks$. To select a minimal set of boundary attachments, we
define the \emph{boundary-hub graph} $H_{\mathrm{hub}}$ from $\Fref$ by keeping
all edges of $\Fref$ outside the core zones, replacing each core zone $\Zone$ by
a zero-cost star joining a virtual hub $h_\Zone$ to the boundary vertices of
$\Zone$, and merging the virtual hubs of all core zones in the same connected
component of $\CoreGraph$ into a single merged hub (since $S_{\mathrm{skel}}$
already connects them). Within each component of $H_{\mathrm{hub}}$, we take an
inclusion-minimal tree $T_{\mathrm{hub}}$ spanning the merged hubs and the
outside marked vertices of $\Marks$. Because suppressing degree-two nonterminals
in a tree with $R$ required terminals leaves only branching vertices of degree
$\ge 3$, the total degree of the $\CountCoreComp$ merged hubs in
$T_{\mathrm{hub}}$ across all components is at most
$2(\CountCoreComp+\CountMarks)$. We define the \emph{attachment edge set}
$P_{\mathrm{att}}\subseteq\InputEdges$ by taking the unique path inside each
core zone $\Zone$ connecting each of these at most
$2(\CountCoreComp+\CountMarks)$ boundary exits, as well as each marked vertex of
$\Marks$ inside a core zone, to $S_{\mathrm{skel}}$. Since each attachment path
stays inside a single zone of cost $O(\Scale\log^3\NumPairs)$ and
$\CountCoreComp$ is a small constant fraction of $\CountCore$, we have
$\cost(P_{\mathrm{att}})\le\CostCore/4+O(\CountMarks\Scale\log^3\NumPairs)$.
\end{enumerate}
Removing the $\CountCore$ core zones from $\Fref$, inserting
$S_{\mathrm{skel}}\cup P_{\mathrm{att}}$, and breaking cycles (plus pruning
unused segments of subdivided edges) yields a new reference forest
$\Fref_{\mathrm{new}}$ that preserves all connections among $\Marks$ and
satisfies
$\cost(\Fref_{\mathrm{new}})\le\costref-\CostCore/2+O(\CountMarks\Scale\log^3\NumPairs)$.
Using $\CostCore=\Omega(\CountCoreEdges\Scale/\log \NumPairs)$ and
$\ForestSaving=\max\bigl(0,\costref-\cost(\Fref_{\mathrm{new}})\bigr)$, and
combining $\SmallPairs$, the discarded sparse-zone pairs
($\CountAux-\CountCoreEdges$), and the dense-core pairs ($\CountCoreEdges$) gives
the single-batch charging bound
(\FullRef{Lemma~\ref*{lem:batch-charging}}{lem:batch-charging}):
\[
 \sum_{i\in\SmallPairs\cup\Retained}\Pay{i}
 =O(\log \NumPairs)\,\ForestSaving
 +O(\log\log \NumPairs)\,\costref
 +O(\CountMarks\Scale\log^4\NumPairs).
\]

\paragraph{5. Strided batch classes and completing the upper bound (\FullRef{Section~\ref*{sec:summation}}{sec:summation}).}
Why can we not process all batches $b=0,1,2,\ldots$ in a single pass with
$\Marks$ equal to all terminal pairs in later batches $b'>b$? Because the very
next batch $b+1$ has scale $2\Scale$ and could contain
$\CountMarks=\Omega(\OPT/\Scale)$ pairs, making the mark overhead
$O(\CountMarks\Scale\log^4\NumPairs)=O(\OPT\log^4\NumPairs)\gg\OPT$. Instead, we
define a class stride $\ClassCount=\Theta(\log\log \NumPairs)$ so that
$2^{\ClassCount}=\Theta(\log^5\NumPairs)$ with a sufficiently large constant,
and we define $\ClassCount$ \emph{strided batch classes} by partitioning the
batch indices $b\in\{0,1,\ldots,\lfloor\log_2\NumPairs\rfloor\}$ according to
their residue $r=b\bmod\ClassCount\in\{0,1,\ldots,\ClassCount-1\}$. Within each
class $r$, we reset $\Fref\gets\Fopt$ at $t=0$ and process the batches
$b=r+t\ClassCount$ in increasing order of $t=0,1,2,\ldots$, where we define the
\emph{future marks} $\FutureMarks{r}{t}$ at step $(r,t)$ to be the original
endpoints of all pairs in strictly later batches $r+t'\ClassCount$ ($t'>t$) of
the same class $r$. Because every such future pair has greedy payment at least
$\Sep=2^{\ClassCount}\Scale$, applying the cycle-rank girth bound to $\Fref$
(as in \FullRef{Lemma~\ref*{lem:batch-accounting}}{lem:batch-accounting})
bounds their number by
$\CountMarks=O\bigl(\frac{\costref}{2^{\ClassCount}\Scale}\log \NumPairs\bigr)$,
so the mark overhead in batch $b$ collapses to
\[
 O(\CountMarks\Scale\log^4\NumPairs)
 =O\!\left(\frac{\log^5\NumPairs}{2^{\ClassCount}}\costref\right)
 \le\costref.
\]
Within each class $r$, the cost reductions
$\BatchSaving{b}=\cost(\Fref_{\mathrm{old}})-\cost(\Fref_{\mathrm{new}})\ge 0$
sum to at most $\OPT$, so $O(\log \NumPairs)\sum_{b\text{ in class }r}\BatchSaving{b}=O(\OPT\log \NumPairs)$
per class and $O(\OPT\log \NumPairs\log\log \NumPairs)$ across all
$\ClassCount=\Theta(\log\log \NumPairs)$ classes; meanwhile, the
$O(\log\log \NumPairs)\,\costref\le O(\log\log \NumPairs)\,\OPT$ term per batch
sums over all $O(\log \NumPairs)$ batches to
$O(\OPT\log \NumPairs\log\log \NumPairs)$. Adding the $<\OPT$ payment of pairs
below $\OPT/\NumPairs$ and the $\le\ALG/32$ payment of
$\FirstPairs\cup\LongPairs$ gives
$\ALG\le\ALG/32+O(\OPT\log \NumPairs\log\log \NumPairs)$, proving
\FullRef{Theorem~\ref*{thm:main}}{thm:main}.

\subsection{Lower Bound: Harmonically Priced Shortcut Edges with Future Edges Present}

To prove the matching $\Omega(\log \NumPairs\log\log \NumPairs)$ lower bound
(\FullRef{Theorem~\ref*{thm:lower}}{thm:lower}), we construct an infinite family
of instances on which a single unit-cost path $\LowerPath$ connects all terminal endpoints
at cost $\OPT\le\LowerVertices-1$, while greedy is forced to purchase the
shortcut edges of a large matching across a harmonic spectrum of prices without
ever purchasing an edge of $\LowerPath$.

\paragraph{1. Doubly exponential scale and harmonic shortcut prices (\FullRef{Section~\ref*{sec:lower}}{sec:lower}).}
Fix a sufficiently large integer parameter $\LowerLevel$, and define the doubly
exponential parameters $\LowerLogSize=2^{\LowerLevel}$,
$\LowerVertices=2^{\LowerLogSize}$, and $\LowerMatchingSize=\LowerVertices/2$.
Let $\LowerPath$ be the unit-cost path on vertices
$0,1,\ldots,\LowerVertices-1$, of total cost $\cost(\LowerPath)=\LowerVertices-1$.
We define an \emph{ordered perfect matching}
$\LowerMatching=(\LowerMatchingEdge{1},\ldots,\LowerMatchingEdge{\LowerMatchingSize})$
to be a perfect matching of $\{0,1,\ldots,\LowerVertices-1\}$ equipped with
distinct edge ranks $1,\ldots,\LowerMatchingSize$, and we define the edges
$\LowerMatchingEdge{i}$ of $\LowerMatching$ as \emph{shortcut edges}. The
online request sequence will present the endpoints of the retained shortcut
edges $\LowerMatchingEdge{i}$ in increasing order of rank $i$, with a decreasing
harmonic price schedule $\LowerPrice{i}=\Theta(\LowerLogSize\LowerMatchingSize/i)$
so that the sum of prices across ranks collects an extra
$\sum_i \frac{1}{i}=\Theta(\LowerLevel)=\Theta(\log\log\NumPairs)$ factor on top
of $\LowerLogSize=\Theta(\log\NumPairs)$.

However, we cannot start this harmonic price schedule at $i=1$: for very small
$i$, $\Theta(\LowerLogSize\LowerMatchingSize/i)$ would exceed the comparison-path
cost $\LowerVertices-1$
(and before enough earlier shortcut edges have been bought, an expensive request
could be bypassed using cheaper future shortcut edges of price $\Theta(\LowerLogSize)$).
We therefore skip a small prefix of ranks by defining a prefix fraction
$\LowerPrefixFraction=\Theta(\LowerLevel/\LowerLogSize)=\Theta(\LowerLevel/2^{\LowerLevel})$
and a starting rank cutoff
$\LowerFirstRank=\lceil\LowerPrefixFraction\LowerMatchingSize\rceil=\Theta(\LowerMatchingSize\LowerLevel/\LowerLogSize)$.
For ranks $i<\LowerFirstRank$ (which will be deleted from the final instance,
but which we pessimistically treat as free in our test graph), we set the
auxiliary price $\LowerPrice{i}=0$; for ranks $i\ge\LowerFirstRank$, we define
the rank fraction
$\LowerRankFraction=i/\LowerMatchingSize\in[\LowerPrefixFraction,1]$ and define
the harmonic auxiliary price
\[
 \LowerPrice{i}
 =\Theta\!\left(\frac{\LowerLogSize\LowerMatchingSize}{i}\right)
 =\Theta\!\left(\frac{\LowerLogSize}{\LowerRankFraction}\right).
\]
Because the ratio of the last rank to the first rank is
$\LowerMatchingSize/\LowerFirstRank=\Theta(1/\LowerPrefixFraction)=\Theta(2^{\LowerLevel}/\LowerLevel)$,
the harmonic sum over $i\in[\LowerFirstRank,\LowerMatchingSize]$ is
$\sum_{i=\LowerFirstRank}^{\LowerMatchingSize}\frac{1}{i}=\ln(\LowerMatchingSize/\LowerFirstRank)+O(1)=\Theta(\LowerLevel)$.
Consequently, we define the total offered shortcut price $\LowerTotalPrice$ as
\[
 \LowerTotalPrice
 =\sum_{i=\LowerFirstRank}^{\LowerMatchingSize}\LowerPrice{i}
 =\Theta(\LowerLogSize\LowerMatchingSize)\sum_{i=\LowerFirstRank}^{\LowerMatchingSize}\frac{1}{i}
 =\Theta(\LowerVertices\LowerLogSize\LowerLevel).
\]
If greedy is forced to pay $\LowerPrice{i}$ for a set of ranks
$i\ge\LowerFirstRank$ whose prices sum to at least $\LowerTotalPrice/2$, then
greedy pays $\ALG\ge\LowerTotalPrice/2=\Omega(\LowerVertices\LowerLogSize\LowerLevel)$,
whereas buying the edges of $\LowerPath$ gives a feasible solution of cost
$\LowerVertices-1$, so $\OPT\le\LowerVertices-1$. Each retained price is at most
$\LowerMaxPrice=O(\LowerLogSize^2/\LowerLevel)$, so retaining total price at
least $\LowerTotalPrice/2$ forces
$\NumPairs\ge\LowerTotalPrice/(2\LowerMaxPrice)
=\Omega(\LowerVertices\LowerLevel^2/\LowerLogSize)$, while
$\NumPairs\le\LowerVertices/2$. Thus $\log \NumPairs=\Theta(\LowerLogSize)$ and
$\log\log \NumPairs=\Theta(\LowerLevel)$. This yields
$\ALG/\OPT=\Omega(\LowerLogSize\LowerLevel)=\Omega(\log \NumPairs\log\log \NumPairs)$.

\paragraph{2. The online obstacle: future shortcut edges and the rank-$i$ test graph.}
Why is it nontrivial to force greedy to buy the shortcut edge
$\LowerMatchingEdge{i}$ at price $\LowerPrice{i}$ when terminal pair $i$
arrives? In the online Steiner Forest problem, the entire input graph
$\InputGraph$ (including all shortcut edges that will be requested at
\emph{future} steps $j>i$) is present and visible to greedy from step $1$. When
terminal pair $i$ arrives, greedy has already purchased the shortcut edges for
earlier requests $j<i$ (so those edges now have residual cost $0$), while all
future shortcut edges $j>i$ are available in $\InputGraph$ at their positive
prices $\LowerPrice{j}$, and all edges of $\LowerPath$ are available at cost
$1$. If the endpoints of $\LowerMatchingEdge{i}$ can be connected by an
alternate path of total cost $\le\LowerPrice{i}$ that mixes edges of
$\LowerPath$, free earlier shortcut edges ($j<i$), and priced future shortcut
edges ($j>i$), then greedy may bypass $\LowerMatchingEdge{i}$!

To rule out any such bypass, for each rank $i\ge\LowerFirstRank$ we define the
\emph{rank-$i$ test graph} on $\{0,1,\ldots,\LowerVertices-1\}$ by deleting
$\LowerMatchingEdge{i}$, assigning test cost $\LowerTestCost{j}=0$ to every
lower-ranked shortcut edge $\LowerMatchingEdge{j}$ ($j<i$), assigning test cost
$\LowerTestCost{j}=\LowerPrice{j}$ to every higher-ranked shortcut edge
$\LowerMatchingEdge{j}$ ($j>i$), and keeping cost $1$ on every edge of
$\LowerPath$. We define rank $i$ to be \emph{bad} if the shortest-path distance
between the endpoints of $\LowerMatchingEdge{i}$ in the rank-$i$ test graph is
at most $\LowerPrice{i}$, and \emph{good} if this distance is strictly greater
than $\LowerPrice{i}$. Deleting all prefix edges ($j<\LowerFirstRank$) and bad
edges in the final instance only removes edges from the rank-$i$ test graph and
thus can only increase the cost of every alternate path between the endpoints of
$\LowerMatchingEdge{i}$.

\paragraph{3. Lazy path exposure and exponential-tilt bound.}
To prove that there exists an ordered perfect matching $\LowerMatching$ whose
good ranks retain at least half of $\LowerTotalPrice$, we analyze a uniformly
random perfect matching $\LowerMatching$ equipped with an independent uniform
permutation of the ranks $1,\ldots,\LowerMatchingSize$, and show that for every
rank $i\ge\LowerFirstRank$ with $\LowerRankFraction=i/\LowerMatchingSize$,
$\Prb(i\textnormal{ is bad})=O(\LowerVertices^{-1/2}/\LowerRankFraction)$.

We prove the same probability bound uniformly for each fixed vertex
$u\in\{0,\ldots,\LowerVertices-1\}$, where we define $\LowerPartner{u}$ to be
the matching partner of $u$ in $\LowerMatching$ and we define
$\LowerDeletedEdge=\{u,\LowerPartner{u}\}$ to be the shortcut edge incident to
$u$, and condition on the event that $\LowerDeletedEdge$ is assigned rank $i$
(so $\LowerDeletedEdge=\LowerMatchingEdge{i}$). Crucially, under this
conditioning the \emph{unranked} matching $\LowerMatching$ remains uniformly
distributed, and \emph{independently} the remaining $\LowerMatchingSize-1$ ranks
$\{1,\ldots,\LowerMatchingSize\}\setminus\{i\}$ are uniformly distributed over
the remaining $\LowerMatchingSize-1$ edges of
$\LowerMatching\setminus\{\LowerDeletedEdge\}$. This decoupling allows us to
expose the geometry of the unranked matching $\LowerMatching$ first and the
ranks of the traversed shortcut edges second. Averaging the uniform bound over
endpoints of the rank-$i$ edge then gives the claimed bad-rank probability,
as in \FullRef{Lemma~\ref*{lem:lower-bad-probability}}{lem:lower-bad-probability}.

We define the test-cost budget
$\LowerRadius=\LowerPrice{i}=\Theta(\LowerLogSize/\LowerRankFraction)$. If
$\LowerDeletedEdge=\{u,\LowerPartner{u}\}$ is bad, the rank-$i$ test graph
contains a simple path from $u$ to $\LowerPartner{u}$ avoiding $\LowerDeletedEdge$
of total test cost at most $\LowerRadius$. Because $\LowerMatching$ is a
matching (so no two shortcut edges share an endpoint, and no shortcut edge in
$\LowerMatching\setminus\{\LowerDeletedEdge\}$ touches $u$ or
$\LowerPartner{u}$), any such simple path must begin and end with edges of
$\LowerPath$ and alternate between nonempty monotone runs along $\LowerPath$ and
traversals of distinct shortcut edges in
$\LowerMatching\setminus\{\LowerDeletedEdge\}$. We define
$\LowerShortcutCount\ge 0$ to be the number of shortcut-edge traversals on the
path (so the path has $\LowerShortcutCount+1$ nonempty monotone runs along
$\LowerPath$), and we define $\LowerPathSteps$ to be the total number of
unit-cost $\LowerPath$-steps across those runs (satisfying
$\LowerShortcutCount+1\le\LowerPathSteps\le\LowerRadius$). We define a
\emph{pattern} by specifying $\LowerShortcutCount$, the $\LowerShortcutCount+1$
positive integer run lengths along $\LowerPath$, and the left/right direction of
each run.

Fix one such pattern and trace the walk it prescribes starting from $u$,
\emph{exposing the random matching $\LowerMatching$ lazily}: we walk along
$\LowerPath$ according to each run without revealing any matching edges, and
expose the matching partner of a run's endpoint only when the pattern calls for
a shortcut-edge traversal (aborting if the walk steps off $\LowerPath$ or
repeats a vertex). If the walk completes all $\LowerShortcutCount+1$ runs and
$\LowerShortcutCount$ shortcut-edge traversals without aborting, ending at some
newly visited vertex $z$, then only $\LowerShortcutCount$ matching edges
($2\LowerShortcutCount$ vertices) have been revealed and none touches $u$ or
$z$. Conditional on this revealed history, the matching on the remaining
$\LowerVertices-2\LowerShortcutCount$ unexposed vertices is still uniform, so
$\LowerPartner{u}$ is equally likely to be any of the
$\LowerVertices-2\LowerShortcutCount-1$ unexposed vertices other than $u$:
\[
 \Prb\bigl(\LowerPartner{u}=z\mid\text{completed non-aborted exposure history}\bigr)
 =\frac{1}{\LowerVertices-2\LowerShortcutCount-1}
 =\Theta\!\left(\frac{1}{\LowerVertices}\right),
\]
since $\LowerShortcutCount\le\LowerRadius\le\LowerLogSize^2\ll\LowerVertices$.

We now combine this $\Theta(1/\LowerVertices)$ unranked endpoint-hitting
probability with an exponential moment (Chernoff/Markov) bound over the
independently assigned ranks of the $\LowerShortcutCount$ traversed shortcut
edges. For any walk of test cost at most $\LowerRadius$, its indicator satisfies
$\mathbf{1}\{\text{cost}\le\LowerRadius\}\le\e^{\LowerTilt\LowerRadius}\e^{-\LowerTilt\cdot\text{cost}}$
for any exponential tilt parameter $\LowerTilt>0$. How should we choose
$\LowerTilt$? Because a fraction $\approx\LowerRankFraction$ of all shortcut
edges are free ($j<i$), each traversed shortcut edge contributes expected weight
at least $\approx\LowerRankFraction$, so to prevent walks with more
shortcut-edge traversals $\LowerShortcutCount$ from blowing up, the sum over
each $\LowerPath$-run of length $t\ge 1$ must supply a damping factor
$O(1/\LowerTilt)<1/(2\LowerRankFraction)$, which forces
$\LowerTilt=\Omega(\LowerRankFraction)$. Remarkably, when we define
$\LowerTilt=\Theta(\LowerRankFraction)$ (with a sufficiently large leading
constant), the $\LowerRankFraction$ in $\LowerTilt$ cancels the
$1/\LowerRankFraction$ in the distance budget
$\LowerRadius=\LowerPrice{i}=\Theta(\LowerLogSize/\LowerRankFraction)$, so the
Markov factor $\e^{\LowerTilt\LowerRadius}=\e^{\Theta(\LowerLogSize)}$ is
independent of $\LowerRankFraction$ and can be kept $\le\LowerVertices^{1/2}$!
Specifically, for a fixed pattern with $\LowerPathSteps$ steps on $\LowerPath$
and $\LowerShortcutCount$ shortcut-edge traversals,
$\e^{-\LowerTilt\cdot\text{cost}}$ factors into $\e^{-\LowerTilt\LowerPathSteps}$
times $\prod_{a=1}^{\LowerShortcutCount}\e^{-\LowerTilt\LowerTestCost{J_a}}$,
where we define $J_1,\ldots,J_{\LowerShortcutCount}$ to be the ranks of the
$\LowerShortcutCount$ traversed shortcut edges:
\begin{itemize}[leftmargin=*,nosep]
\item \emph{Mean shortcut weight $\LowerMeanWeight$ and the prefix cutoff $\LowerFirstRank$:}
We define $\LowerMeanWeight=\frac{1}{\LowerMatchingSize-1}\sum_{j\ne i}\e^{-\LowerTilt\LowerTestCost{j}}$
to be the average of $\e^{-\LowerTilt\LowerTestCost{j}}$ across all
$\LowerMatchingSize-1$ ranks $j\ne i$. The $i-1$ earlier ranks $j<i$ have
$\LowerTestCost{j}=0$ and contribute $\frac{i-1}{\LowerMatchingSize-1}\le\LowerRankFraction$
to $\LowerMeanWeight$. Every future rank $j>i$ has price at least
$\LowerPrice{\LowerMatchingSize}=\Theta(\LowerLogSize)$; since
$i\ge\LowerFirstRank$ implies
$\LowerRankFraction\ge\LowerPrefixFraction=\Theta(\LowerLevel/\LowerLogSize)$,
we have $\LowerTilt\LowerTestCost{j}\ge\LowerTilt\LowerPrice{\LowerMatchingSize}=\Theta(\LowerRankFraction\LowerLogSize)=\Omega(\LowerLevel)$.
With a sufficiently large constant in $\LowerPrefixFraction$, every future
shortcut edge $j>i$ is damped by
$\e^{-\Omega(\LowerLevel)}\le 1/\LowerLogSize\ll\LowerRankFraction$, so
$\LowerMeanWeight\le 2\LowerRankFraction$.
\item \emph{Sampling without replacement:}
Conditioned on the unranked walk reaching $\LowerPartner{u}$, the ranks
$J_1,\ldots,J_{\LowerShortcutCount}$ are sampled \emph{without replacement} from
$\{1,\ldots,\LowerMatchingSize\}\setminus\{i\}$. By Maclaurin's inequality,
sampling without replacement is dominated by sampling with replacement:
$\E\bigl[\prod_{a=1}^{\LowerShortcutCount}\e^{-\LowerTilt\LowerTestCost{J_a}}\bigr]\le\LowerMeanWeight^{\LowerShortcutCount}\le(2\LowerRankFraction)^{\LowerShortcutCount}$.
\item \emph{Summing the geometric series over all patterns:}
We define $\LowerRunSum=\sum_{t\ge 1}2\e^{-\LowerTilt t}=\frac{2}{\e^{\LowerTilt}-1}\le\frac{2}{\LowerTilt}$
to be the single-run sum over all lengths $t\ge 1$ and both directions along
$\LowerPath$. Choosing the constant in $\LowerTilt=\Theta(\LowerRankFraction)$
large enough gives $\LowerRunSum\le\frac{1}{4\LowerRankFraction}$, so
$\LowerRunSum\LowerMeanWeight\le\frac{1}{2}<1$ and summing over all patterns
yields $\sum_{\LowerShortcutCount\ge 0}\LowerRunSum^{\LowerShortcutCount+1}\LowerMeanWeight^{\LowerShortcutCount}=O(1/\LowerRankFraction)$.
\end{itemize}
Finally, choosing the leading constant in
$\LowerPrice{i}=\Theta(\LowerLogSize/\LowerRankFraction)$ small enough relative
to $\LowerTilt=\Theta(\LowerRankFraction)$ so that
$\LowerTilt\LowerRadius\le\frac{1}{2}\ln 2\cdot\LowerLogSize$ (and hence
$\e^{\LowerTilt\LowerRadius}\le 2^{\LowerLogSize/2}=\LowerVertices^{1/2}$) gives
\[
 \Prb(i\textnormal{ is bad})
 =O\!\left(\frac{\e^{\LowerTilt\LowerRadius}}{\LowerVertices}\sum_{\LowerShortcutCount\ge 0}\LowerRunSum^{\LowerShortcutCount+1}\LowerMeanWeight^{\LowerShortcutCount}\right)
 =O\!\left(\frac{\LowerVertices^{-1/2}}{\LowerRankFraction}\right).
\]

\paragraph{4. Deterministic selection and tie-breaking-independent greedy execution.}
We define $\LowerBadPrice=\sum_{i\ge\LowerFirstRank:\,i\textnormal{ bad}}\LowerPrice{i}$
to be the total price lost to bad ranks. By linearity of expectation,
\[
 \E[\LowerBadPrice]
 =\sum_{i=\LowerFirstRank}^{\LowerMatchingSize}\LowerPrice{i}\,\Prb(i\textnormal{ is bad})
 =O(\LowerLogSize\LowerVertices^{-1/2}\LowerMatchingSize^2)\sum_{i=\LowerFirstRank}^{\LowerMatchingSize}\frac{1}{i^2}
 =O\!\left(\frac{\LowerLogSize\LowerVertices^{1/2}}{\LowerPrefixFraction}\right)
 <\frac{\LowerTotalPrice}{2},
\]
so there exists an ordered matching whose good ranks $i\ge\LowerFirstRank$ have
total price at least $\LowerTotalPrice/2$; enumerating ordered matchings
lexicographically and selecting the first such matching makes the construction
completely deterministic. To form the final simple graph $\InputGraph$ of
maximum degree $3$, we keep $\LowerPath$ and only the good shortcut edges of
rank $i\ge\LowerFirstRank$, subdividing each retained good shortcut edge
$\LowerMatchingEdge{i}=\{u,v\}$ with a private degree-two vertex
$\LowerSubdivision{i}$ into two edges $\{u,\LowerSubdivision{i}\}$ and
$\{\LowerSubdivision{i},v\}$ of cost $\LowerPrice{i}/2$ each, and present the
retained good pairs in increasing order of rank $i$. By induction on $i$, every
alternate path between $u$ and $v$ avoiding $\LowerSubdivision{i}$ in the actual
residual graph is also a path of the same cost in the rank-$i$ test graph and
thus has cost strictly greater than $\LowerPrice{i}$, so greedy uniquely buys
the two edges incident to $\LowerSubdivision{i}$ at cost $\LowerPrice{i}$
regardless of tie-breaking, paying
$\ALG\ge\LowerTotalPrice/2=\Omega(\OPT\log \NumPairs\log\log \NumPairs)$ and
establishing \FullRef{Theorem~\ref*{thm:lower}}{thm:lower} and
\FullRef{Corollary~\ref*{cor:tight}}{cor:tight}.
The formal proof of \FullRef{Corollary~\ref*{cor:tight}}{cor:tight} extends this
family to every sufficiently large request count by choosing a consecutive
parameter level and padding with repeated served pairs; the consecutive
levels change $\log\NumPairs$ and $\log\log\NumPairs$ by only constant factors.


%% file: preliminaries.tex
\section{Preliminaries}
\label{sec:prelim}
Let $\InputGraph=(\InputVertices,\InputEdges)$ be a finite undirected graph with strictly positive edge costs
$\cost:\InputEdges\to\PositiveReals$, and let $\TerminalPair{i}_{i=1}^{\NumPairs}$ be a sequence of
terminal pairs, where $\NumPairs\ge1$, such that the endpoints of every pair are
connected by a path in $\InputGraph$. For an edge set $\EdgeSet\subseteq\InputEdges$, write
$\cost(\EdgeSet)=\sum_{e\in \EdgeSet}\cost(e)$.

A \emph{subdivision} replaces an edge $e=uv$ by a $u$--$v$ path whose
internal vertices are new and have degree two. The new edges have strictly
positive costs summing to $\cost(e)$, so distances between original
vertices are preserved. Subdivision vertices are not terminal vertices.
Whenever we subdivide an edge for the analysis, all subgraphs and paths using
that edge are interpreted in the same subdivided graph; original vertices
and edges retain their meaning before subdivision.

For $1\le i\le \NumPairs$, write $\Bought{i}$ for the edges purchased by greedy
to connect the $i$th terminal pair. Let
$\BoughtThrough{i}=\bigcup_{j=1}^{i}\Bought{j}$ be the edges purchased through
step $i$, with $\BoughtThrough{0}=\varnothing$.
For $1\le i\le \NumPairs+1$, the residual cost of $e$ is zero if
$e\in \BoughtThrough{i-1}$ and $\cost(e)$ otherwise; let $\Residual{i}$ denote
shortest-path distance with these residual costs. For $1\le i\le \NumPairs$,
$\Residual{i}$ gives the residual distances before greedy connects the $i$th terminal pair,
and $\Residual{i+1}$ gives them afterward. Distinct vertices may have zero
residual distance. In particular, $\Residual{\NumPairs+1}$ gives the residual
distances after all terminal pairs have been connected.
Greedy buys the unpurchased edges of a shortest $\Source{i}$--$\Target{i}$ path and
pays $\Pay{i}=\Residual{i}\TerminalPair{i}$. Its total cost is
$\ALG=\sum_{i=1}^{\NumPairs} \Pay{i}=\cost(\BoughtThrough{\NumPairs})$.

Let $\Fopt$ be a minimum-cost forest connecting every terminal pair in the
complete sequence, and let $\OPT=\cost(\Fopt)$. We restrict attention to
the nontrivial case $\OPT>0$ (and hence $\ALG>0$); if $\OPT=0$, greedy
also pays zero.

All unqualified logarithms are natural. We use $\log_2$ explicitly where
base two is convenient. The shifted logarithms in \FullRef{Theorem~\ref*{thm:main}}{thm:main}
handle small $\NumPairs$. Terminal pairs with zero payment contribute nothing
to $\ALG$ and may remain in the execution.


%% file: upper_bound.tex
\section{The Upper Bound}
\label{sec:upper}

Fix the completed greedy execution on $\InputGraph=(\InputVertices,\InputEdges)$
and an optimal offline Steiner forest $\Fopt$ of cost $\OPT=\cost(\Fopt)$.
Our goal is to bound the total greedy expenditure
$\ALG=\sum_{i=1}^{\NumPairs}\Pay{i}$ by
$O(\OPT\log(2\NumPairs)\log\log(4\NumPairs))$.
Throughout this section, we write
\[
 \CycleFactor(n)=\lceil\log_2(8n)\rceil+1
 \qquad\text{and}\qquad
 \LogPairs=\CycleFactor(\NumPairs)=\lceil\log_2(8\NumPairs)\rceil+1=\Theta(\log(2\NumPairs)),
\]
so that $\log_2\LogPairs=\Theta(\log\log(4\NumPairs))$ and
$1+\ln\NumPairs\le1+\log_2\NumPairs\le\lceil\log_2\NumPairs\rceil+2\le\LogPairs$.

\subsection{Architecture of the Analysis and Cost Batches}
\label{sec:upper-objects}

Why does the classical analysis of greedy give $O(\log^2\NumPairs)$ rather than
$O(\log\NumPairs\log\log\NumPairs)$, and how do we improve it?
When terminal pair $\TerminalPair{i}$ arrives, greedy pays
$\Pay{i}=\Residual{i}\TerminalPair{i}$ to connect $\Source{i}$ and $\Target{i}$,
and afterward their residual distance is $\Residual{i+1}\TerminalPair{i}=0$.
Suppose we group greedy's payments into $O(\log\NumPairs)$ doubling cost scales
and consider a single scale $\Scale$, consisting of all pairs with
$\Pay{i}\in[\Scale,2\Scale)$. If we add a zero-cost edge between $\Source{i}$ and
$\Target{i}$ to $\Fopt$ for each such pair, every cycle created in $\Fopt$ must
contain $\Fopt$-edges of total cost at least $\Scale$: right before the last pair
on the cycle arrived, all earlier pairs on the cycle had already been connected
by greedy, so their endpoints were at residual distance zero. Because a graph of
total edge cost $\OPT$ with no cycles shorter than $\Scale$ can contain at most
$O(\frac{\OPT}{\Scale}\log\NumPairs)$ added edges, greedy can pay at most
$O(\OPT\log\NumPairs)$ at scale $\Scale$. Summing this bound over all
$O(\log\NumPairs)$ scales yields $O(\OPT\log^2\NumPairs)$.

The reason this classical argument loses a full $O(\log\NumPairs)$ factor at every
scale is that it compares every scale against the same static forest $\Fopt$.
Instead, we maintain an evolving forest $\Fref$ of original graph edges,
initialized to $\Fopt$, and update $\Fref$ as we process the scales. Whenever a
scale $\Scale$ has a dense cluster of greedy connections between expensive regions
of $\Fref$, we replace those regions of $\Fref$ by a much cheaper subgraph built
from the edges greedy already purchased. This replacement permanently reduces the
cost of $\Fref$ by a nonnegative saving
\[
 \ForestSaving=\cost(\Fref_{\mathrm{old}})-\cost(\Fref_{\mathrm{new}})\ge0,
\]
and because the cost of $\Fref$ starts at $\OPT$ and never becomes negative, the
sum of these cost reductions within each strided class, between resets, is at most $\OPT$. Across all $\ClassCount$ classes, the total is at most $\ClassCount\OPT$.

Turning this idea into a proof requires solving three concrete problems, which we
address one by one in the subsections that follow:
\begin{enumerate}[label=\textnormal{\arabic*.},leftmargin=*,nosep]
\item In greedy's execution, previously purchased edges are reused for free.
Thus, when greedy pays $\Pay{i}\in[\Scale,2\Scale)$ to connect $\Source{i}$ and
$\Target{i}$, the path between $\Source{i}$ and $\Target{i}$ in greedy's forest
may pass through a huge greedy component of previously purchased edges whose original edge cost
is much larger than $\Scale$, making it too expensive to insert into $\Fref$.
In \FullRef{Section~\ref*{sec:history-idea}}{sec:history-idea}, we track the growing components of greedy's
forest and replace such a pair $\TerminalPair{i}$ by a nearby pair of vertices
$(\AuxLeft{i},\AuxRight{i})$ from the same greedy component that was also
separated by residual distance at least $\Scale/2$ before step $i$, but is
connected in greedy's forest by a path $\AuxPath{i}$ of original cost at most
$2^8\LogPairs^3\Scale$. A small fraction of pairs have no nearby partner in their
greedy component; we show that their payments across the entire execution sum to
at most $\ALG/32$.
\item In \FullRef{Sections~\ref*{sec:construction}}{sec:construction} and~\ref{sec:replacement}, we use these
cheap greedy-forest paths $\AuxPath{i}$ (of original cost at most
$2^8\LogPairs^3\Scale$) to charge the remaining pairs at scale $\Scale$ to
$\Fref$. Trees of $\Fref$ costing less than $2^{16}\LogPairs^3\Scale$ are small:
each such tree has cost $\mathrm{poly}(\LogPairs)\,\Scale$, so the cycle-length
bound limits the number of pairs inside it with only an
$O(\log\LogPairs)=O(\log\log\NumPairs)$ overhead rather than $O(\log\NumPairs)$.
Trees of $\Fref$ costing at least $2^{16}\LogPairs^3\Scale$ are partitioned into
connected pieces (\emph{zones}) of uniform cost
$\Theta(\LogPairs^3\Scale)$. Sparse connections between zones are again charged
directly to $\costref$ with only an $O(\log\LogPairs)$ overhead, whereas densely
connected zones are replaced using the paths $\AuxPath{i}$ (each costing at most
$\frac{1}{2^8}$ of a zone), paying for those connections at $O(\LogPairs)$ times
the resulting cost reduction $\ForestSaving$.
\item Replacing zones of $\Fref$ at scale $\Scale$ must not disconnect terminal
pairs that belong to larger, not-yet-processed scales in the current strided class. When updating $\Fref$, we
therefore mark the $\CountMarks$ endpoints of those future pairs and add paths
inside the zones to preserve all existing connections between marked vertices, which adds
$O(\LogPairs^3)\,\CountMarks\Scale$ to the new forest (and hence an extra
$O(\LogPairs^4)\,\CountMarks\Scale$ to the $O(\LogPairs)\,\ForestSaving$
charging bound). In \FullRef{Section~\ref*{sec:summation}}{sec:summation}, we group the
scales into $\ClassCount=\Theta(\log\LogPairs)$ interleaved classes so that
consecutive scales within the same class jump by a factor of
$2^{\ClassCount}=\Theta(\LogPairs^5)$. Because every future pair in the same
class has payment at least $2^{\ClassCount}\Scale$, there are so few future
endpoints $\CountMarks$ that reconnecting them costs at most $\costref$.
\end{enumerate}

We now define the reference forest, the doubling cost batches, and the division of
each batch into small and large reference trees; we analyze a single batch
$\CurrentBatch$ of scale $\Scale>0$ before combining all batches at the end of
the section.

\begin{definition}[Reference forest]\label{def:reference-forest}
The \emph{reference forest} $\Fref$ is a forest of original edges of $\InputGraph$
maintained by the analysis. We initialize $\Fref=\Fopt$ and update it as the
analysis proceeds, preserving the terminal-pair connections still needed.
\end{definition}

Because $\Fopt$ connects every terminal pair, greedy pays $\Pay{i}\le\OPT$ for
every $1\le i\le\NumPairs$. Pairs with payment less than $\OPT/\NumPairs$ cost
less than $\OPT$ in total, and the remaining pairs fall into logarithmically many
doubling batches.

\begin{definition}[Cost batch]\label{def:batch}
The \emph{low-payment set} consists of all indices $i$ with
$\Pay{i}<\OPT/\NumPairs$. A \emph{cost batch} at scale $\Scale>0$ is a subset of
terminal-pair indices $\CurrentBatch\subseteq\{1,\ldots,\NumPairs\}$ whose greedy
payments satisfy
\[
 \Scale\le\Pay{i}<2\Scale
 \qquad\text{for all }i\in\CurrentBatch.
\]
Partitioning $[\OPT/\NumPairs,\OPT]$ into doubling intervals
$[2^b\OPT/\NumPairs,\,2^{b+1}\OPT/\NumPairs)$ for
$0\le b\le\lfloor\log_2\NumPairs\rfloor$ partitions all remaining indices into a
family $\BatchFamily$ of at most $\lfloor\log_2\NumPairs\rfloor+1\le\LogPairs$
cost batches.
\end{definition}

For a fixed batch $\CurrentBatch\in\BatchFamily$ of scale $\Scale$, we classify
the trees of the current reference forest $\Fref$ and the pairs of $\CurrentBatch$
using the scale threshold $2^{16}\LogPairs^3\Scale$.

\begin{definition}[Small and large reference trees]
\label{def:ref-tree}
For a cost batch $\CurrentBatch$ of scale $\Scale$, a tree $\EdgeSet$ of the
reference forest $\Fref$ is \emph{small} if
$\cost(\EdgeSet)<2^{16}\LogPairs^3\Scale$ and \emph{large} if
$\cost(\EdgeSet)\ge2^{16}\LogPairs^3\Scale$.
\end{definition}

\begin{definition}[Small and large pairs]\label{def:pair-sizes}
For a cost batch $\CurrentBatch$ of scale $\Scale$ whose terminal pairs are
connected in $\Fref$, we partition
\[
 \CurrentBatch=\SmallPairs\mathbin{\dot\cup}\LargePairs,
\]
where $\SmallPairs$ (\emph{small pairs}) contains the indices $i\in\CurrentBatch$
whose endpoints $\Source{i},\Target{i}$ lie in a small tree of $\Fref$, and
$\LargePairs$ (\emph{large pairs}) contains those whose endpoints lie in a large
tree of $\Fref$.
\end{definition}

\FloatBarrier
\subsection{Approaching Greedy: Component Lifetimes and Auxiliary Pairs}
\label{sec:history-idea}
\label{sec:auxiliary-idea}
\label{sec:lifetimes}
\label{sec:connections}

In a cost batch $\CurrentBatch$ of scale $\Scale$, the small terminal pairs
$\SmallPairs$ (whose endpoints lie in reference trees of $\Fref$ costing less
than $2^{16}\LogPairs^3\Scale$) will be charged directly to their reference
trees in \FullRef{Section~\ref*{sec:construction}}{sec:construction}. For the large terminal pairs
$\LargePairs$, we track the forest of edges purchased by greedy to set aside a small fraction
of corner-case terminal pairs whose payments sum to at most $\ALG/32$ across all
batches, and for every remaining terminal pair $i$ we extract an auxiliary vertex pair
$(\AuxLeft{i},\AuxRight{i})$ that:
\begin{enumerate}[label=\textnormal{(\roman*)},nosep]
\item is separated by residual distance at least $\Scale/2$ right before greedy
connects pair $i$, and connected by purchased edges right after; and
\item is connected in the original graph by a realizing path $\AuxPath{i}$
of original cost at most $2^8\LogPairs^3\Scale$.
\end{enumerate}
In \FullRef{Section~\ref*{sec:replacement}}{sec:replacement}, we divide these retained pairs according to
the density of their occurrences in the reference forest $\Fref$: those falling
in sparse parts of $\Fref$ are charged directly to $\costref$, while those in
dense parts are accounted for by replacing the expensive pieces of $\Fref$ with
the cheap paths $\AuxPath{i}$.

Because a residual shortest path never leaves and re-enters a greedy component of
already purchased edges through positive-cost edges, the edges purchased by greedy form a forest at
every step. As greedy components grow and merge, we assign each positive-cost
greedy component a persistent label so we can track which earlier terminal pairs
already reside in it. Crucially, whenever components merge, we keep the label of
the component with the maximum original cost; thus, every time a label is
retired, the merged component's cost at least doubles.

\begin{definition}[Greedy component labels]
\label{def:history}
Give each positive-cost greedy component a persistent label. When greedy
connects a terminal pair $i$ with $\Pay{i}>0$, examine the existing positive-cost
greedy components touched by the newly purchased edges $\Bought{i}$. If none are
touched, create a new label $\PairLabel{i}$ and call $i$ a \emph{birth pair}.
Otherwise, let $\PairLabel{i}$ be the label of a touched component of maximum
original cost (breaking ties in favor of the smaller label), assign
pair $i$ to $\PairLabel{i}$, and retire the labels of all other touched
components.
\end{definition}

A label names a growing greedy component rather than a fixed tree. To record the
final extent of that component, we store its largest tree before retirement (or
at the end of the execution), and for each nonbirth pair we identify which of
its two endpoints was farther from the surviving component before the merge
(\FullRef{Figure~\ref*{fig:component-history}}{fig:component-history}).

\begin{definition}[Lifetime tree and far endpoint]
\label{def:lifetime}
For each greedy component label $\ell$, its \emph{lifetime tree}
$\LifeTree{\ell}$ is the greedy tree carrying label $\ell$ immediately before
$\ell$ retires, or at the end of the execution if $\ell$ never retires.
For every nonbirth pair $i$ with $\Pay{i}>0$, let $\OldTree{i}$ be the
\emph{old surviving tree}: the greedy component carrying label $\PairLabel{i}$
immediately before step $i$. Its \emph{far endpoint}
$\Far{i}\in\{\Source{i},\Target{i}\}$ is the endpoint of pair $i$ maximizing
residual distance to $\OldTree{i}$ before step $i$:
\[
 \Far{i}\in\arg\max_{v\in\{\Source{i},\Target{i}\}}
 \min_{u\in V(\OldTree{i})}\Residual{i}(v,u).
\]
\end{definition}

\input{figures/component-history}

All earlier terminal pairs assigned to a label $\ell$ remain inside its growing
component while $\ell$ is alive. The next lemma shows that for any nonbirth pair
$i$, its far endpoint $\Far{i}$ is at residual distance at least $\Pay{i}/2$ from
every earlier terminal pair with the same label.

\begin{lemma}\label{lem:endpoint}
For every nonbirth pair $i$ with $\Pay{i}>0$, its far endpoint $\Far{i}$ satisfies
$\Residual{i}(\Far{i},\OldTree{i})\ge\Pay{i}/2$. Consequently, for every earlier
pair $j<i$ with $\PairLabel{j}=\PairLabel{i}$, both endpoints of pair $j$ lie in
$\OldTree{i}$, and either endpoint $w\in\{\Source{j},\Target{j}\}$ satisfies
\[
 \Residual{i}(\Far{i},w)\ge\Pay{i}/2,
 \qquad
 \Residual{i+1}(\Far{i},w)=0.
\]
\end{lemma}
\begin{proof}
Splitting greedy's step-$i$ residual shortest path at its first and last vertices
in $V(\OldTree{i})$ yields two endpoint-to-$\OldTree{i}$ segments whose step-$i$
residual costs sum to $\Pay{i}$ (since any subpath between two vertices of
$\OldTree{i}$ has residual cost $0$). Each segment is a shortest residual path
from its endpoint to $\OldTree{i}$, so the longer segment has residual cost at
least $\Pay{i}/2$, giving $\Residual{i}(\Far{i},\OldTree{i})\ge\Pay{i}/2$.
Every earlier pair $j<i$ with $\PairLabel{j}=\PairLabel{i}$ was connected into
the component carrying label $\PairLabel{i}$ by step $j\le i-1$, so its endpoints
belong to $V(\OldTree{i})$ before step $i$ and are connected to $\Far{i}$ by the
purchased edges $\BoughtThrough{i}$ after step $i$.
\end{proof}

\FullRef{Lemma~\ref*{lem:endpoint}}{lem:endpoint} lets us replace a large pair $i\in\LargePairs$ whose
lifetime tree $\LifeTree{\PairLabel{i}}$ is expensive by pairing its far endpoint
$\Far{i}$ with an earlier far endpoint $\Far{j}$ ($j<i$) carrying the same label:
their residual distance before step $i$ is still at least $\Pay{i}/2\ge\Scale/2$,
and they are connected inside $\LifeTree{\PairLabel{i}}$ after step $i$. We first
define the three properties that make a replacement vertex pair \emph{helpful} for
a large terminal pair $i\in\LargePairs$, and then specify how
\FullRef{Algorithm~\ref*{alg:extract-aux}}{alg:extract-aux} (\textsc{ExtractAuxiliaryPairs}) either sets $i$
aside as a corner case or assigns it an \emph{auxiliary pair}
$(\AuxLeft{i},\AuxRight{i})$ that is helpful.

\begin{definition}[Helpful pair and realizing path]
\label{def:auxiliary-pair}\label{def:realizing-path}
Fix a cost batch $\CurrentBatch$ of scale $\Scale$ and a reference forest
$\Fref$. For a large terminal pair $i\in\LargePairs$, a \emph{helpful pair} is a
pair of original vertices $(x,z)$ together with a path $\pi$ in $\InputGraph$ (its
\emph{realizing path}) satisfying:
\begin{enumerate}[label=\textnormal{(\roman*)},nosep]
\item $\Residual{i}(x,z)\ge\Scale/2$ and
$\Residual{i+1}(x,z)=0$.
\item $\cost(\pi)\le2^8\LogPairs^3\Scale$.
\item Both $x$ and $z$ belong to large trees of $\Fref$ (possibly
different large trees, each of cost at least $2^{16}\LogPairs^3\Scale$).
\end{enumerate}
\end{definition}

For each large terminal pair $i\in\LargePairs$, \FullRef{Algorithm~\ref*{alg:extract-aux}}{alg:extract-aux}
either sets $i$ aside as a corner case or assigns it an \emph{auxiliary pair}
$(\AuxLeft{i},\AuxRight{i})$ with realizing path $\AuxPath{i}$ by checking whether
its lifetime tree $\LifeTree{\PairLabel{i}}$ already supplies a short path between
$\Source{i}$ and $\Target{i}$:
\begin{itemize}[leftmargin=*,nosep]
\item If $i$ is a birth pair or $\cost(\LifeTree{\PairLabel{i}})\le2^6\LogPairs^2\Pay{i}$,
the terminal pair's own endpoints $\TerminalPair{i}$ are already connected in
$\LifeTree{\PairLabel{i}}$ (specifically, along $\Bought{i}$ if $i$ is a birth
pair) by a path of original cost at most
$2^6\LogPairs^2\Pay{i}<2^7\LogPairs^2\Scale\le2^8\LogPairs^3\Scale$, so we keep
$(\AuxLeft{i},\AuxRight{i})=\TerminalPair{i}$.
\item Otherwise ($i$ is a nonbirth terminal pair with
$\cost(\LifeTree{\PairLabel{i}})>2^6\LogPairs^2\Pay{i}$),
$\LifeTree{\PairLabel{i}}$ is too expensive to bound the
$\Source{i}$--$\Target{i}$ path directly. In this case,
\FullRef{Algorithm~\ref*{alg:extract-aux}}{alg:extract-aux} pairs the far endpoint $\Far{i}$ with the
closest earlier far endpoint $\Far{j}$ ($j<i$) among earlier nonbirth terminal pairs in
$\LargePairs$ with the same label $\PairLabel{j}=\PairLabel{i}$,
measuring original-cost tree distance $\Residual{\LifeTree{\PairLabel{i}}}$
along the lifetime tree $\LifeTree{\PairLabel{i}}$. Two corner cases are set
aside so that every retained pair has a partner within tree distance
$2^8\LogPairs^3\Scale$: $\FirstPairs$ collects any such terminal pair $i$ that is the first
nonbirth pair of its label $\PairLabel{i}$ in $\LargePairs$, and $\LongPairs$
collects those whose nearest earlier far endpoint $\Far{j}$ in $\LifeTree{\PairLabel{i}}$
is at tree distance $\ProxyLength{i}>2^8\LogPairs^3\Scale$. We denote the set of
remaining (\emph{retained}) large terminal pairs by
$\Retained=\LargePairs\setminus(\FirstPairs\cup\LongPairs)$.
\end{itemize}

\begin{algobox}[htbp]{Extract Auxiliary Pairs\label{alg:extract-aux}\label{def:proxy-parent}}
\KwIn{Batch $\CurrentBatch$ of scale $\Scale$, reference forest $\Fref$, greedy component labels $\PairLabel{i}$, lifetime trees $\LifeTree{\ell}$, and far endpoints $\Far{i}$.}
\KwOut{Partition $\CurrentBatch=\SmallPairs\mathbin{\dot\cup}\FirstPairs\mathbin{\dot\cup}\LongPairs\mathbin{\dot\cup}\Retained$, and auxiliary pairs $\{(\AuxLeft{i},\AuxRight{i},\AuxPath{i})\}_{i\in\Retained}$.}
\AlgoProc{ExtractAuxiliaryPairs}{\CurrentBatch,\Scale,\Fref,\{\PairLabel{i},\LifeTree{\ell},\Far{i}\}}
\AlgoLine \AlgoInd{1}$\SmallPairs\gets\{i\in\CurrentBatch:\Source{i},\Target{i}\text{ lie in a tree of }\Fref\text{ of cost }<2^{16}\LogPairs^3\Scale\}$,\quad $\LargePairs\gets\CurrentBatch\setminus\SmallPairs$
\AlgoLine \AlgoInd{1}$\FirstPairs\gets\varnothing$,\quad $\LongPairs\gets\varnothing$,\quad $\Retained\gets\varnothing$
\AlgoLine \AlgoInd{1}\textbf{for each} $i\in\LargePairs$ \textbf{in increasing order do}
\AlgoLine \AlgoInd{2}\textbf{if} $i$ is a birth pair \textbf{or} $\cost(\LifeTree{\PairLabel{i}})\le2^6\LogPairs^2\Pay{i}$ \textbf{then}
\AlgoLine\label{line:extract-direct} \AlgoInd{3}$\Retained\gets\Retained\cup\{i\}$,\quad $(\AuxLeft{i},\AuxRight{i})\gets\TerminalPair{i}$,\quad $\AuxPath{i}\gets\text{unique }\Source{i}\text{--}\Target{i}\text{ path in }\LifeTree{\PairLabel{i}}$
\AlgoLine\label{line:extract-else} \AlgoInd{2}\textbf{else}\AlgoComment{Nonbirth pair with $\cost(\LifeTree{\PairLabel{i}})>2^6\LogPairs^2\Pay{i}$}
\AlgoLine \AlgoInd{3}\textbf{if} no nonbirth $j<i$ in $\LargePairs$ has $\PairLabel{j}=\PairLabel{i}$ \textbf{then}
\AlgoLine\label{line:extract-first} \AlgoInd{4}$\FirstPairs\gets\FirstPairs\cup\{i\}$
\AlgoLine \AlgoInd{3}\textbf{else}
\AlgoLine \AlgoInd{4}$j\gets\arg\min\bigl\{\Residual{\LifeTree{\PairLabel{i}}}(\Far{i},\Far{j}):\text{nonbirth }j<i\text{ in }\LargePairs\text{ with }\PairLabel{j}=\PairLabel{i}\bigr\}$
\AlgoLine \AlgoInd{4}$\ProxyLength{i}\gets\Residual{\LifeTree{\PairLabel{i}}}(\Far{i},\Far{j})$
\AlgoLine\label{line:extract-long} \AlgoInd{4}\textbf{if} $\ProxyLength{i}>2^8\LogPairs^3\Scale$ \textbf{then} $\LongPairs\gets\LongPairs\cup\{i\}$
\AlgoLine\label{line:extract-proxy} \AlgoInd{4}\textbf{else} $\Retained\gets\Retained\cup\{i\}$,\quad $(\AuxLeft{i},\AuxRight{i})\gets(\Far{i},\Far{j})$,\quad $\AuxPath{i}\gets\text{unique }\Far{i}\text{--}\Far{j}\text{ path in }\LifeTree{\PairLabel{i}}$
\AlgoLine \AlgoInd{1}\textbf{return} $(\SmallPairs,\FirstPairs,\LongPairs,\Retained),\ \{(\AuxLeft{i},\AuxRight{i},\AuxPath{i})\}_{i\in\Retained}$
\end{algobox}

\begin{lemma}\label{lem:selection}
For every index $i\in\Retained$, the auxiliary pair $(\AuxLeft{i},\AuxRight{i})$ and
realizing path $\AuxPath{i}$ returned by \FullRef{Algorithm~\ref*{alg:extract-aux}}{alg:extract-aux} form a
helpful pair for terminal pair $i$ (\FullRef{Definition~\ref*{def:auxiliary-pair}}{def:auxiliary-pair}).
\end{lemma}
\begin{proof}
If $i\in\Retained$ is added in \FullRef{Line~\ref*{line:extract-direct}}{line:extract-direct} (when $i$ is a birth pair or
$\cost(\LifeTree{\PairLabel{i}})\le2^6\LogPairs^2\Pay{i}$), then
$(\AuxLeft{i},\AuxRight{i})=\TerminalPair{i}$ has
$\Residual{i}\TerminalPair{i}=\Pay{i}\ge\Scale>\Scale/2$ and
$\Residual{i+1}\TerminalPair{i}=0$. If $i$ is a birth pair, $\Bought{i}$ uses no
previously purchased positive-cost edges, so its original cost is
$\Pay{i}<2\Scale\le2^8\LogPairs^3\Scale$; otherwise
$\cost(\AuxPath{i})\le\cost(\LifeTree{\PairLabel{i}})\le2^6\LogPairs^2\Pay{i}<2^7\LogPairs^2\Scale\le2^8\LogPairs^3\Scale$.
Since $i\in\LargePairs$, both endpoints lie in a large tree of $\Fref$.

If $i\in\Retained$ is added in \FullRef{Line~\ref*{line:extract-proxy}}{line:extract-proxy}, then its chosen earlier terminal pair
$j<i$ belongs to $\LargePairs$ and shares label $\PairLabel{j}=\PairLabel{i}$.
By \FullRef{Lemma~\ref*{lem:endpoint}}{lem:endpoint}, $(\AuxLeft{i},\AuxRight{i})=(\Far{i},\Far{j})$
satisfies $\Residual{i}(\Far{i},\Far{j})\ge\Pay{i}/2\ge\Scale/2$ and
$\Residual{i+1}(\Far{i},\Far{j})=0$. \FullRef{Line~\ref*{line:extract-proxy}}{line:extract-proxy} of
\FullRef{Algorithm~\ref*{alg:extract-aux}}{alg:extract-aux} ensures
$\cost(\AuxPath{i})=\ProxyLength{i}\le2^8\LogPairs^3\Scale$, and because both
$i$ and $j$ belong to $\LargePairs$, both $\Far{i}$ and
$\Far{j}$ lie in large trees of $\Fref$ (even if $j$ itself
was placed in $\FirstPairs\cup\LongPairs$).
\end{proof}

It remains to show that the terminal pairs excluded in \FullRef{Lines~\ref*{line:extract-first}}{line:extract-first} and~\ref{line:extract-long} of
\FullRef{Algorithm~\ref*{alg:extract-aux}}{alg:extract-aux} contribute at most $\ALG/32$ across all batches.
We need two clean structural facts about trees: first, our maximum-cost merge
rule bounds the total cost of all lifetime trees $\LifeTree{\ell}$ by
$\LogPairs\cdot\ALG$; second, in any tree of total edge cost $\MetricCost$, the sum of
distances from each vertex in a sequence to its nearest earlier vertex is at most
$2\MetricCost(1+\ln\PointCount)$.

\begin{lemma}\label{lem:mass}
Across the entire greedy execution, the lifetime trees satisfy
\[
 \sum_\ell\cost(\LifeTree{\ell})
 \le\ALG(1+\log_2\NumPairs)
 \le\LogPairs\cdot\ALG.
\]
\end{lemma}
\begin{proof}
The edges $\Bought{i}$ purchased at step $i$ (with $\Pay{i}>0$) enter a
component of cost at least $\Pay{i}$. By \FullRef{Definition~\ref*{def:history}}{def:history}, a label
retires only when its component merges with another component of equal or
greater original cost, so the merged component has at least twice the
retiring component's cost. Hence the edges of $\Bought{i}$ can belong to at most
$\log_2(\ALG/\Pay{i})$ retiring lifetime trees plus one final surviving tree,
and writing $\PositivePairs\le\NumPairs$ for the number of positive-payment pairs
and $\PaymentWeight{i}=\Pay{i}/\ALG$ (so $\sum_{i:\Pay{i}>0}\PaymentWeight{i}=1$) gives
\begin{align*}
 \sum_\ell\cost(\LifeTree{\ell})
 &\le\sum_{i:\Pay{i}>0}\Pay{i}\bigl(1+\log_2(\ALG/\Pay{i})\bigr) \\
 &=\ALG\Bigl(1+\sum_{i:\Pay{i}>0}\PaymentWeight{i}\log_2(1/\PaymentWeight{i})\Bigr)
 && \text{(since $\sum_{i:\Pay{i}>0}\PaymentWeight{i}=1$)} \\
 &\le\ALG(1+\log_2\PositivePairs)
 && \text{(since $\ln x\le x-1$ on $x=1/(\PositivePairs\PaymentWeight{i})$)} \\
 &\le\ALG(1+\log_2\NumPairs)
 \le\LogPairs\cdot\ALG.
 && \text{(since $\PositivePairs\le\NumPairs$ and $1+\log_2\NumPairs\le\LogPairs$)} \qedhere
\end{align*}
\end{proof}

\begin{lemma}\label{lem:nn}
Let $\MetricTree$ be a finite tree with nonnegative edge costs of total cost $\MetricCost$, and let
$\MetricPoint{1},\ldots,\MetricPoint{\PointCount}$ be any sequence of
$\PointCount\ge1$ vertices in $\MetricTree$. For $2\le i\le\PointCount$, let
$\InsertionDistance{i}=\min_{1\le j<i}\Residual{\MetricTree}(\MetricPoint{i},\MetricPoint{j})$
be the shortest distance in $\MetricTree$ from $\MetricPoint{i}$ to
$\{\MetricPoint{1},\ldots,\MetricPoint{i-1}\}$. Then
\[
 \sum_{i=2}^{\PointCount}\InsertionDistance{i}
 \le2\MetricCost\ln\PointCount
 \le2\MetricCost(1+\ln\PointCount).
\]
\end{lemma}
\begin{proof}
For any $r>0$, suppose $j\ge1$ of the values
$\InsertionDistance{2},\ldots,\InsertionDistance{\PointCount}$ are at least $r$.
The corresponding $j$ points together with $\MetricPoint{1}$ form $j+1$ points in
$\MetricTree$ at pairwise tree distance at least $r$. Since an Euler tour of
$\MetricTree$ traverses each edge twice and has total length $2\MetricCost$,
visiting these $j+1$ points in tour order gives $(j+1)r\le2\MetricCost$.
Thus the $j$th largest value among
$\InsertionDistance{2},\ldots,\InsertionDistance{\PointCount}$ is at most
$2\MetricCost/(j+1)$, and summing over $1\le j\le\PointCount-1$ yields
\begin{align*}
 \sum_{i=2}^{\PointCount}\InsertionDistance{i}
 &\le2\MetricCost\sum_{j=1}^{\PointCount-1}\frac{1}{j+1} \\
 &\le2\MetricCost\int_1^{\PointCount}\frac{\dd x}{x}
 =2\MetricCost\ln\PointCount.
 && \qedhere
\end{align*}
\end{proof}

\begin{lemma}\label{lem:nn-batch}
For every cost batch $\CurrentBatch\in\BatchFamily$, the distances $\ProxyLength{i}$ computed in \FullRef{Algorithm~\ref*{alg:extract-aux}}{alg:extract-aux} satisfy
\[
 \sum_{i\in\LongPairs}\ProxyLength{i}\le2\LogPairs\sum_\ell\cost(\LifeTree{\ell}).
\]
\end{lemma}
\begin{proof}
For each label $\ell$, let $i_1<\cdots<i_{\PointCount_\ell}$ ($\PointCount_\ell\le\NumPairs$) be the nonbirth terminal pairs in $\LargePairs$ with $\PairLabel{i}=\ell$. Every $i_t\in\LongPairs$ ($t\ge2$) has $\ProxyLength{i_t}=\min_{1\le s<t}\Residual{\LifeTree{\ell}}(\Far{i_t},\Far{i_s})$, so
\begin{align*}
 \sum_{i\in\LongPairs}\ProxyLength{i}
 &=\sum_\ell\sum_{i\in\LongPairs:\PairLabel{i}=\ell}\ProxyLength{i} \\
 &\le\sum_{\ell:\PointCount_\ell\ge2} 2\cost(\LifeTree{\ell})\ln\PointCount_\ell
 && \text{(by \FullRef{Lemma~\ref*{lem:nn}}{lem:nn} on $\{\Far{i_t}\}$ in $\LifeTree{\ell}$)} \\
 &\le2\LogPairs\sum_\ell\cost(\LifeTree{\ell}).
 && \text{(since $\ln\PointCount_\ell\le\ln\NumPairs\le\LogPairs$)} \qedhere
\end{align*}
\end{proof}

\begin{lemma}\label{lem:excluded-payments}
Across the family $\BatchFamily$ of all cost batches, the excluded terminal-pair
sets $\FirstPairs$ and $\LongPairs$ produced by \FullRef{Algorithm~\ref*{alg:extract-aux}}{alg:extract-aux}
satisfy
\[
 \sum_{\CurrentBatch\in\BatchFamily}\sum_{i\in\FirstPairs}\Pay{i}\le\frac{\ALG}{64}
 \qquad\text{and}\qquad
 \sum_{\CurrentBatch\in\BatchFamily}\sum_{i\in\LongPairs}\Pay{i}\le\frac{\ALG}{64}.
\]
\end{lemma}
\begin{proof}
In each batch $\CurrentBatch\in\BatchFamily$, \FullRef{Line~\ref*{line:extract-first}}{line:extract-first} places at most one nonbirth pair $i$ per label $\ell$ into $\FirstPairs$:
\begin{align*}
 \sum_{\CurrentBatch\in\BatchFamily}\sum_{i\in\FirstPairs}\Pay{i}
 &\le\sum_{\CurrentBatch\in\BatchFamily}\sum_\ell\frac{\cost(\LifeTree{\ell})}{2^6\LogPairs^2}
 && \text{(by \FullRef{Lines~\ref*{line:extract-else}}{line:extract-else} and~\ref{line:extract-first})} \\
 &=\frac{|\BatchFamily|}{2^6\LogPairs^2}\sum_\ell\cost(\LifeTree{\ell}) \\
 &\le\frac{\LogPairs\cdot\LogPairs\cdot\ALG}{2^6\LogPairs^2}
 =\frac{\ALG}{64}.
 && \text{(by \FullRef{Definition~\ref*{def:batch}}{def:batch} and \FullRef{Lemma~\ref*{lem:mass}}{lem:mass})}
\end{align*}
Similarly, for $\LongPairs$:
\begin{align*}
 \sum_{\CurrentBatch\in\BatchFamily}\sum_{i\in\LongPairs}\Pay{i}
 &\le\sum_{\CurrentBatch\in\BatchFamily}\sum_{i\in\LongPairs}2\Scale
 && \text{(by \FullRef{Definition~\ref*{def:batch}}{def:batch})} \\
 &\le\frac{1}{2^7\LogPairs^3}\sum_{\CurrentBatch\in\BatchFamily}\sum_{i\in\LongPairs}\ProxyLength{i}
 && \text{(by \FullRef{Line~\ref*{line:extract-long}}{line:extract-long}, $\ProxyLength{i}>2^8\LogPairs^3\Scale$ for $i\in\LongPairs$)} \\
 &\le\frac{1}{2^7\LogPairs^3}\sum_{\CurrentBatch\in\BatchFamily}2\LogPairs\sum_\ell\cost(\LifeTree{\ell})
 && \text{(by \FullRef{Lemma~\ref*{lem:nn-batch}}{lem:nn-batch})} \\
 &\le\frac{2\LogPairs^3\ALG}{2^7\LogPairs^3}
 =\frac{\ALG}{64}.
 && \text{(by \FullRef{Definition~\ref*{def:batch}}{def:batch} and \FullRef{Lemma~\ref*{lem:mass}}{lem:mass})} \qedhere
\end{align*}
\end{proof}

\FloatBarrier
\subsection{Cycle Rank and Small-Tree Charging}
\label{sec:construction}
\label{sec:girth}

Because a multigraph without short cycles cannot have too many edges relative to
its total edge length, we first relate the minimum cycle length of a general
multigraph to its \emph{cycle rank} $\CycleRank=|E|-|V|+c$ (the number of edges
minus vertices plus connected components, where loops and parallel edges are
allowed), and then apply this bound to the small trees of $\Fref$.

\begin{lemma}\label{lem:girth}
Consider a finite undirected multigraph with nonnegative edge lengths,
total edge length $\TestCost$, and cycle rank $\CycleRank\ge1$. For
$\CycleFactor(\CycleRank)=\lceil\log_2(8\CycleRank)\rceil+1$, the multigraph
contains a cycle of length at most
\[
 \frac{4\CycleFactor(\CycleRank)}{\CycleRank}\TestCost.
\]
Consequently, if every cycle has length at least $\Sep>0$, then
$\CycleRank\Sep\le4\CycleFactor(\CycleRank)\TestCost$.
\end{lemma}
\begin{proof}
Repeatedly delete bridges and isolated vertices and suppress degree-two vertices
(summing the lengths of their two incident edges). This preserves cycle rank and
does not increase total length. Choose a connected component $\KernelGraph$ of
the resulting multigraph whose ratio of total length $\KernelCost$ to cycle rank
$\KernelRank\ge1$ is at most $\TestCost/\CycleRank$. If $\KernelGraph$ is a
single loop, it has $\KernelRank=1$ and length at most $\TestCost/\CycleRank$,
proving the bound. Otherwise $\KernelGraph$ has minimum degree at least $3$, so
its vertex and edge counts $\KernelVertices,\KernelEdges$ satisfy
$3\KernelVertices\le2\KernelEdges=2(\KernelRank+\KernelVertices-1)$, giving
$\KernelVertices\le2\KernelRank-2\le2\CycleRank$ and $\KernelEdges\ge\KernelRank$.

Set $\WalkSteps=\CycleFactor(\CycleRank)$ and consider a $\WalkSteps$-step
nonbacktracking random walk on $\KernelGraph$: pick the first directed edge
uniformly at random among all $2\KernelEdges$ oriented edges, and at each
subsequent step move along a uniformly chosen outgoing edge other than the
reverse of the just-traversed edge. Because every oriented edge at a vertex $v$
has $\deg(v)-1\ge2$ valid predecessors, the uniform distribution over oriented
edges is stationary, so the walk's expected length is
$\WalkSteps\KernelCost/\KernelEdges$. By Markov's inequality, a walk is
short (of length at most $2\WalkSteps\KernelCost/\KernelEdges$) with
probability at least $1/2$.

Suppose toward a contradiction that every cycle in $\KernelGraph$ has length
strictly greater than $4\WalkSteps\KernelCost/\KernelEdges$. Every short
nonbacktracking walk is vertex-simple, since a repeated vertex would give a
cycle of length no greater than the walk's length. Two distinct simple walks
with the same endpoints contain a cycle in their union. Thus no two distinct
short walks can share the same initial oriented edge and the same final vertex
(otherwise their union would contain a cycle of length at most
$4\WalkSteps\KernelCost/\KernelEdges$). Since $\KernelGraph$ has minimum degree
at least $3$, each step after the first has at least $2$ choices, so any
particular $\WalkSteps$-step walk has conditional probability at most
$2^{-(\WalkSteps-1)}$ given its first edge. Union-bounding over the
$\KernelVertices\le2\CycleRank$ possible final vertices gives
\begin{align*}
 \Prb(\text{short}\mid\text{first edge})
 &\le\KernelVertices\,2^{-(\WalkSteps-1)} \\
 &\le\frac{2\CycleRank}{8\CycleRank}
 =\frac{1}{4},
 && \text{(since $\KernelVertices\le2\CycleRank$ and $\WalkSteps=\lceil\log_2(8\CycleRank)\rceil+1$)}
\end{align*}
contradicting the $1/2$ lower bound from Markov's inequality. Thus $\KernelGraph$
(and hence the original multigraph) contains a cycle of length at most
\begin{align*}
 \frac{4\WalkSteps\KernelCost}{\KernelEdges}
 &\le\frac{4\WalkSteps\KernelCost}{\KernelRank}
 && \text{(since $\KernelEdges\ge\KernelRank$)} \\
 &\le\frac{4\CycleFactor(\CycleRank)}{\CycleRank}\TestCost.
 && \text{(since $\KernelCost/\KernelRank\le\TestCost/\CycleRank$)} \qedhere
\end{align*}
\end{proof}

\FullRef{Lemma~\ref*{lem:girth}}{lem:girth} immediately bounds the number of terminal pairs with
payment at least $\Sep$ that any reference forest $\Fref$ can connect.

\begin{lemma}\label{lem:threshold}
Let $\ThresholdIndices\subseteq\{1,\ldots,\NumPairs\}$ be a set of
$\SelectedCount\ge1$ terminal-pair indices such that every $i\in\ThresholdIndices$
has greedy payment $\Pay{i}\ge\Sep>0$ and is connected in $\Fref$. Then
\[
 \SelectedCount\Sep\le4\CycleFactor(\SelectedCount)\,\costref.
\]
\end{lemma}
\begin{proof}
Add a zero-length edge $\AuxEdge{i}=\TerminalPair{i}$ to $\Fref$ for each
$i\in\ThresholdIndices$. Since $\Fref$ connects $\Source{i}$ and $\Target{i}$,
each added edge joins two vertices in the same tree of $\Fref$, so the resulting
multigraph has cycle rank $\CycleRank=\SelectedCount$ and total edge length
$\TestCost=\costref$. In any cycle, let $\AuxEdge{i}=\TerminalPair{i}$ be the
added edge with the largest index $i$; every other added edge $\AuxEdge{j}$
($j<i$) on the cycle has $\Residual{i}\TerminalPair{j}=0$ before step $i$, so
the $\Fref$-edges on the cycle have total cost at least
$\Residual{i}\TerminalPair{i}=\Pay{i}\ge\Sep$. Applying \FullRef{Lemma~\ref*{lem:girth}}{lem:girth}
yields $\SelectedCount\Sep\le4\CycleFactor(\SelectedCount)\,\costref$.
\end{proof}

We now apply \FullRef{Lemma~\ref*{lem:threshold}}{lem:threshold} to bound the total payment of the small
pairs $\SmallPairs$ in a batch. Whenever $\ParallelCount$ pairs (or parallel
auxiliary edges in \FullRef{Section~\ref*{sec:replacement}}{sec:replacement}) are supported by reference
subtrees of cost $O(M\Scale)$ with cycle threshold $\Omega(\Scale)$, the cycle-rank
bound yields an inequality of the form $\ParallelCount-1\le48M\CycleFactor(\ParallelCount)$,
where $\CycleFactor(\ParallelCount)=\lceil\log_2(8\ParallelCount)\rceil+1$.
We first record the numerical cap that unrolls this inequality.

\begin{lemma}\label{lem:numerical}
For any real number $M\ge2^8$ and $\CycleFactor(\ParallelCount)=\lceil\log_2(8\ParallelCount)\rceil+1$,
every integer $\ParallelCount\ge1$ satisfying
$\ParallelCount-1\le48M\CycleFactor(\ParallelCount)$ satisfies
\[
 \ParallelCount\le\lceil2^{12}M\log_2(2^9M)\rceil.
\]
\end{lemma}
\begin{proof}
Set $\ZoneLog=\log_2(2^9M)\ge17$ and $\ParallelCutoff=2^{12}M\ZoneLog$. We have
\begin{align*}
 48M\CycleFactor(\ParallelCutoff)
 &\le48M(\log_2\ParallelCutoff+5)
 && \text{(since $\CycleFactor(n)\le\log_2 n+5$)} \\
 &=48M(\ZoneLog+8+\log_2\ZoneLog)
 && \text{(since $\ParallelCutoff=2^3\cdot2^9M\ZoneLog$)} \\
 &\le96M\ZoneLog
 <\ParallelCutoff-1.
 && \text{(since $8+\log_2\ZoneLog\le\ZoneLog$ for $\ZoneLog\ge17$)}
\end{align*}
Since $\ParallelCount-1-48M(\log_2\ParallelCount+5)$ is strictly increasing for
$\ParallelCount\ge\ParallelCutoff$, no integer $\ParallelCount>\lceil\ParallelCutoff\rceil$
can satisfy $\ParallelCount-1\le48M\CycleFactor(\ParallelCount)$.
\end{proof}

\begin{lemma}\label{lem:small}
For any cost batch $\CurrentBatch$ of scale $\Scale>0$ and any reference forest
$\Fref$ connecting the pairs in $\CurrentBatch$, the small pairs $\SmallPairs$
(whose endpoints lie in reference trees of $\Fref$ of cost less than
$2^{16}\LogPairs^3\Scale$) satisfy
\[
 \sum_{i\in\SmallPairs}\Pay{i}
 \le288\CycleFactor(\LogPairs)\,\costref
 =O(\log\LogPairs)\,\costref,
\]
where $\CycleFactor(\LogPairs)=\lceil\log_2(8\LogPairs)\rceil+1=\Theta(\log\LogPairs)$.
\end{lemma}
\begin{proof}
Let $\SmallTree$ be a small tree of $\Fref$ (so
$\cost(\SmallTree)<2^{16}\LogPairs^3\Scale$) that connects $\SelectedCount\ge1$
pairs of $\SmallPairs$. Because each such pair has payment $\Pay{i}\ge\Scale$,
\begin{align*}
 \SelectedCount\Scale
 &\le4\CycleFactor(\SelectedCount)\,\cost(\SmallTree)
 && \text{(by \FullRef{Lemma~\ref*{lem:threshold}}{lem:threshold} on $\SmallTree$ with $\Sep=\Scale$)} \\
 &<4\cdot2^{16}\LogPairs^3\Scale\,\CycleFactor(\SelectedCount).
 && \text{(since $\cost(\SmallTree)<2^{16}\LogPairs^3\Scale$)}
\end{align*}
Dividing by $\Scale$ gives
$\SelectedCount-1<\SelectedCount\le48(2^{16}\LogPairs^3)\CycleFactor(\SelectedCount)$,
so with $M=2^{16}\LogPairs^3\ge2^8$,
\begin{align*}
 \SelectedCount
 &\le\bigl\lceil2^{12}(2^{16}\LogPairs^3)\log_2(2^9\cdot2^{16}\LogPairs^3)\bigr\rceil
 && \text{(by \FullRef{Lemma~\ref*{lem:numerical}}{lem:numerical})} \\
 &=\bigl\lceil2^{28}\LogPairs^3(25+3\log_2\LogPairs)\bigr\rceil \\
 &\le2^{34}\LogPairs^4,
 && \text{(since $\LogPairs\ge4$)}
\end{align*}
and therefore
\begin{align*}
 \CycleFactor(\SelectedCount)
 &\le\lceil\log_2(8\cdot2^{34}\LogPairs^4)\rceil+1 \\
 &\le38+4\lceil\log_2\LogPairs\rceil \\
 &\le36\CycleFactor(\LogPairs).
 && \text{(since $\CycleFactor(\LogPairs)\ge4+\lceil\log_2\LogPairs\rceil$)}
\end{align*}
Summing over all small trees $\SmallTree$ of $\Fref$ yields
\begin{align*}
 \sum_{i\in\SmallPairs}\Pay{i}
 &\le\sum_{\SmallTree}2\SelectedCount\Scale
 && \text{(by \FullRef{Definition~\ref*{def:batch}}{def:batch}, $\Pay{i}<2\Scale$)} \\
 &\le\sum_{\SmallTree}8\CycleFactor(\SelectedCount)\,\cost(\SmallTree)
 && \text{(by \FullRef{Lemma~\ref*{lem:threshold}}{lem:threshold} on $\SmallTree$)} \\
 &\le\sum_{\SmallTree}288\CycleFactor(\LogPairs)\,\cost(\SmallTree)
 && \text{(since $\CycleFactor(\SelectedCount)\le36\CycleFactor(\LogPairs)$)} \\
 &\le288\CycleFactor(\LogPairs)\,\costref.
 && \text{(since the small trees are edge-disjoint in $\Fref$)} \qedhere
\end{align*}
\end{proof}

\FloatBarrier
\subsection{Approaching \texorpdfstring{$\OPT$}{OPT}: Zone Decomposition, Dense Core, and Forest Surgery}
\label{sec:replacement}

We have accounted for the small-tree pairs $\SmallPairs$ and the excluded
terminal pairs $\FirstPairs\cup\LongPairs$. It remains to account for the
$\CountAux=|\Retained|$ retained large pairs in the current batch $\CurrentBatch$.
Each $i\in\Retained$ has payment $\Pay{i}<2\Scale$ and supplies an auxiliary pair
$(\AuxLeft{i},\AuxRight{i})$ in large trees of $\Fref$ together with a realizing
path $\AuxPath{i}$ of original cost at most $2^8\LogPairs^3\Scale$ (\FullRef{Lemma~\ref*{lem:selection}}{lem:selection}).

Because a large tree of $\Fref$ can have arbitrary cost above
$2^{16}\LogPairs^3\Scale$, we first partition every large tree into connected
zones of uniform scale $[2^{16}\LogPairs^3\Scale,\,3\cdot2^{16}\LogPairs^3\Scale]$.
Note that the minimum zone cost $2^{16}\LogPairs^3\Scale$ is $2^8$ times
the maximum realizing-path cost $2^8\LogPairs^3\Scale$; this $2^8$-factor gap is
what makes replacing a zone by a realizing path profitable.

\begin{definition}[Reference zones]\label{def:zones}
Allowing edge subdivisions (replacing an edge $e$ by a path of positive-cost
segments summing to $\cost(e)$), a \emph{zone decomposition} of the large trees
of $\Fref$ at scale $\Scale$ is a partition of their edges into connected subtrees
called \emph{zones} such that every zone $\Zone$ satisfies
\[
 2^{16}\LogPairs^3\Scale\le\cost(\Zone)\le3\cdot2^{16}\LogPairs^3\Scale.
\]
Zones are edge-disjoint and share only \emph{boundary vertices} (vertices shared
by two or more zones, or artificial subdivision vertices).
\end{definition}

\input{figures/zone-partition}

\begin{lemma}\label{lem:zones}
Every large tree $\PartitionTree$ of $\Fref$ (with
$\cost(\PartitionTree)\ge2^{16}\LogPairs^3\Scale$) admits a zone decomposition into
connected edge-disjoint zones each costing between $2^{16}\LogPairs^3\Scale$ and
$3\cdot2^{16}\LogPairs^3\Scale$ (\FullRef{Figure~\ref*{fig:zone-partition}}{fig:zone-partition}). Consequently, across the entire reference forest
$\Fref$, the total number of zones is at most
\[
 \frac{\costref}{2^{16}\LogPairs^3\Scale}.
\]
\end{lemma}
\begin{proof}
Set $M=2^{16}\LogPairs^3$, so the target zone cost range is $[M\Scale,3M\Scale]$.
Subdivide any edge of $\PartitionTree$ heavier than
$M\Scale$ into segments of cost at most $M\Scale$, root $\PartitionTree$
arbitrarily, and process vertices bottom-up from leaves to the root. Each child
passes upward a connected leftover subtree containing that child of cost less than
$M\Scale$; adjoining the parent edge increases its cost to less than $2M\Scale$. If this
piece has cost at least $M\Scale$, output it as a zone; otherwise accumulate such
small pieces at the parent until their combined cost first reaches $M\Scale$,
and output that union as a zone of cost in $[M\Scale,2M\Scale)$. Pass the remaining leftover
(of cost less than $M\Scale$) upward. Finally, at the root, merge any nonempty leftover into
an adjacent output zone (which exists since $\cost(\PartitionTree)\ge M\Scale$),
producing a zone of cost in $[M\Scale,3M\Scale]$. Since the zones are edge-disjoint and each
costs at least $M\Scale$, there are at most $\costref/(M\Scale)$ zones in total.
\end{proof}

We now build the charging graph $\ChargingGraph$ on the zones of the large trees
of $\Fref$ and the $\CountAux=|\Retained|$
auxiliary pairs $\{(\AuxLeft{i},\AuxRight{i})\}_{i\in\Retained}$.

\begin{definition}[Charging graph]\label{def:charging-graph}
Given the zones of the large trees of $\Fref$ and the auxiliary pairs
$\{(\AuxLeft{i},\AuxRight{i})\}_{i\in\Retained}$, the \emph{charging graph}
$\ChargingGraph$ is constructed as follows (\FullRef{Figure~\ref*{fig:zone-copies}}{fig:zone-copies}):
\begin{enumerate}[label=\textnormal{\arabic*.},leftmargin=*,nosep]
\item Take disjoint copies of the zones, retaining original edge costs on all
tree edges. If a boundary vertex belongs to multiple zones, its copies remain
distinct, and we fix one copy to receive all auxiliary endpoints at that vertex.
\item For each $i\in\Retained$, add a zero-length \emph{auxiliary edge}
$\AuxEdge{i}$ between the assigned copies of $\AuxLeft{i}$ and $\AuxRight{i}$.
\end{enumerate}
\end{definition}

\input{figures/zone-copies}

\begin{lemma}\label{lem:chronology}
Every cycle in the charging graph $\ChargingGraph$ has total tree-edge cost at least $\Scale/2$.
\end{lemma}
\begin{proof}
Since the zone copies are acyclic, any cycle $C$ in $\ChargingGraph$ contains at
least one zero-length auxiliary edge. Let $\AuxEdge{i}=(\AuxLeft{i},\AuxRight{i})$
be the auxiliary edge on $C$ with the largest index $i\in\Retained$. By
\FullRef{Lemma~\ref*{lem:selection}}{lem:selection}, every other auxiliary edge $\AuxEdge{j}$ ($j<i$) on
$C$ satisfies $\Residual{i}(\AuxLeft{j},\AuxRight{j})\le\Residual{j+1}(\AuxLeft{j},\AuxRight{j})=0$
before step $i$. Projecting $C\setminus\{\AuxEdge{i}\}$ to $\InputGraph$
therefore yields a walk between $\AuxLeft{i}$ and $\AuxRight{i}$ whose residual
cost before step $i$ comes only from the tree edges of $C$
(\FullRef{Figure~\ref*{fig:cycle-certificate}}{fig:cycle-certificate}), so
\begin{align*}
 \cost(C)
 &\ge\Residual{i}(\AuxLeft{i},\AuxRight{i})
 && \text{(since $\Residual{i}(\AuxLeft{j},\AuxRight{j})=0$ for all $j<i$ on $C$)} \\
 &\ge\frac{\Scale}{2}.
 && \text{(by \FullRef{Lemma~\ref*{lem:selection}}{lem:selection})} \qedhere
\end{align*}
\end{proof}

\input{figures/cycle-certificate}

To analyze how densely the $\CountAux$ auxiliary edges link the zones of
$\Fref$, we contract each zone copy in $\ChargingGraph$ to a single vertex and
peel away sparse zones.

\begin{definition}[Zone multigraph]
\label{def:zone-multigraph}
The \emph{zone multigraph} $\ZoneGraph$ is the unweighted multigraph obtained by
contracting each zone copy in the charging graph $\ChargingGraph$ to a single
vertex and keeping one edge for each of the $\CountAux$ auxiliary edges
(including loops and parallel edges). Each edge of $\ZoneGraph$ inherits the
index $i\in\Retained$ and original-cost realizing path $\AuxPath{i}$ of its
auxiliary pair.
\end{definition}

\begin{definition}[Dense zone core]
\label{def:replacement-records}
Define the multiplicity cap $\ParallelCap$ and core degree threshold
$\CoreDegree$ by
\begin{equation}\label{eq:core-thresholds}
 \ParallelCap=\bigl\lceil2^{12}(2^{16}\LogPairs^3)\log_2(2^9\cdot2^{16}\LogPairs^3)\bigr\rceil,
 \qquad
 \CoreDegree=(2^{13}-1)\ParallelCap.
\end{equation}
The \emph{dense zone core} $\CoreGraph$ is obtained from $\ZoneGraph$ by deleting
all self-loops and then repeatedly removing any vertex of current degree strictly
less than $\CoreDegree$ together with its incident edges until no such vertex
remains. We write $\CountCore$ for the number of vertices (core zones),
$\CountCoreEdges$ for the number of edges, and $\CountCoreComp$ for the number of
connected components of $\CoreGraph$.
\end{definition}

We first show that the cycle-rank bound caps the number of loops and parallel
edges in $\ZoneGraph$ by $\ParallelCap$. Consequently, all auxiliary edges
discarded in forming $\CoreGraph$ from $\ZoneGraph$ have total greedy payment at
most $O(\log\LogPairs)\,\costref$, while every zone remaining in $\CoreGraph$ has
at least $2^{13}-1$ distinct neighboring core zones.

\begin{lemma}\label{lem:sparse}
With $\ParallelCap$ and $\CoreDegree=(2^{13}-1)\ParallelCap$ defined in
\eqref{eq:core-thresholds}:
\begin{enumerate}[label=\textnormal{(\roman*)},nosep]
\item Every vertex of $\ZoneGraph$ has at most $\ParallelCap$ self-loops, and
every pair of distinct vertices is joined by at most $\ParallelCap$ parallel
edges.
\item The number of auxiliary edges discarded in forming $\CoreGraph$ from
$\ZoneGraph$ satisfies
\[
 \CountAux-\CountCoreEdges
 \le\frac{\ParallelCap+\CoreDegree}{2^{16}\LogPairs^3\Scale}\costref
 \le\frac{2^{28}\CycleFactor(\LogPairs)}{\Scale}\costref,
\]
and the greedy payments of these discarded edges total at most
$2^{29}\CycleFactor(\LogPairs)\,\costref=O(\log\LogPairs)\,\costref$, where
$\CycleFactor(\LogPairs)=\lceil\log_2(8\LogPairs)\rceil+1=\Theta(\log\LogPairs)$.
\item Whenever $\CoreGraph$ is nonempty, it has minimum degree at least
$\CoreDegree=(2^{13}-1)\ParallelCap$ and at most $\ParallelCap$ parallel edges between
any two vertices, so every vertex in $\CoreGraph$ has at least $2^{13}-1$ distinct
neighbors.
\end{enumerate}
\end{lemma}
\begin{proof}
Set $M=2^{16}\LogPairs^3$ (so every zone has cost in $[M\Scale,3M\Scale]$). We
bound the multiplicities in $\ZoneGraph$ using the charging graph $\ChargingGraph$:
\begin{itemize}[leftmargin=*,nosep]
\item If $\ParallelCount\ge2$ parallel edges join two distinct zones in $\ZoneGraph$, the
subgraph of $\ChargingGraph$ consisting of those two zone trees and the
$\ParallelCount$ zero-length auxiliary edges has cycle rank
$\CycleRank=\ParallelCount-1\ge1$ and total tree cost $\TestCost\le6M\Scale$, so
\begin{align*}
 (\ParallelCount-1)\frac{\Scale}{2}
 &\le4\CycleFactor(\ParallelCount-1)\,\TestCost
 && \text{(by \FullRef{Lemmas~\ref*{lem:chronology}}{lem:chronology} and~\ref{lem:girth})} \\
 &\le24M\Scale\,\CycleFactor(\ParallelCount).
 && \text{(since $\TestCost\le6M\Scale$)}
\end{align*}
Dividing by $\Scale/2$ gives $\ParallelCount-1\le48M\CycleFactor(\ParallelCount)$,
which implies $\ParallelCount\le\ParallelCap$ by \FullRef{Lemma~\ref*{lem:numerical}}{lem:numerical}.
\item If $\ParallelCount\ge1$ loops are incident to one zone in $\ZoneGraph$, that zone
tree and its $\ParallelCount$ auxiliary edges have cycle rank
$\CycleRank=\ParallelCount$ and tree cost $\TestCost\le3M\Scale$, so
\begin{align*}
 \ParallelCount\frac{\Scale}{2}
 &\le4\CycleFactor(\ParallelCount)\,\TestCost
 && \text{(by \FullRef{Lemmas~\ref*{lem:chronology}}{lem:chronology} and~\ref{lem:girth})} \\
 &\le12M\Scale\,\CycleFactor(\ParallelCount),
 && \text{(since $\TestCost\le3M\Scale$)}
\end{align*}
giving $\ParallelCount-1<\ParallelCount\le24M\CycleFactor(\ParallelCount)\le48M\CycleFactor(\ParallelCount)$
and again $\ParallelCount\le\ParallelCap$ by \FullRef{Lemma~\ref*{lem:numerical}}{lem:numerical}.
\end{itemize}
This proves~(i).

For~(ii), \FullRef{Lemma~\ref*{lem:zones}}{lem:zones} shows that $\ZoneGraph$ has at most
$\costref/(M\Scale)$ vertices. Deleting loops removes at most $\ParallelCap$
edges per vertex by~(i), and each vertex peeled in forming $\CoreGraph$ has degree strictly
less than $\CoreDegree$ at the moment of removal, removing fewer than
$\CoreDegree$ edges. Hence each vertex of $\ZoneGraph$ accounts for at most
$\ParallelCap+\CoreDegree=2^{13}\ParallelCap$ discarded edges. Notice that the
factor $M=2^{16}\LogPairs^3$ inside $\ParallelCap$ cancels the $M$ in the
denominator of the zone count:
\begin{align*}
 \frac{\ParallelCap+\CoreDegree}{M}
 &=\frac{2^{13}\bigl\lceil2^{12}M\log_2(2^9M)\bigr\rceil}{M}
 && \text{(by \eqref{eq:core-thresholds})} \\
 &\le2^{13}\bigl(2^{12}(25+3\log_2\LogPairs)+1\bigr)
 && \text{(since $M=2^{16}\LogPairs^3\ge1$)} \\
 &\le2^{25}(26+3\log_2\LogPairs) \\
 &\le2^{28}\CycleFactor(\LogPairs).
 && \text{(since $\CycleFactor(\LogPairs)\ge4+\lceil\log_2\LogPairs\rceil$)}
\end{align*}
Multiplying by $\costref/\Scale$ gives the bound on $\CountAux-\CountCoreEdges$,
and multiplying by $2\Scale$ bounds the associated greedy payments. Part~(iii)
follows immediately from~(i) and the stopping condition of the peeling process.
\end{proof}

It remains to account for the $\CountCoreEdges$ auxiliary edges inside the dense
core $\CoreGraph$. By \FullRef{Lemma~\ref*{lem:sparse}(iii)}{lem:sparse}, every core zone in
$\CoreGraph$ has at least $2^{13}-1$ distinct neighboring core zones. To replace
these densely connected core zones in $\Fref$ by a cheaper subgraph without
disconnecting the rest of $\Fref$ or a given set $\Marks$ of
$\CountMarks=|\Marks|$ marked original vertices, \FullRef{Algorithm~\ref*{alg:replace-zones}}{alg:replace-zones}
(\textsc{ReplaceDenseZones}) constructs two subgraphs (shown in
\FullRef{Figure~\ref*{fig:marked-replacement}}{fig:marked-replacement}):
\begin{enumerate}[label=\textnormal{\arabic*.},leftmargin=*,nosep]
\item \emph{Replacement skeleton $S_{\mathrm{skel}}$.} Because every core zone
has at least $2^{13}-1$ distinct neighbors in $\CoreGraph$, we can choose a
small dominating set of \emph{center zones} $\Centers\subseteq V(\CoreGraph)$
with $|\Centers|\le\CountCore/2^9$ (every core zone outside $\Centers$ has a
neighbor in $\Centers$). Each noncenter core zone is hooked to an adjacent
center zone by one \emph{assignment edge} in $E_{\mathrm{assign}}\subseteq E(\CoreGraph)$
to form a star around each center, and the stars in each component of
$\CoreGraph$ are linked by a spanning forest of \emph{interstar edges}
$E_{\mathrm{inter}}\subseteq E(\CoreGraph)$. In the original graph $\InputGraph$,
the \emph{replacement skeleton} $S_{\mathrm{skel}}$ consists of the realizing
paths $\AuxPath{i}$ of all edges in $E_{\mathrm{assign}}\cup E_{\mathrm{inter}}$,
together with the full zone trees of the center zones and of the endpoint zones
of $E_{\mathrm{inter}}$. Because each realizing path costs at most $\frac{1}{2^8}$
of a zone and at most $3|\Centers|$ full zone trees are kept, $S_{\mathrm{skel}}$
connects all core zones in each component of $\CoreGraph$ at less than
$\frac{1}{32}$ of their original cost.
\item \emph{Boundary-hub graph $H_{\mathrm{hub}}$ and attachment paths $P_{\mathrm{att}}$.}
Replacing the core zones by $S_{\mathrm{skel}}$ could sever paths in $\Fref$
that pass through core zones on their way between external reference edges or
marked vertices in $\Marks$. To identify which attachments must be restored, we
form an auxiliary \emph{boundary-hub graph} $H_{\mathrm{hub}}$ from $\Fref$ by
keeping all reference edges outside the core zones, replacing each core zone
$\Zone$ by a zero-cost star joining a new hub vertex $h_\Zone$ to that zone's
boundary vertices, and merging the hubs of zones that belong to the same
connected component of $\CoreGraph$ (\FullRef{Figure~\ref*{fig:marked-replacement}(b)}{fig:marked-replacement}).
In each component of $H_{\mathrm{hub}}$, taking a minimal tree $T_{\mathrm{hub}}$
spanning the merged hubs and the marks of $\Marks$ outside the core zones
identifies the exact boundary exits that must be reconnected; connecting each
such boundary exit and each mark of $\Marks$ inside a core zone to
$S_{\mathrm{skel}}$ by a path inside its zone forms the attachment subgraph
$P_{\mathrm{att}}$ (\FullRef{Figure~\ref*{fig:marked-replacement}(c)}{fig:marked-replacement}).
\end{enumerate}

\input{figures/marked-replacement}

\begin{algobox}[htbp]{Replace Dense Zones\label{alg:replace-zones}}
\KwIn{Reference forest $\Fref$, batch scale $\Scale$, auxiliary pairs with realizing paths $\{(\AuxLeft{i},\AuxRight{i},\AuxPath{i})\}_{i\in\Retained}$, and marked vertices $\Marks\subseteq\InputVertices$.}
\KwOut{Updated reference forest $\Fref_{\mathrm{new}}$ preserving connectivity on $\Marks$, and cost saving $\ForestSaving=\costref-\cost(\Fref_{\mathrm{new}})\ge0$.}
\AlgoProc{ReplaceDenseZones}{\Fref,\Scale,\{(\AuxLeft{i},\AuxRight{i},\AuxPath{i})\}_{i\in\Retained},\Marks}
\AlgoLine\label{line:replace-zones} \AlgoInd{1}Decompose large trees of $\Fref$ into zones of cost $[2^{16}\LogPairs^3\Scale,\,3\cdot2^{16}\LogPairs^3\Scale]$ (\FullRef{Lemma~\ref*{lem:zones}}{lem:zones})
\AlgoLine \AlgoInd{1}Form the charging graph $\ChargingGraph$ on disjoint zone copies with auxiliary edges $\{\AuxEdge{i}\}_{i\in\Retained}$, and contract each zone copy to a single vertex to obtain the zone multigraph $\ZoneGraph$ (\FullRef{Definitions~\ref*{def:charging-graph}}{def:charging-graph} and~\FullRef{\ref*{def:zone-multigraph}}{def:zone-multigraph})
\AlgoLine \AlgoInd{1}$\CoreGraph\gets\ZoneGraph$ with all loops deleted
\AlgoLine\label{line:replace-peel} \AlgoInd{1}\textbf{while} there exists $v\in V(\CoreGraph)$ with $\deg_{\CoreGraph}(v)<\CoreDegree=(2^{13}-1)\ParallelCap$ \textbf{do}\AlgoComment{Peel sparse zones}
\AlgoLine \AlgoInd{2}$\CoreGraph\gets\CoreGraph\setminus\{v\}$
\AlgoLine \AlgoInd{1}\textbf{if} $E(\CoreGraph)=\varnothing$ \textbf{then return} $\Fref_{\mathrm{new}}\gets\Fref,\ \ForestSaving\gets0$
\AlgoLine \AlgoInd{1}Choose a dominating set $\Centers\subseteq V(\CoreGraph)$ of center zones with $|\Centers|\le\CountCore/2^9$
\AlgoLine \AlgoInd{1}Choose assignment edges $E_{\mathrm{assign}}\subseteq E(\CoreGraph)$ hooking each $\Zone\in V(\CoreGraph)\setminus\Centers$ to an adjacent center, and a spanning forest $E_{\mathrm{inter}}\subseteq E(\CoreGraph)$ across the stars in each component of $\CoreGraph$
\AlgoLine\label{line:replace-skel} \AlgoInd{1}$S_{\mathrm{skel}}\gets\bigl(\bigcup_{i\in E_{\mathrm{assign}}\cup E_{\mathrm{inter}}}\AuxPath{i}\bigr)\cup\bigl(\bigcup_{\Zone\text{ in }\Centers\text{ or incident to }E_{\mathrm{inter}}}E(\Zone)\bigr)$\AlgoComment{Replacement skeleton}
\AlgoLine\label{line:replace-hub} \AlgoInd{1}$H_{\mathrm{hub}}\gets$ boundary-hub graph of $\Fref$ with one merged hub per component of $\CoreGraph$
\AlgoLine\label{line:replace-tree} \AlgoInd{1}$T_{\mathrm{hub}}\gets$ union of minimal trees in $H_{\mathrm{hub}}$ spanning the merged hubs and outside marks in each component
\AlgoLine\label{line:replace-att} \AlgoInd{1}$P_{\mathrm{att}}\gets$ paths inside core zones connecting each hub-incident edge of $T_{\mathrm{hub}}$ and internal mark to $S_{\mathrm{skel}}$
\AlgoLine \AlgoInd{1}$F'\gets$ forest obtained from $\bigl(\Fref\setminus\bigcup_{\Zone\in V(\CoreGraph)}E(\Zone)\bigr)\cup S_{\mathrm{skel}}\cup P_{\mathrm{att}}$ by breaking cycles and pruning subdivision leaves
\AlgoLine \AlgoInd{1}\textbf{if} $\cost(F')<\costref$ \textbf{then} $\Fref_{\mathrm{new}}\gets F'$ \textbf{else} $\Fref_{\mathrm{new}}\gets\Fref$
\AlgoLine \AlgoInd{1}\textbf{return} $\Fref_{\mathrm{new}},\ \ForestSaving\gets\costref-\cost(\Fref_{\mathrm{new}})$
\end{algobox}

\begin{lemma}\label{lem:replacement}
For any cost batch $\CurrentBatch$ of scale $\Scale$ with $\CountAux=|\Retained|$
retained auxiliary pairs and any set $\Marks$ of $\CountMarks=|\Marks|$ marked
original vertices, \FullRef{Algorithm~\ref*{alg:replace-zones}}{alg:replace-zones} returns a valid reference
forest $\Fref_{\mathrm{new}}$ of original edges with
$\cost(\Fref_{\mathrm{new}})\le\costref$ that preserves all connections in
$\Fref$ between vertices of $\Marks$, and its cost saving
$\ForestSaving=\costref-\cost(\Fref_{\mathrm{new}})\ge0$ satisfies
\begin{equation}\label{eq:replacement}
 \CountAux\Scale
 \le32\LogPairs\,\ForestSaving
 +2^{28}\CycleFactor(\LogPairs)\,\costref
 +2^{24}\LogPairs^4\,\CountMarks\Scale.
\end{equation}
\end{lemma}
\begin{proof}
Set $M=2^{16}\LogPairs^3$, so every zone has cost in $[M\Scale,3M\Scale]$. Let
$\CostCore\ge M\Scale\CountCore$ be the total original edge cost of the
$\CountCore$ core zones in $\CoreGraph$. If $\CoreGraph$ has no edges
($\CountCoreEdges=0$), \FullRef{Lemma~\ref*{lem:sparse}(ii)}{lem:sparse} gives
$\CountAux\Scale\le2^{28}\CycleFactor(\LogPairs)\,\costref$, so \eqref{eq:replacement}
holds with $\ForestSaving=0$. Assume now that $\CountCoreEdges>0$. By
\FullRef{Lemma~\ref*{lem:sparse}(iii)}{lem:sparse}, every vertex in $\CoreGraph$ has at least $2^{13}-1$
distinct neighbors, so each of the $\CountCoreComp$ connected components of
$\CoreGraph$ has at least $2^{13}$ vertices and $\CountCoreComp\le\CountCore/2^{13}$.

\paragraph{Cost and connectivity of the replacement skeleton $S_{\mathrm{skel}}$.}
Sample each vertex of $\CoreGraph$ independently with probability
$\ln(2^{13})/2^{13}$, and add any vertex that has neither itself
nor a neighbor sampled. The resulting set $\Centers$ dominates $\CoreGraph$ and
has expected size at most
$\CountCore\bigl(\ln(2^{13})/2^{13}+\e^{-\ln(2^{13})}\bigr)=\CountCore(13\ln 2+1)/2^{13}<\CountCore/2^9$,
so a dominating set $\Centers$ of size $|\Centers|\le\CountCore/2^9$ exists.
In \FullRef{Line~\ref*{line:replace-skel}}{line:replace-skel}, the replacement skeleton $S_{\mathrm{skel}}$ includes $|E_{\mathrm{assign}}|=\CountCore-|\Centers|$ assignment
realizing paths and $|E_{\mathrm{inter}}|=|\Centers|-\CountCoreComp$ interstar realizing paths (fewer
than $\CountCore$ realizing paths in total, each costing at most
$2^8\LogPairs^3\Scale=\frac{1}{2^8}M\Scale$), together with the full zone trees of
the $|\Centers|$ center zones and the at most $2(|\Centers|-\CountCoreComp)$ endpoint
zones of $E_{\mathrm{inter}}$ (at most $3|\Centers|$ zone trees in total, each costing
at most $3M\Scale$). Because every core zone incident to two or more edges of
$E_{\mathrm{assign}}\cup E_{\mathrm{inter}}$ is either a center zone or an endpoint of $E_{\mathrm{inter}}$ (whose full zone
tree is retained to connect its realizing-path endpoints), $S_{\mathrm{skel}}$
is connected within each component of $\CoreGraph$ and visits every
core zone, and its total cost satisfies
\begin{align*}
 \cost(S_{\mathrm{skel}})
 &\le\CountCore\cdot\frac{M\Scale}{2^8}+3|\Centers|\cdot3M\Scale
 && \text{(by \FullRef{Lemma~\ref*{lem:selection}}{lem:selection} and \FullRef{Lemma~\ref*{lem:zones}}{lem:zones})} \\
 &\le M\Scale\CountCore\left(\frac{1}{2^8}+\frac{9}{2^9}\right)
 && \text{(since $|\Centers|\le\CountCore/2^9$)} \\
 &=\frac{11}{2^9}M\Scale\CountCore
 <\frac{\CostCore}{32}.
 && \text{(since $\CostCore\ge M\Scale\CountCore$)}
\end{align*}

\paragraph{Cost of the attachment paths $P_{\mathrm{att}}$ and validity of $F'$.}
In the boundary-hub graph $H_{\mathrm{hub}}$ (\FullRef{Line~\ref*{line:replace-hub}}{line:replace-hub}), merging the zone hubs within each of the
$\CountCoreComp$ connected components of $\CoreGraph$ produces $\CountCoreComp$
merged hubs and preserves all connections in $\Fref$. Let $\OutsideMarks$ be the
number of marks in $\Marks$ that lie outside the core zones (so
$\CountMarks-\OutsideMarks$ marks lie inside core zones). In each component of
$H_{\mathrm{hub}}$ with $\RequiredCount\ge1$ required terminals (merged hubs plus
outside marks), take a minimal tree spanning those terminals (\FullRef{Line~\ref*{line:replace-tree}}{line:replace-tree});
suppressing degree-two nonterminals preserves all terminal degrees and leaves
$\BranchCount$ branching vertices (each of degree at least $3$), so the total
terminal degree in that tree is at most
$2(\RequiredCount+\BranchCount-1)-3\BranchCount\le2\RequiredCount-2<2\RequiredCount$.
Summing over all components of $H_{\mathrm{hub}}$, the total number of tree edges of $T_{\mathrm{hub}}$ incident to the
$\CountCoreComp$ merged hubs is at most $2(\CountCoreComp+\OutsideMarks)$. Each
such incidence exits a core zone through a boundary vertex and is connected
to $S_{\mathrm{skel}}$ inside that zone at cost at most $3M\Scale$;
likewise, each of the $\CountMarks-\OutsideMarks$ internal marks is connected to
$S_{\mathrm{skel}}$ inside its core zone at cost at most $3M\Scale$. Thus the
attachment paths $P_{\mathrm{att}}$ added in \FullRef{Line~\ref*{line:replace-att}}{line:replace-att} satisfy
\begin{align*}
 \cost(P_{\mathrm{att}})
 &\le3M\Scale\bigl[2(\CountCoreComp+\OutsideMarks)+\CountMarks-\OutsideMarks\bigr]
 && \text{(by \FullRef{Lemma~\ref*{lem:zones}}{lem:zones}, each zone has cost $\le3M\Scale$)} \\
 &\le6M\Scale(\CountCoreComp+\CountMarks)
 && \text{(since $\OutsideMarks\le\CountMarks$)} \\
 &\le\frac{6}{2^{13}}\CostCore+6\cdot2^{16}\LogPairs^3\,\CountMarks\Scale.
 && \text{(since $M\Scale\CountCoreComp\le M\Scale\CountCore/2^{13}\le\CostCore/2^{13}$)}
\end{align*}
Since $S_{\mathrm{skel}}$ is connected within each core component, routing
through $S_{\mathrm{skel}}$ and $P_{\mathrm{att}}$ in $\InputGraph$ preserves all
connections in $\Fref$ between marks in $\Marks$. Break cycles while preserving
connected components, then iteratively prune unmarked artificial subdivision
leaves. In the globally subdivided graph, artificial vertices have degree two,
are not in $\Marks$, and are not endpoints or branching points of any
$\AuxPath{i}$. Pruning therefore removes partially retained subdivided edges;
recombining the fully retained chains produces a forest $F'$ of whole original
edges satisfying
\begin{align*}
 \cost(F')
 &\le(\costref-\CostCore)+\cost(S_{\mathrm{skel}})+\cost(P_{\mathrm{att}}) \\
 &\le\costref-\CostCore+\left(\frac{1}{32}+\frac{6}{2^{13}}\right)\CostCore+6\cdot2^{16}\LogPairs^3\,\CountMarks\Scale \\
 &\le\costref-\frac{\CostCore}{2}+2^{19}\LogPairs^3\,\CountMarks\Scale.
\end{align*}

\paragraph{Relating the cost saving $\ForestSaving$ to $\CountAux\Scale$.}
Consider the subgraph of the charging graph $\ChargingGraph$ formed by the
$\CountCore$ core zone trees and the $\CountCoreEdges$ zero-length auxiliary
edges of $\CoreGraph$. Because contracting trees preserves cycle rank,
contracting the $\CountCore$ zone trees leaves $\CoreGraph$, which has
$\CountCore$ vertices, $\CountCoreEdges$ edges, $\CountCoreComp$ connected
components, and minimum degree $\CoreDegree\ge4$ (so
$2\CountCoreEdges\ge\CoreDegree\CountCore$). Thus this charging subgraph has
cycle rank
$\CoreRank=\CountCoreEdges-\CountCore+\CountCoreComp\ge(1-2/\CoreDegree)\CountCoreEdges\ge\CountCoreEdges/2$
and total tree cost $\CostCore$, and therefore
\begin{align*}
 \CostCore
 &\ge\frac{\CoreRank(\Scale/2)}{4\CycleFactor(\CoreRank)}
 && \text{(by \FullRef{Lemmas~\ref*{lem:chronology}}{lem:chronology} and~\ref{lem:girth})} \\
 &\ge\frac{\CountCoreEdges\Scale}{16\LogPairs}.
 && \text{(since $\CoreRank\ge\CountCoreEdges/2$ and $\CycleFactor(\CoreRank)\le\LogPairs$)}
\end{align*}
Consequently, the cost saving $\ForestSaving=\max\bigl(0,\costref-\cost(F')\bigr)$
satisfies
\begin{align*}
 \ForestSaving
 &\ge\frac{\CostCore}{2}-2^{19}\LogPairs^3\,\CountMarks\Scale
 && \text{(from the bound on $\cost(F')$)} \\
 &\ge\frac{\CountCoreEdges\Scale}{32\LogPairs}-2^{19}\LogPairs^3\,\CountMarks\Scale,
 && \text{(since $\CostCore\ge\CountCoreEdges\Scale/(16\LogPairs)$)}
\end{align*}
and hence
\begin{align*}
 \CountAux\Scale
 &=\CountCoreEdges\Scale+(\CountAux-\CountCoreEdges)\Scale \\
 &\le32\LogPairs\,\ForestSaving+2^{24}\LogPairs^4\,\CountMarks\Scale+2^{28}\CycleFactor(\LogPairs)\,\costref.
 && \text{(by the bound on $\ForestSaving$ and \FullRef{Lemma~\ref*{lem:sparse}(ii)}{lem:sparse})} \qedhere
\end{align*}
\end{proof}

Combining the small-tree payment bound (\FullRef{Lemma~\ref*{lem:small}}{lem:small}) with the forest
replacement bound for $\Retained$ (\FullRef{Lemma~\ref*{lem:replacement}}{lem:replacement}, where each of the
$\CountAux=|\Retained|$ retained pairs has payment $\Pay{i}<2\Scale$) immediately
gives our single-batch charging bound.

\begin{lemma}\label{lem:batch-charging}
For any cost batch $\CurrentBatch$ of scale $\Scale$, any reference forest $\Fref$
connecting the pairs in $\CurrentBatch$, and any set $\Marks$ of
$\CountMarks=|\Marks|$ marked original vertices, the update of
\FullRef{Algorithm~\ref*{alg:replace-zones}}{alg:replace-zones} preserves all connections in $\Fref$ between
vertices of $\Marks$, does not increase $\costref$, and satisfies
\[
 \sum_{i\in\SmallPairs\cup\Retained}\Pay{i}
 \le64\LogPairs\,\ForestSaving
 +2^{30}\CycleFactor(\LogPairs)\,\costref
 +2^{25}\LogPairs^4\,\CountMarks\Scale,
\]
where $\ForestSaving=\costref-\cost(\Fref_{\mathrm{new}})\ge0$ is the cost
decrease of the reference forest.
\end{lemma}
\input{figures/batch-schedule}
\begin{proof}
We have
\begin{align*}
 \sum_{i\in\SmallPairs\cup\Retained}\Pay{i}
 &\le\sum_{i\in\SmallPairs}\Pay{i}+2\CountAux\Scale
 && \text{(by \FullRef{Definition~\ref*{def:batch}}{def:batch})} \\
 &\le64\LogPairs\,\ForestSaving+\bigl(288+2\cdot2^{28}\bigr)\CycleFactor(\LogPairs)\,\costref+2^{25}\LogPairs^4\,\CountMarks\Scale
 && \text{(\FullRef{Lemmas~\ref*{lem:small}}{lem:small},~\ref{lem:replacement})} \\
 &\le64\LogPairs\,\ForestSaving+2^{30}\CycleFactor(\LogPairs)\,\costref+2^{25}\LogPairs^4\,\CountMarks\Scale.
 && \qedhere
\end{align*}
\end{proof}

\FloatBarrier
\subsection{Strided Batch Classes and Proof of \texorpdfstring{\FullRef{Theorem~\ref*{thm:main}}{thm:main}}{Theorem 1}}
\label{sec:accounting-procedure}
\label{sec:summation}

\FullRef{Lemma~\ref*{lem:batch-charging}}{lem:batch-charging} bounds the payments in a single batch by
$64\LogPairs\,\ForestSaving+O(\log\LogPairs)\,\costref$, plus the overhead
$2^{25}\LogPairs^4\,\CountMarks\Scale$ for preserving connectivity among the
$\CountMarks$ marked vertices. If we processed consecutive doubling batches
$b=0,1,2,\ldots$ with a single evolving reference forest, the mark overhead would
exceed $\costref$. Instead, we now index the batches in $\BatchFamily$ by
$b\in\{0,1,\ldots,\lfloor\log_2\NumPairs\rfloor\}$ with scale
\[
 \BatchScale{b}=2^b\frac{\OPT}{\NumPairs},
 \qquad
 \Batch{b}=\bigl\{i:\BatchScale{b}\le\Pay{i}<2\BatchScale{b}\bigr\},
\]
and partition them into $\ClassCount=\Theta(\log\LogPairs)$ strided batch
classes so that consecutive batches in the same class have scales separated by
a factor of $2^{\ClassCount}=\Theta(\LogPairs^5)$ (\FullRef{Figure~\ref*{fig:batch-schedule}}{fig:batch-schedule}).

\begin{definition}[Strided batch classes]\label{def:batch-classes}\label{def:parameters}
Define the number of batch classes $\ClassCount$ by
\[
 \ClassCount
 =\bigl\lceil\log_2(2^{28}\LogPairs^5)\bigr\rceil
 =28+\lceil5\log_2\LogPairs\rceil
 =\Theta(\log\LogPairs)=\Theta(\log\log(4\NumPairs)),
\]
so that consecutive scales in a class differ by a factor of
$2^{\ClassCount}\ge2^{28}\LogPairs^5$. For each residue class
$r\in\{0,1,\ldots,\ClassCount-1\}$, \emph{class $r$} consists of the batches
$\Batch{b}$ with $b=r+t\ClassCount\le\lfloor\log_2\NumPairs\rfloor$ in increasing
order of $t\ge0$. At step $(r,t)$, the \emph{future marks} $\FutureMarks{r}{t}$
are the distinct original endpoints of all terminal pairs belonging to later
batches $\Batch{r+t'\ClassCount}$ ($t'>t$) of the same class $r$.
\end{definition}

\FullRef{Algorithm~\ref*{alg:accounting}}{alg:accounting} (\textsc{StridedBatchAccounting}) gives the
complete retrospective accounting procedure: it processes the $\ClassCount$
classes one by one, resetting $\Fref\gets\Fopt$ at the beginning of each class
and evolving $\Fref$ through the batches $b=r+t\ClassCount$ of that class.

\begin{algobox}[htbp]{Strided Batch Accounting\label{alg:accounting}\label{proc:accounting}}
\KwIn{Graph $\InputGraph$, terminal pairs $\TerminalPair{i}_{i=1}^{\NumPairs}$, greedy edge purchases $\{\Bought{i}\}_{i=1}^{\NumPairs}$, and optimal offline forest $\Fopt$.}
\KwOut{Complete partition of batch indices and reference-forest savings $\{\BatchSaving{b}\}$.}
\AlgoProc{StridedBatchAccounting}{\InputGraph,\TerminalPair{i}_{i=1}^{\NumPairs},\{\Bought{i}\}_{i=1}^{\NumPairs},\Fopt}
\AlgoLine \AlgoInd{1}Compute greedy component labels $\PairLabel{i}$, lifetime trees $\LifeTree{\ell}$, and far endpoints $\Far{i}$ from $\{\Bought{i}\}_{i=1}^{\NumPairs}$ (\FullRef{Definition~\ref*{def:lifetime}}{def:lifetime})
\AlgoLine \AlgoInd{1}$\Batch{b}\gets\{i:\BatchScale{b}\le\Pay{i}<2\BatchScale{b}\}$ for $0\le b\le\lfloor\log_2\NumPairs\rfloor$,\quad $\ClassCount\gets\lceil\log_2(2^{28}\LogPairs^5)\rceil$
\AlgoLine \AlgoInd{1}\textbf{for} $r=0,1,\ldots,\ClassCount-1$ \textbf{do}
\AlgoLine \AlgoInd{2}$\Fref\gets\Fopt$
\AlgoLine \AlgoInd{2}\textbf{for} $t=0,1,\ldots$ \textbf{while} $b=r+t\ClassCount\le\lfloor\log_2\NumPairs\rfloor$ \textbf{do}
\AlgoLine \AlgoInd{3}$\FutureMarks{r}{t}\gets\bigcup_{\substack{t'>t\\r+t'\ClassCount\le\lfloor\log_2\NumPairs\rfloor}}\bigcup_{i\in\Batch{r+t'\ClassCount}}\{\Source{i},\Target{i}\}$\AlgoComment{Marks for later batches in class $r$}
\AlgoLine \AlgoInd{3}$(\SmallPairs,\FirstPairs,\LongPairs,\Retained),\{(\AuxLeft{i},\AuxRight{i},\AuxPath{i})\}_{i\in\Retained}\gets\FullRef{\textsc{ExtractAuxiliaryPairs}}{alg:extract-aux}\bigl(\Batch{b},\BatchScale{b},\Fref,\{\PairLabel{i},\LifeTree{\ell},\Far{i}\}\bigr)$
\AlgoLine \AlgoInd{3}$\Fref,\BatchSaving{b}\gets\FullRef{\textsc{ReplaceDenseZones}}{alg:replace-zones}\bigl(\Fref,\BatchScale{b},\{(\AuxLeft{i},\AuxRight{i},\AuxPath{i})\}_{i\in\Retained},\FutureMarks{r}{t}\bigr)$
\AlgoLine \AlgoInd{1}\textbf{return} $\{(\SmallPairs,\FirstPairs,\LongPairs,\Retained),\BatchSaving{b}\}_{b=0}^{\lfloor\log_2\NumPairs\rfloor}$
\end{algobox}

The next lemma shows that the scale gap $2^{\ClassCount}\ge2^{28}\LogPairs^5$
between consecutive batches in class $r$ makes the number of future marks
$\CountMarks=|\FutureMarks{r}{t}|$ so small that the mark reconnection term in
\FullRef{Lemma~\ref*{lem:batch-charging}}{lem:batch-charging} is at most $\costref$.

\begin{lemma}\label{lem:batch-accounting}
At every step $(r,t)$ of \FullRef{Algorithm~\ref*{alg:accounting}}{alg:accounting} (processing batch
$\Batch{b}$ with $b=r+t\ClassCount$), the reference forest $\Fref$ connects all
terminal pairs in the unprocessed batches $\Batch{r+t'\ClassCount}$ ($t'\ge t$)
of class $r$, its cost $\costref\le\OPT$ never increases, and the payments in
$\SmallPairs\cup\Retained$ for batch $\Batch{b}$ satisfy
\begin{equation}\label{eq:batch}
 \sum_{i\in\SmallPairs\cup\Retained}\Pay{i}
 \le64\LogPairs\,\BatchSaving{b}
 +\bigl(2^{30}\CycleFactor(\LogPairs)+1\bigr)\costref,
\end{equation}
where $\BatchSaving{b}=\costref-\cost(\Fref_{\mathrm{new}})\ge0$ is the cost
decrease of $\Fref$ recorded at batch $\Batch{b}$.
\end{lemma}
\begin{proof}
We proceed by induction on $t\ge0$ within class $r$. At $t=0$, $\Fref=\Fopt$
has cost $\OPT$ and connects every terminal pair. Suppose $\Fref$ connects all
pairs in the unprocessed batches $\Batch{r+t'\ClassCount}$ ($t'\ge t$) of class
$r$, and let $\FutureCount$ be the total number of terminal pairs in the strictly
later batches ($t'>t$) of class $r$. Every such pair lies in a batch of scale at
least $2^{\ClassCount}\BatchScale{b}$ and therefore has greedy payment at least
$\Sep=2^{\ClassCount}\BatchScale{b}$. If $\FutureCount\ge1$, each future pair
contributes at most two marked endpoints to $\FutureMarks{r}{t}$, so
$\CountMarks=|\FutureMarks{r}{t}|\le2\FutureCount$ and
\begin{align*}
 \CountMarks\BatchScale{b}
 &\le2\FutureCount\BatchScale{b} \\
 &\le\frac{8\CycleFactor(\FutureCount)}{2^{\ClassCount}}\costref
 && \text{(by \FullRef{Lemma~\ref*{lem:threshold}}{lem:threshold} with $\Sep=2^{\ClassCount}\BatchScale{b}$)} \\
 &\le\frac{8\LogPairs}{2^{\ClassCount}}\costref;
 && \text{(since $\FutureCount\le\NumPairs$)}
\end{align*}
the same bound holds trivially if $\FutureCount=0$ ($\CountMarks=0$).
Consequently, the mark overhead in \FullRef{Lemma~\ref*{lem:batch-charging}}{lem:batch-charging} satisfies
\begin{align*}
 2^{25}\LogPairs^4\,\CountMarks\BatchScale{b}
 &\le\frac{2^{28}\LogPairs^5}{2^{\ClassCount}}\costref \\
 &\le\costref.
 && \text{(by \FullRef{Definition~\ref*{def:batch-classes}}{def:batch-classes}, $2^{\ClassCount}\ge2^{28}\LogPairs^5$)}
\end{align*}
Substituting this into \FullRef{Lemma~\ref*{lem:batch-charging}}{lem:batch-charging} proves \eqref{eq:batch}.
Finally, \FullRef{Lemma~\ref*{lem:replacement}}{lem:replacement} guarantees that $\Fref_{\mathrm{new}}$
has cost at most $\costref$ and preserves connectivity between all pairs of marks
in $\FutureMarks{r}{t}$, which maintains the inductive invariant for $t+1$.
\end{proof}

We now assemble the pieces to prove \FullRef{Theorem~\ref*{thm:main}}{thm:main}.

\begin{proof}[Proof of \FullRef{Theorem~\ref*{thm:main}}{thm:main}]
Run \FullRef{Algorithm~\ref*{alg:accounting}}{alg:accounting} on the completed greedy execution. Every
terminal pair $i\in\{1,\ldots,\NumPairs\}$ belongs either to the low-payment set
($\Pay{i}<\OPT/\NumPairs$) or to a unique batch $\Batch{b}\in\BatchFamily$, where
$\Batch{b}$ is partitioned into
$\SmallPairs\mathbin{\dot\cup}\FirstPairs\mathbin{\dot\cup}\LongPairs\mathbin{\dot\cup}\Retained$.
Within each class $r\in\{0,\ldots,\ClassCount-1\}$, $\Fref$ starts at $\Fopt$
(cost $\OPT$) and its cost never increases, so $\costref\le\OPT$ at every batch
and $\sum_{b\text{ in class }r}\BatchSaving{b}=\OPT-\cost(\Fref_{\mathrm{final}})\le\OPT$.
Decomposing $\ALG=\sum_{i=1}^{\NumPairs}\Pay{i}$ across these sets gives
\begin{align*}
 \ALG
 &=\sum_{i:\Pay{i}<\OPT/\NumPairs}\Pay{i}
 +\sum_{\CurrentBatch\in\BatchFamily}\sum_{i\in\FirstPairs\cup\LongPairs}\Pay{i}
 +\sum_{r=0}^{\ClassCount-1}\sum_{b\text{ in }r}\sum_{i\in\SmallPairs\cup\Retained}\Pay{i} \\
 &\le\OPT+\frac{\ALG}{32}+\sum_{r=0}^{\ClassCount-1}\sum_{b\text{ in }r}\sum_{i\in\SmallPairs\cup\Retained}\Pay{i}
 && \text{(\FullRef{Def.~\ref*{def:batch}}{def:batch}, \FullRef{Lemma~\ref*{lem:excluded-payments}}{lem:excluded-payments})} \\
 &\le\OPT+\frac{\ALG}{32}+\sum_{r=0}^{\ClassCount-1}\sum_{b\text{ in }r}\Bigl(64\LogPairs\,\BatchSaving{b}+\bigl(2^{30}\CycleFactor(\LogPairs)+1\bigr)\costref\Bigr)
 && \text{(by \FullRef{Lemma~\ref*{lem:batch-accounting}}{lem:batch-accounting})} \\
 &\le\OPT+\frac{\ALG}{32}+64\ClassCount\LogPairs\,\OPT+\bigl(2^{30}\CycleFactor(\LogPairs)+1\bigr)\LogPairs\,\OPT.
 && \text{(since $|\BatchFamily|\le\LogPairs$)}
\end{align*}
Subtracting $\ALG/32$ from both sides and multiplying by $32/31$, and noting that
$\LogPairs=\Theta(\log(2\NumPairs))$ while both
$\ClassCount=28+\lceil5\log_2\LogPairs\rceil$ and
$\CycleFactor(\LogPairs)=\lceil\log_2(8\LogPairs)\rceil+1$ are
$\Theta(\log\log(4\NumPairs))$, we obtain
\begin{align*}
 \ALG
 &\le\frac{32}{31}\,\OPT\,\Bigl(1+64\ClassCount\LogPairs+\bigl(2^{30}\CycleFactor(\LogPairs)+1\bigr)\LogPairs\Bigr) \\
 &=O\bigl(\OPT\log(2\NumPairs)\log\log(4\NumPairs)\bigr).
 \qedhere
\end{align*}
\end{proof}

%% file: figures/component-history.tex
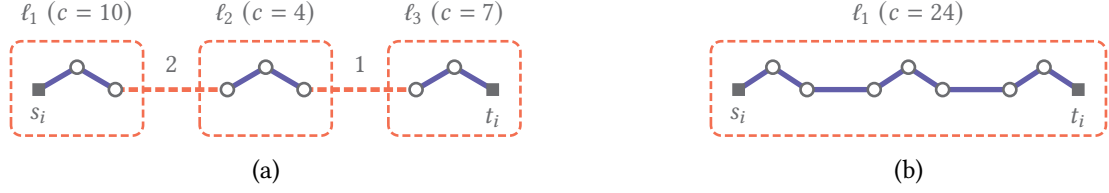
\begin{figure}[t]
\centering
\begin{minipage}[b]{0.53\textwidth}
\centering
\begin{tikzpicture}[scale=0.78]
\def\dem{Red!70}
\def\demi{Blue!70}
\def\ter{White}
\def\col{Black!70}
\def\noder{0.11cm}

\foreach \x/\name/\lbl/\cst in {0/a/\ell_1/10, 3.2/b/\ell_2/4, 6.4/c/\ell_3/7}{
  \coordinate (\name1) at (\x-0.65,0);
  \coordinate (\name2) at (\x,0.38);
  \coordinate (\name3) at (\x+0.65,0);
  \draw[\demi, line width=2.0pt] (\name1) -- (\name2) -- (\name3);
  \draw[\dem, line width=1.0pt, dash pattern=on 3pt off 1.5pt, rounded corners=4pt]
    (\x-1.12,-0.78) rectangle (\x+1.12,0.82);
  \node[\col, above=2pt] at (\x,0.82) {\small{$\lbl\ (c=\cst)$}};
}
\draw[\dem, line width=1.8pt, dash pattern=on 3.5pt off 1.5pt] (a3) -- node[\col, above=2pt] {\small{$2$}} (b1);
\draw[\dem, line width=1.8pt, dash pattern=on 3.5pt off 1.5pt] (b3) -- node[\col, above=2pt] {\small{$1$}} (c1);

\foreach \v in {a2,a3,b1,b2,b3,c1,c2}{
  \draw[\col, fill=\ter, line width=1.0pt] (\v) circle (\noder);
}
\draw[\col, fill=\col, line width=1.0pt] (a1) [square=\noder];
\node[\col, below=3pt] at (a1) {\small{$\Source{i}$}};
\draw[\col, fill=\col, line width=1.0pt] (c3) [square=\noder];
\node[\col, below=3pt] at (c3) {\small{$\Target{i}$}};
\end{tikzpicture}
\par\vspace{2pt}{\small (a)}
\end{minipage}
\hfill
\begin{minipage}[b]{0.44\textwidth}
\centering
\begin{tikzpicture}[scale=0.78]
\def\dem{Red!70}
\def\demi{Blue!70}
\def\ter{White}
\def\col{Black!70}
\def\noder{0.11cm}

\foreach \x/\name in {0/a, 2.3/b, 4.6/c}{
  \coordinate (\name1) at (\x-0.58,0);
  \coordinate (\name2) at (\x,0.38);
  \coordinate (\name3) at (\x+0.58,0);
  \draw[\demi, line width=2.0pt] (\name1) -- (\name2) -- (\name3);
}
\draw[\demi, line width=2.0pt] (a3) -- (b1);
\draw[\demi, line width=2.0pt] (b3) -- (c1);
\draw[\dem, line width=1.0pt, dash pattern=on 3pt off 1.5pt, rounded corners=4pt]
  (-1.05,-0.78) rectangle (5.65,0.82);
\node[\col, above=2pt] at (2.3,0.82) {\small{$\ell_1\ (c=24)$}};

\foreach \v in {a2,a3,b1,b2,b3,c1,c2}{
  \draw[\col, fill=\ter, line width=1.0pt] (\v) circle (\noder);
}
\draw[\col, fill=\col, line width=1.0pt] (a1) [square=\noder];
\node[\col, below=3pt] at (a1) {\small{$\Source{i}$}};
\draw[\col, fill=\col, line width=1.0pt] (c3) [square=\noder];
\node[\col, below=3pt] at (c3) {\small{$\Target{i}$}};
\end{tikzpicture}
\par\vspace{2pt}{\small (b)}
\end{minipage}
\caption{Greedy component labels at step $i$.
(a) Before step $i$, three existing greedy components carrying labels $\ell_1,\ell_2,\ell_3$ of original costs $10,4,7$ (solid blue edges inside dashed red boundaries) are linked by the newly purchased edges $\Bought{i}$ of costs $2$ and $1$ (dashed red edges) between $\Source{i}$ and $\Target{i}$.
(b) After merging, the single greedy component has total cost $10+4+7+3=24$ and inherits the label $\PairLabel{i}=\ell_1$ of the maximum-cost old component $\OldTree{i}$, while labels $\ell_2$ and $\ell_3$ retire.}
\label{fig:component-history}
\end{figure}

%% file: figures/zone-partition.tex
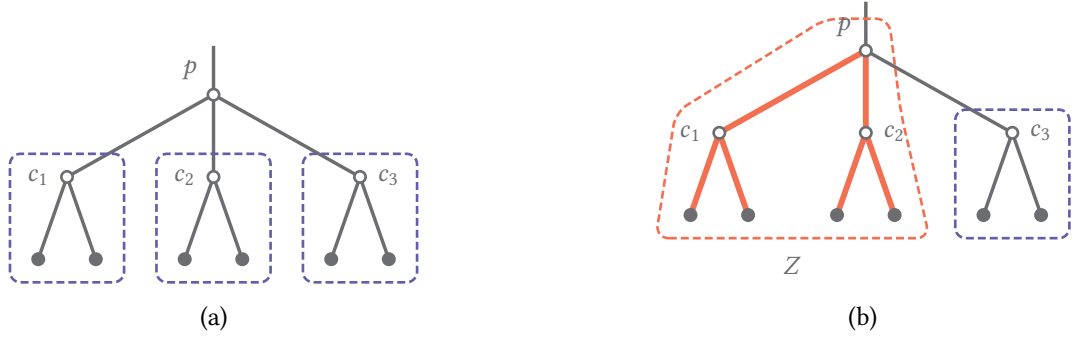
\begin{figure}[t]
\centering
\begin{minipage}[b]{0.48\textwidth}
\centering
\begin{tikzpicture}[
  scale=0.68,
  level distance=1.6cm,
  level 1/.style={sibling distance=2.85cm},
  level 2/.style={sibling distance=1.12cm},
  edge from parent/.style={draw=\col, line width=1.3pt}
]
\def\dem{Red!70}
\def\demi{Blue!70}
\def\ter{White}
\def\col{Black!70}
\def\noder{0.11cm}

\node[coordinate] (p) {}
  child {node[coordinate] (c1) {}
    child {node[coordinate] (l1) {}}
    child {node[coordinate] (l2) {}}
  }
  child {node[coordinate] (c2) {}
    child {node[coordinate] (l3) {}}
    child {node[coordinate] (l4) {}}
  }
  child {node[coordinate] (c3) {}
    child {node[coordinate] (l5) {}}
    child {node[coordinate] (l6) {}}
  };

\draw[\col, line width=1.3pt] (p) -- ++(0,0.95);

\foreach \c/\la/\lb in {c1/l1/l2, c2/l3/l4, c3/l5/l6}{
  \draw[\demi, line width=1.0pt, dash pattern=on 3pt off 1.5pt, rounded corners=4pt]
    ($(\la)+(-0.55,-0.45)$) rectangle ($(\lb)+(0.55,2.05)$);
}

\foreach \v in {p,c1,c2,c3}{
  \draw[\col, fill=\ter, line width=1.0pt] (\v) circle (\noder);
}
\foreach \v in {l1,l2,l3,l4,l5,l6}{
  \draw[\col, fill=\col, line width=1.0pt] (\v) circle (\noder);
}

\node[\col, above left=3pt] at (p) {\small{$p$}};
\node[\col, left=3pt] at (c1) {\small{$c_1$}};
\node[\col, left=3pt] at (c2) {\small{$c_2$}};
\node[\col, right=3pt] at (c3) {\small{$c_3$}};
\end{tikzpicture}
\par\vspace{2pt}{\small (a)}
\end{minipage}
\hfill
\begin{minipage}[b]{0.48\textwidth}
\centering
\begin{tikzpicture}[
  scale=0.68,
  level distance=1.6cm,
  level 1/.style={sibling distance=2.85cm},
  level 2/.style={sibling distance=1.12cm},
  edge from parent/.style={draw=\col, line width=1.3pt}
]
\def\dem{Red!70}
\def\demi{Blue!70}
\def\ter{White}
\def\col{Black!70}
\def\noder{0.11cm}

\node[coordinate] (p) {}
  child {node[coordinate] (c1) {}
    child {node[coordinate] (l1) {}}
    child {node[coordinate] (l2) {}}
  }
  child {node[coordinate] (c2) {}
    child {node[coordinate] (l3) {}}
    child {node[coordinate] (l4) {}}
  }
  child {node[coordinate] (c3) {}
    child {node[coordinate] (l5) {}}
    child {node[coordinate] (l6) {}}
  };

\draw[\col, line width=1.3pt] (p) -- ++(0,0.95);
\draw[\dem, line width=2.2pt] (p) -- (c1);
\draw[\dem, line width=2.2pt] (c1) -- (l1);
\draw[\dem, line width=2.2pt] (c1) -- (l2);
\draw[\dem, line width=2.2pt] (p) -- (c2);
\draw[\dem, line width=2.2pt] (c2) -- (l3);
\draw[\dem, line width=2.2pt] (c2) -- (l4);

\draw[\dem, line width=1.0pt, dash pattern=on 3pt off 1.5pt, rounded corners=5pt]
  ($(l1)+(-0.68,-0.45)$) -- ($(l4)+(0.68,-0.45)$) -- ($(c2)+(0.68,0.40)$) -- ($(p)+(0.45,0.62)$) -- ($(p)+(-0.78,0.62)$) -- ($(c1)+(-0.85,0.40)$) -- cycle;
\node[\col, below=3pt] at ($(l2)!0.5!(l3)+(0,-0.45)$) {\small{$\Zone$}};

\draw[\demi, line width=1.0pt, dash pattern=on 3pt off 1.5pt, rounded corners=4pt]
  ($(l5)+(-0.55,-0.45)$) rectangle ($(l6)+(0.55,2.05)$);

\foreach \v in {p,c1,c2,c3}{
  \draw[\col, fill=\ter, line width=1.0pt] (\v) circle (\noder);
}
\foreach \v in {l1,l2,l3,l4,l5,l6}{
  \draw[\col, fill=\col, line width=1.0pt] (\v) circle (\noder);
}

\node[\col, above left=2pt] at (p) {\small{$p$}};
\node[\col, left=3pt] at (c1) {\small{$c_1$}};
\node[\col, right=3pt] at (c2) {\small{$c_2$}};
\node[\col, right=3pt] at (c3) {\small{$c_3$}};
\end{tikzpicture}
\par\vspace{2pt}{\small (b)}
\end{minipage}
\caption{Bottom-up zone decomposition of a large reference tree $\PartitionTree$ with $M=2^{16}\LogPairs^3$ (\FullRef{Lemma~\ref*{lem:zones}}{lem:zones}).
(a) Each child $c_1,c_2,c_3$ of an internal vertex $p$ passes upward a leftover subtree of cost $<M\Scale$ (dashed blue contours) along a parent edge of cost $\le M\Scale$.
(b) Accumulating child pieces at $p$ outputs a connected zone $\Zone$ of cost in $[M\Scale,2M\Scale)$ (bold red edges inside the dashed red boundary) and passes the remaining leftover of cost $<M\Scale$ upward; merging a final root leftover yields zone costs in $[M\Scale,3M\Scale]$.}
\label{fig:zone-partition}
\end{figure}

%% file: figures/zone-copies.tex
\begin{figure}[t]
\centering
\begin{minipage}[b]{0.42\textwidth}
\centering
\begin{tikzpicture}[scale=0.8]
\def\dem{Red!70}
\def\demi{Blue!70}
\def\ter{White}
\def\col{Black!70}
\def\noder{0.11cm}

\coordinate (v) at (0,0);
\coordinate (u) at (2.1,0);
\coordinate (w) at (4.2,0);

\draw[\dem, line width=2.0pt] (v) -- node[\col, above=3pt] {\small{$\ZoneTree{1}$}} (u);
\draw[\demi, line width=2.0pt] (u) -- node[\col, above=3pt] {\small{$\ZoneTree{2}$}} (w);

\draw[\col, fill=\col, line width=1.0pt] (v) circle (\noder);
\draw[\col, fill=\ter, line width=1.0pt] (u) circle (\noder);
\draw[\col, fill=\col, line width=1.0pt] (w) circle (\noder);

\node[\col, below=3pt] at (v) {\small{$v$}};
\node[\col, below=3pt] at (u) {\small{$u$}};
\node[\col, below=3pt] at (w) {\small{$w$}};
\end{tikzpicture}
\par\vspace{2pt}{\small (a)}
\end{minipage}
\hfill
\begin{minipage}[b]{0.54\textwidth}
\centering
\begin{tikzpicture}[scale=0.8]
\def\dem{Red!70}
\def\demi{Blue!70}
\def\ter{White}
\def\col{Black!70}
\def\noder{0.11cm}

\coordinate (v1) at (0,0);
\coordinate (u1) at (2.0,0);
\coordinate (u2) at (3.5,0);
\coordinate (w2) at (5.5,0);

\draw[\dem, line width=2.0pt] (v1) -- node[\col, above=2pt] {\small{$\ZoneCopy{1}$}} (u1);
\draw[\demi, line width=2.0pt] (u2) -- node[\col, above=2pt] {\small{$\ZoneCopy{2}$}} (w2);
\draw[\col, line width=1.4pt, dash pattern=on 3.5pt off 1.5pt] (v1) to[out=58, in=122] node[\col, above=3pt] {\small{$\AuxEdge{i}\ (0)$}} (w2);

\draw[\col, fill=\col, line width=1.0pt] (v1) circle (\noder);
\draw[\col, fill=\ter, line width=1.0pt] (u1) circle (\noder);
\draw[\col, fill=\ter, line width=1.0pt] (u2) circle (\noder);
\draw[\col, fill=\col, line width=1.0pt] (w2) circle (\noder);

\node[\col, below=3pt] at (v1) {\small{$\ZoneVertex{v}{1}$}};
\node[\col, below=3pt] at (u1) {\small{$\ZoneVertex{u}{1}$}};
\node[\col, below=3pt] at (u2) {\small{$\ZoneVertex{u}{2}$}};
\node[\col, below=3pt] at (w2) {\small{$\ZoneVertex{w}{2}$}};
\end{tikzpicture}
\par\vspace{2pt}{\small (b)}
\end{minipage}
\caption{Constructing the charging graph $\ChargingGraph$ (\FullRef{Definition~\ref*{def:charging-graph}}{def:charging-graph}).
(a) Two edge-disjoint zones $\ZoneTree{1}$ (red) and $\ZoneTree{2}$ (blue) of $\Fref$ share a boundary vertex $u$.
(b) In $\ChargingGraph$, the zone copies $\ZoneCopy{1}$ and $\ZoneCopy{2}$ keep their original edge costs while the two copies $\ZoneVertex{u}{1}$ and $\ZoneVertex{u}{2}$ of $u$ remain distinct with no free edge between them; each auxiliary pair adds a zero-length dashed auxiliary edge $\AuxEdge{i}$ between the assigned copies of its endpoints.}
\label{fig:zone-copies}
\end{figure}
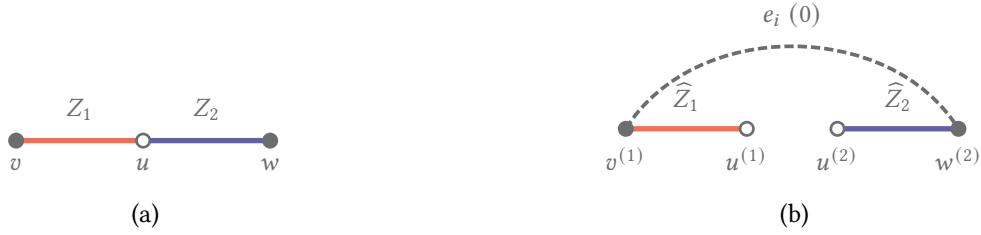

%% file: figures/cycle-certificate.tex
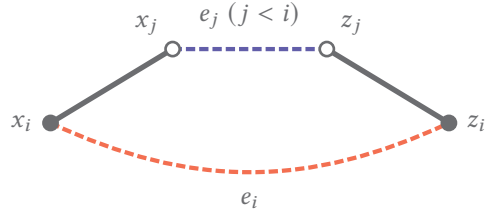
\begin{figure}[t]
\centering
\begin{tikzpicture}[scale=0.85]
\def\dem{Red!70}
\def\demi{Blue!70}
\def\ter{White}
\def\col{Black!70}
\def\noder{0.11cm}

\coordinate (xi) at (0,0);
\coordinate (xj) at (1.9,1.15);
\coordinate (zj) at (4.3,1.15);
\coordinate (zi) at (6.2,0);

\draw[\col, line width=2.0pt] (xi) -- (xj);
\draw[\demi, line width=1.6pt, dash pattern=on 3.5pt off 1.5pt] (xj) -- node[\col, above=3pt] {\small{$\AuxEdge{j}\ (j<i)$}} (zj);
\draw[\col, line width=2.0pt] (zj) -- (zi);
\draw[\dem, line width=1.6pt, dash pattern=on 3.5pt off 1.5pt] (xi) to[bend right=25] node[\col, below=3pt] {\small{$\AuxEdge{i}$}} (zi);

\draw[\col, fill=\col, line width=1.0pt] (xi) circle (\noder);
\draw[\col, fill=\ter, line width=1.0pt] (xj) circle (\noder);
\draw[\col, fill=\ter, line width=1.0pt] (zj) circle (\noder);
\draw[\col, fill=\col, line width=1.0pt] (zi) circle (\noder);

\node[\col, left=3pt] at (xi) {\small{$\AuxLeft{i}$}};
\node[\col, above left=2pt] at (xj) {\small{$\AuxLeft{j}$}};
\node[\col, above right=2pt] at (zj) {\small{$\AuxRight{j}$}};
\node[\col, right=3pt] at (zi) {\small{$\AuxRight{i}$}};
\end{tikzpicture}
\caption{Cycle lower bound in the charging graph $\ChargingGraph$ (\FullRef{Lemma~\ref*{lem:chronology}}{lem:chronology}). Deleting the highest-index auxiliary edge $\AuxEdge{i}=(\AuxLeft{i},\AuxRight{i})$ (dashed red) from a cycle in $\ChargingGraph$ leaves a walk of zone-tree segments (solid bold) and earlier auxiliary edges $\AuxEdge{j}$ with $j<i$ (dashed blue). Because $\Residual{i}(\AuxLeft{j},\AuxRight{j})=0$ before step $i$, the tree segments on the cycle have total original cost at least $\Residual{i}(\AuxLeft{i},\AuxRight{i})\ge\Scale/2$.}
\label{fig:cycle-certificate}
\end{figure}

%% file: figures/marked-replacement.tex
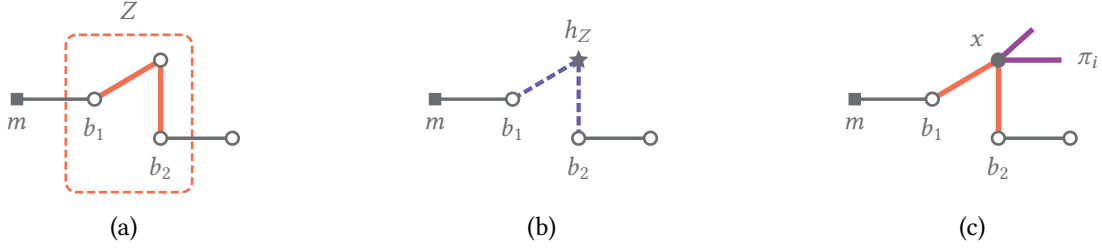
\begin{figure}[t]
\centering
\begin{minipage}[b]{0.31\textwidth}
\centering
\begin{tikzpicture}[scale=0.76]
\def\dem{Red!70}
\def\demi{Blue!70}
\def\ter{White}
\def\col{Black!70}
\def\treeone{Purple}
\def\noder{0.11cm}

\coordinate (m) at (-0.2,0);
\coordinate (b1) at (1.15,0);
\coordinate (u) at (2.3,0.68);
\coordinate (b2) at (2.3,-0.68);
\coordinate (o) at (3.55,-0.68);

\path (-0.5,-1.65) rectangle (3.85,1.45);
\draw[\dem, line width=1.0pt, dash pattern=on 3pt off 1.5pt, rounded corners=4pt]
  (0.65,-1.62) rectangle (2.85,1.12);
\node[\col, above=2pt] at (1.75,1.12) {\small{$\Zone$}};

\draw[\col, line width=1.3pt] (m) -- (b1);
\draw[\dem, line width=2.0pt] (b1) -- (u) -- (b2);
\draw[\col, line width=1.3pt] (b2) -- (o);

\draw[\col, fill=\col, line width=1.0pt] (m) [square=\noder];
\draw[\col, fill=\ter, line width=1.0pt] (b1) circle (\noder);
\draw[\col, fill=\ter, line width=1.0pt] (u) circle (\noder);
\draw[\col, fill=\ter, line width=1.0pt] (b2) circle (\noder);
\draw[\col, fill=\ter, line width=1.0pt] (o) circle (\noder);

\node[\col, below=3pt] at (m) {\small{$m$}};
\node[\col, below=3pt] at (b1) {\small{$b_1$}};
\node[\col, below=3pt] at (b2) {\small{$b_2$}};
\end{tikzpicture}
\par\vspace{2pt}{\small (a)}
\end{minipage}
\hfill
\begin{minipage}[b]{0.31\textwidth}
\centering
\begin{tikzpicture}[scale=0.76]
\def\dem{Red!70}
\def\demi{Blue!70}
\def\ter{White}
\def\col{Black!70}
\def\treeone{Purple}
\def\noder{0.11cm}

\coordinate (m) at (-0.2,0);
\coordinate (b1) at (1.15,0);
\coordinate (h) at (2.3,0.68);
\coordinate (b2) at (2.3,-0.68);
\coordinate (o) at (3.55,-0.68);
\path (-0.5,-1.65) rectangle (3.85,1.45);

\draw[\col, line width=1.3pt] (m) -- (b1);
\draw[\demi, line width=1.6pt, dash pattern=on 3.5pt off 1.5pt] (b1) -- (h) -- (b2);
\draw[\col, line width=1.3pt] (b2) -- (o);

\draw[\col, fill=\col, line width=1.0pt] (m) [square=\noder];
\draw[\col, fill=\ter, line width=1.0pt] (b1) circle (\noder);
\draw[\col, fill=\col, line width=1.0pt] (h) [star=\noder];
\draw[\col, fill=\ter, line width=1.0pt] (b2) circle (\noder);
\draw[\col, fill=\ter, line width=1.0pt] (o) circle (\noder);

\node[\col, below=3pt] at (m) {\small{$m$}};
\node[\col, below=3pt] at (b1) {\small{$b_1$}};
\node[\col, above=3pt] at (h) {\small{$h_{\Zone}$}};
\node[\col, below=3pt] at (b2) {\small{$b_2$}};
\end{tikzpicture}
\par\vspace{2pt}{\small (b)}
\end{minipage}
\hfill
\begin{minipage}[b]{0.33\textwidth}
\centering
\begin{tikzpicture}[scale=0.76]
\def\dem{Red!70}
\def\demi{Blue!70}
\def\ter{White}
\def\col{Black!70}
\def\treeone{Purple}
\def\noder{0.11cm}

\coordinate (m) at (-0.2,0);
\coordinate (b1) at (1.15,0);
\coordinate (x) at (2.3,0.68);
\coordinate (b2) at (2.3,-0.68);
\coordinate (o) at (3.55,-0.68);
\coordinate (p1) at (3.4,0.68);
\coordinate (p2) at (2.9,1.22);
\path (-0.5,-1.65) rectangle (3.85,1.45);

\draw[\col, line width=1.3pt] (m) -- (b1);
\draw[\dem, line width=2.0pt] (b1) -- (x) -- (b2);
\draw[\col, line width=1.3pt] (b2) -- (o);
\draw[\treeone, line width=2.0pt] (x) -- (p1);
\draw[\treeone, line width=2.0pt] (x) -- (p2);

\draw[\col, fill=\col, line width=1.0pt] (m) [square=\noder];
\draw[\col, fill=\ter, line width=1.0pt] (b1) circle (\noder);
\draw[\col, fill=\col, line width=1.0pt] (x) circle (\noder);
\draw[\col, fill=\ter, line width=1.0pt] (b2) circle (\noder);
\draw[\col, fill=\ter, line width=1.0pt] (o) circle (\noder);

\node[\col, below=3pt] at (m) {\small{$m$}};
\node[\col, below=3pt] at (b1) {\small{$b_1$}};
\node[\col, above left=2pt] at (x) {\small{$x$}};
\node[\col, below=3pt] at (b2) {\small{$b_2$}};
\node[\col, right=2pt] at (p1) {\small{$\AuxPath{i}$}};
\end{tikzpicture}
\par\vspace{2pt}{\small (c)}
\end{minipage}
\caption{Restoring marked connectivity locally at a core zone $\Zone$ in \FullRef{Algorithm~\ref*{alg:replace-zones}}{alg:replace-zones}.
(a) A core zone $\Zone$ (bold red edges inside the dashed boundary) meets the rest of $\Fref$ at boundary vertices $b_1,b_2$, with a marked vertex $m\in\Marks$ outside.
(b) In the intermediate boundary-hub graph, $\Zone$ is replaced by a hub $h_{\Zone}$ with zero-length bookkeeping links (dashed blue) to $b_1$ and $b_2$, while $b_1$ and $b_2$ remain distinct vertices.
(c) Each retained hub incidence $(b_r,h_{\Zone})$ is realized in $\InputGraph$ by a path inside $\Zone$ (bold red) from $b_r$ to a skeleton vertex $x\in V(\Zone)$ on the realizing paths $\AuxPath{i}$ (bold purple).}
\label{fig:marked-replacement}
\end{figure}

%% file: figures/batch-schedule.tex
\begin{figure}[t]
\centering
\begin{tikzpicture}[
  scale=0.82,
  batch/.style={draw=Black!70, line width=1.1pt, dash pattern=on 3pt off 1.5pt, rounded corners=3pt, minimum width=1.3cm, minimum height=0.62cm},
  arrow/.style={-Stealth, draw=Black!70, line width=1.2pt},
  font=\small
]
\def\col{Black!70}
\def\dem{Red!70}
\def\demi{Blue!70}
\def\demii{Green!70}

\foreach \row/\r/\c/\first/\second/\third in {0/0/\dem/0/3/6, 1/1/\demi/1/4/7, 2/2/\demii/2/5/8}{
  \node[\col, anchor=east] at (-1.4,-1.35*\row) {$r=\r:$};
  \node[\col] (init\row) at (-0.5,-1.35*\row) {$\Fopt$};
  \node[batch, draw=\c, text=\col] (x\row) at (1.4,-1.35*\row) {$\Batch{\first}$};
  \node[batch, draw=\c, text=\col] (y\row) at (4.4,-1.35*\row) {$\Batch{\second}$};
  \node[batch, draw=\c, text=\col] (z\row) at (7.4,-1.35*\row) {$\Batch{\third}$};
  \node[\col] (dots\row) at (9.1,-1.35*\row) {$\cdots$};
  \draw[arrow] (init\row) -- (x\row);
  \draw[arrow] (x\row) -- node[\col, above=2pt] {\footnotesize{$\times 2^{\ClassCount}$}} (y\row);
  \draw[arrow] (y\row) -- node[\col, above=2pt] {\footnotesize{$\times 2^{\ClassCount}$}} (z\row);
  \draw[arrow] (z\row) -- (dots\row);
}
\end{tikzpicture}
\caption{Strided batch classes $\Batch{r+t\ClassCount}$ illustrated for $\ClassCount=3$ (\FullRef{Definition~\ref*{def:batch-classes}}{def:batch-classes} and \FullRef{Algorithm~\ref*{alg:accounting}}{alg:accounting}). Each residue class $r\in\{0,1,\ldots,\ClassCount-1\}$ resets its reference forest to $\Fopt$ and processes every $\ClassCount$-th batch in order of increasing scale, so consecutive batches in the same class have scales separated by a factor of $2^{\ClassCount}$, and the reference-forest cost decreases $\BatchSaving{b}$ within each row sum to at most $\OPT$.}
\label{fig:batch-schedule}
\end{figure}

%% file: lower_bound.tex
\section{The Lower Bound}
\label{sec:lower}

We prove \FullRef{Theorem~\ref*{thm:lower}}{thm:lower} by giving greedy many expensive
connections whose endpoints all lie on one inexpensive path. This path
will serve as a cheap offline solution, whereas greedy will buy
separate shortcuts. The main difficulty is
ensuring that a shortcut is still the cheapest choice when all future
shortcuts are already available.

We first describe the path and candidate shortcuts, then give a
distance test that identifies shortcuts whose prices greedy can be
forced to pay. A probabilistic argument shows that enough shortcuts
pass this test. Only then do we select a fixed matching and assemble
the actual online instance.

\subsection{The comparison path and candidate shortcuts}

Let $\LowerVertices$ be an even integer, to be chosen below, and let
$\LowerPath$ be the path on vertices $0,1,\ldots,\LowerVertices-1$ in
that order, with every edge of cost one. Thus
$\cost(\LowerPath)=\LowerVertices-1$. Since all terminal endpoints will
be vertices of this path, it will give the bound
$\OPT\le\LowerVertices-1$.

To create candidate terminal pairs, partition these vertices into
$\LowerMatchingSize=\LowerVertices/2$ disjoint unordered pairs. Write
$\LowerMatching$ for this set of pairs. Equivalently, $\LowerMatching$
is a perfect matching of the \emph{complete graph on the vertex set of
$\LowerPath$}: a pair may join any two path vertices, regardless of
their distance along $\LowerPath$. For each pair we add a candidate
shortcut edge. We add only these matching edges, not all edges of the
complete graph.

Give the matching edges distinct ranks $1,\ldots,\LowerMatchingSize$,
and write $\LowerMatchingEdge{i}$ for the edge of rank $i$. This is an
\emph{ordered perfect matching}. Ranks specify the order in which
candidate pairs will be considered; after some pairs are excluded,
the remaining pairs will still be requested in increasing rank.

The auxiliary graph $\LowerAuxiliaryGraph$ consists of the path
$\LowerPath$ together with all candidate matching edges. If a matching
pair joins consecutive path vertices, its shortcut and the unit-cost
path edge are kept as distinct parallel edges. This auxiliary
multigraph is used to test candidates; we will construct a simple
input graph after choosing which candidates to retain.

\subsection{Prices and the test for a forced purchase}

We want the prices paid by greedy to have a large harmonic sum. To set
their scale, fix an integer $\LowerLevel\ge32$ and now choose
\[
 \LowerLogSize=2^{\LowerLevel},\qquad
 \LowerVertices=2^{\LowerLogSize}.
\]
Thus $\LowerLogSize=\log_2\LowerVertices$. We will use prices
proportional to $1/i$, but only after a prefix of ranks. Excluding this
prefix limits the largest price, which lets us bound the probability
of a cheaper alternate route. Set
\begin{equation}
\label{eq:lower-parameters}
 \LowerPrefixFraction=\frac{256\LowerLevel}{\LowerLogSize},
 \qquad
 \LowerFirstRank=\lceil\LowerPrefixFraction\LowerMatchingSize\rceil.
\end{equation}
The auxiliary price of rank $i$ is
\begin{equation}
\label{eq:lower-prices}
 \LowerPrice{i}=
 \begin{cases}
  0,&i<\LowerFirstRank,\\[2pt]
  \displaystyle\frac{\LowerLogSize\LowerMatchingSize}{1024i},&i\ge \LowerFirstRank.
 \end{cases}
\end{equation}
The zero-price prefix is an auxiliary device for the analysis; none of
its edges will be included in the actual instance.

Before retaining a positive-price shortcut $\LowerMatchingEdge{i}=\{u,v\}$,
we need to rule out every alternate route of cost at most
$\LowerPrice{i}$. Earlier shortcuts may already be free, and later
shortcuts are available at their original prices because the input
graph is known to greedy from the start. Accordingly, for
$i\ge\LowerFirstRank$ form the \emph{test graph}
$\LowerTestGraph{i}$ from $\LowerAuxiliaryGraph$ as follows:
\begin{enumerate}[label=(\roman*),nosep]
\item Delete $\LowerMatchingEdge{i}$, so that only alternate routes
between its endpoints remain.
\item Give every matching edge of rank $j<i$ cost zero.
\item Keep every matching edge of rank $j>i$ at cost $\LowerPrice{j}$,
and every edge of $\LowerPath$ at cost one.
\end{enumerate}
Call rank $i$ \emph{bad} if the distance between the endpoints of
$\LowerMatchingEdge{i}$ in $\LowerTestGraph{i}$ is at most
$\LowerPrice{i}$, and \emph{good} if that distance is strictly greater
than $\LowerPrice{i}$. \FullRef{Figure~\ref*{fig:lower-test-graph}}{fig:lower-test-graph} shows this test.

\input{figures/lower-test-graph}

The test deliberately allows every other candidate shortcut,
including candidates that we will eventually exclude. Removing those
extra edges can only increase alternate-path distances. Thus a good
shortcut remains strictly cheaper than any alternate route when only
retained candidates are present and the earlier ones have been bought.
We will use this observation to prove greedy's behavior after defining
the final instance.

\subsection{Why few shortcuts fail the test}

We now show that some ordered matching has a large total price of
good edges. For this existence argument, choose a uniform perfect
matching $\LowerMatching$ of the complete graph on the path vertices
and, independently, assign its edges a uniform permutation of the
$\LowerMatchingSize$ ranks. Write $\LowerPartner{u}$ for the matching
partner of vertex $u$.

We will count alternate paths in $\LowerTestGraph{i}$, weighting each
matching traversal according to its test cost. Since distinct matching
edges receive ranks sampled without replacement, the product of
their weights is controlled by the following inequality.

\begin{lemma}
\label{lem:lower-sampling}
Let $\SamplePopulation\ge1$, let $\SampleValue{1},\ldots,\SampleValue{\SamplePopulation}$ be nonnegative, and let $\SampleIndex{1},\ldots,\SampleIndex{\SampleSize}$ be a
uniform ordered sample of $\SampleSize$ distinct indices, where $0\le \SampleSize\le \SamplePopulation$.
Then
\[
 \E\!\left[\prod_{a=1}^{\SampleSize} \SampleValue{\SampleIndex{a}}\right]
 \le \left(\frac{\SampleValue{1}+\cdots+\SampleValue{\SamplePopulation}}{\SamplePopulation}\right)^{\SampleSize}.
\]
The empty product is one.
\end{lemma}

\begin{proof}
The cases $\SampleSize=0,1$ are immediate. For $\SampleSize\ge2$, let $\ElementarySymmetric{\SampleSize}(\SampleVector)$ be the sum
of the products over all $\SampleSize$-element index sets. The expectation on the
left is $\ElementarySymmetric{\SampleSize}(\SampleVector)/\binom{\SamplePopulation}{\SampleSize}$. We show that at fixed sum of the $\SampleValue{j}$,
this polynomial is maximized by equal coordinates.

The simplex of nonnegative vectors with the given sum is compact.
Among the maximizers of $\ElementarySymmetric{\SampleSize}$, choose one minimizing the sum of the
squared coordinates. Holding all coordinates except $\SampleValue{a},\SampleValue{b}$ fixed,
we can write
\[
 \ElementarySymmetric{\SampleSize}=\PolyConstant+(\SampleValue{a}+\SampleValue{b})\PolyLinearCoeff+\SampleValue{a}\SampleValue{b}\PolyProductCoeff,
 \qquad \PolyProductCoeff\ge0.
\]
Replacing $\SampleValue{a},\SampleValue{b}$ by their average preserves their sum and does not
decrease $\ElementarySymmetric{\SampleSize}$. If they are unequal, this replacement strictly decreases
their sum of squares, contrary to the choice of maximizer. Thus all
coordinates are equal. Evaluating $\ElementarySymmetric{\SampleSize}$ there proves the inequality.
\end{proof}

An alternate simple path between the endpoints of a deleted matching
edge consists of matching traversals separated by nonempty monotone
runs along $\LowerPath$. The first and last runs are nonempty because
the deleted edge is the only matching edge incident to either endpoint.
Intermediate runs are nonempty because distinct matching edges have
disjoint endpoints. A simple run on a path is monotone.

A \emph{pattern} specifies the number $\LowerShortcutCount$ of matching
traversals, the positive lengths of the $\LowerShortcutCount+1$ path
runs, and the left or right direction of each run. To follow a pattern
from a vertex $u$, expose a matching partner only when a matching
traversal is required. Abort if a path boundary is crossed or any
vertex is repeated. If the walk completes, write $z$ for its final
vertex. This exploration uses the unranked matching only.

The next lemma bounds the contribution of one pattern, allowing an
arbitrary nonnegative weight for each traversed matching rank. It
separates the random matching geometry from the choice of weights we
will use to count cheap paths.

\begin{lemma}
\label{lem:lower-pattern-contribution}
Fix a vertex $u$ and condition its matching edge
$\LowerDeletedEdge=\{u,\LowerPartner{u}\}$ to have rank $i$.
Give each other rank $j\ne i$ a nonnegative weight $\SampleValue{j}$.
For a fixed pattern with $0\le\LowerShortcutCount\le\LowerMatchingSize-1$
matching traversals, assign contribution zero if the exploration aborts
or ends at $z\ne\LowerPartner{u}$, and otherwise assign the product of
the weights of its traversed matching ranks. Its expected contribution
is at most
\[
 \frac{1}{\LowerVertices-2\LowerShortcutCount-1}
 \left(\frac{1}{\LowerMatchingSize-1}
       \sum_{j\ne i}\SampleValue{j}\right)^{\LowerShortcutCount}.
\]
\end{lemma}

\begin{proof}
For every unranked matching, the probability that the edge incident
to $u$ has rank $i$ is $1/\LowerMatchingSize$. Thus this conditioning
leaves the unranked matching uniform; after choosing it, the other
ranks are assigned uniformly to its remaining edges.

Consider a completed, non-aborted exposure history. Exactly
$\LowerShortcutCount$ disjoint matching edges have been revealed, all
of their $2\LowerShortcutCount$ endpoints have been visited, and none
of these edges meets $u$: a traversal returning to $u$ would abort.
The final vertex $z$ is new, hence unexposed. Since the exploration
consults only the revealed matching pairs, the matching on the
unexposed vertices remains uniform. In particular,
\begin{equation}
\label{eq:lower-hidden-partner}
 \Prb\bigl(\LowerPartner{u}=z\mid\text{the completed exposed history}\bigr)
 =\frac{1}{\LowerVertices-2\LowerShortcutCount-1}.
\end{equation}
Any two choices of the partner of $u$ are put in bijection by
interchanging their vertex names in the unexposed matching, leaving
the revealed pairs unchanged.

An earlier run may have passed $\LowerPartner{u}$ without exposing
its partner. This imposes no additional conditioning: if a matching
traversal were attempted there, it would return to $u$ and abort;
if the vertex were only passed, the event $\LowerPartner{u}=z$ would
exclude that completion. The exploration never consults an unexposed
partner.

For each full matching with $\LowerPartner{u}=z$, the traversed matching
edges are distinct and different from $\LowerDeletedEdge$. Their ranks
are therefore a uniform sample without replacement from the other
$\LowerMatchingSize-1$ ranks, independently of the unranked topology.
Apply \FullRef{Lemma~\ref*{lem:lower-sampling}}{lem:lower-sampling} with population size
$\SamplePopulation=\LowerMatchingSize-1$, sample size
$\LowerShortcutCount$, and values $\SampleValue{j}$ for
$j\ne i$. It bounds the expected product of rank weights by the
$\LowerShortcutCount$th power of their average. Multiplying by
\eqref{eq:lower-hidden-partner} and averaging over completed histories
proves the claim; aborted histories contribute zero.
\end{proof}

\FullRef{Figure~\ref*{fig:lower-path-pattern}}{fig:lower-path-pattern} illustrates the runs and exposed
matching pairs used in this lemma.

\input{figures/lower-path-pattern}

We now choose the rank weights to penalize expensive matching
traversals, and sum the resulting bound over all patterns. This gives
the probability estimate needed for the construction.

\begin{lemma}
\label{lem:lower-bad-probability}
For each $i\ge\LowerFirstRank$, writing
$\LowerRankFraction=i/\LowerMatchingSize$ for its normalized rank,
\begin{equation}
\label{eq:lower-bad-probability}
 \Prb(i\text{ is bad})
 \le\frac{\LowerVertices^{-31/32}}{2\LowerRankFraction}.
\end{equation}
\end{lemma}

\begin{proof}
Fix a vertex $u$ and condition
$\LowerDeletedEdge=\{u,\LowerPartner{u}\}$ to have rank $i$, as in
\FullRef{Lemma~\ref*{lem:lower-pattern-contribution}}{lem:lower-pattern-contribution}. If this edge is bad, there
is a simple alternate path of test cost at most $\LowerPrice{i}$:
cycles can be erased from a shortest walk because all test costs are
nonnegative. Let $\LowerPathSteps$ be the total number of unit-cost
steps on $\LowerPath$. A pattern for such a path must satisfy
\begin{equation}
\label{eq:lower-short-pattern}
 \LowerShortcutCount+1\le\LowerPathSteps\le\LowerPrice{i}.
\end{equation}

To count these cheap paths, give a path of test cost $d$ the
\emph{exponential weight} $\e^{-\LowerTilt d}$, where
$\LowerTilt>0$ is a parameter we will choose below. The reason for
this weight is the inequality
\[
 \mathbf{1}_{\{d\le\LowerPrice{i}\}}
 \le\e^{\LowerTilt\LowerPrice{i}}\e^{-\LowerTilt d}.
\]
Thus the total weight of paths, multiplied by
$\e^{\LowerTilt\LowerPrice{i}}$, bounds the number of cheap paths.
Weights multiply across edges, making them suitable for
\FullRef{Lemma~\ref*{lem:lower-pattern-contribution}}{lem:lower-pattern-contribution}. Increasing $\LowerTilt$
suppresses costly paths more strongly but enlarges the prefactor;
we will balance these effects after summing the patterns.

In $\LowerTestGraph{i}$, an earlier matching edge has cost zero and
hence weight one, while a later edge of rank $j$ has weight
$\e^{-\LowerTilt\LowerPrice{j}}$. Denote the \emph{average weight of a
matching traversal} by
\[
 \LowerMeanWeight=
 \frac{i-1+\sum_{j>i}\e^{-\LowerTilt\LowerPrice{j}}}
      {\LowerMatchingSize-1}.
\]
Apply \FullRef{Lemma~\ref*{lem:lower-pattern-contribution}}{lem:lower-pattern-contribution} with precisely these
nonnegative rank weights. The $\LowerPathSteps$ unit-cost path edges
supply an additional factor $\e^{-\LowerTilt\LowerPathSteps}$.
Consequently, for any fixed pattern with
$\LowerShortcutCount\le\LowerMatchingSize-1$, its expected exponential
contribution, counted as zero unless it completes and reaches
$\LowerPartner{u}$, is at most
\begin{equation}
\label{eq:lower-pattern-weight}
 \frac{\e^{-\LowerTilt\LowerPathSteps}
       \LowerMeanWeight^{\LowerShortcutCount}}
      {\LowerVertices-2\LowerShortcutCount-1}.
\end{equation}

We next sum over the run lengths and directions. The total weight
of one nonempty path run, allowing both directions and every positive
length, is
\[
 \LowerRunSum=\sum_{t\ge1}2\e^{-\LowerTilt t}
 =\frac{2}{\e^{\LowerTilt}-1}\le\frac{2}{\LowerTilt}.
\]
Allowing runs that leave the finite path only increases this sum.
For patterns satisfying \eqref{eq:lower-short-pattern}, we have
$\LowerPrice{i}\le\LowerLogSize^2$ and
$\LowerVertices\ge4\LowerLogSize^2+2$, so the denominator in
\eqref{eq:lower-pattern-weight} is at least $\LowerVertices/2$.
Use this bound first, then enlarge the sum to all run lengths and all
$\LowerShortcutCount\ge0$. A union bound gives
\begin{equation}
\label{eq:lower-pattern-sum}
 \Prb(\LowerDeletedEdge\text{ is bad}\mid\operatorname{rank}(\LowerDeletedEdge)=i)
 \le\frac{2\e^{\LowerTilt\LowerPrice{i}}}{\LowerVertices}
      \sum_{\LowerShortcutCount\ge0}
      \LowerRunSum^{\LowerShortcutCount+1}
      \LowerMeanWeight^{\LowerShortcutCount}.
\end{equation}
Each extra matching traversal and subsequent path run contributes a
factor $\LowerRunSum\LowerMeanWeight$. We therefore choose
$\LowerTilt$ to make this factor less than one, so the series converges.

Take $\LowerTilt=16\LowerRankFraction$. This gives
$\LowerRunSum\le1/(8\LowerRankFraction)$, and gives every future
matching edge weight at most $\e^{-4\LowerLevel}$, because
\[
 \LowerTilt\LowerPrice{j}
 \ge\LowerTilt\LowerPrice{\LowerMatchingSize}
 =\frac{\LowerRankFraction\LowerLogSize}{64}
 \ge4\LowerLevel
 \qquad(j>i).
\]
Here $\LowerRankFraction\ge\LowerPrefixFraction$.
Since $\e^{-4\LowerLevel}\le1/\LowerLogSize\le\LowerRankFraction$
and $(i-1)/(\LowerMatchingSize-1)\le\LowerRankFraction$, we obtain
\begin{equation}
\label{eq:lower-beta}
 \LowerMeanWeight
 \le\LowerRankFraction+\e^{-4\LowerLevel}
 \le2\LowerRankFraction.
\end{equation}
Thus $\LowerRunSum\LowerMeanWeight\le1/4$, and
\[
 \sum_{\LowerShortcutCount\ge0}
 \LowerRunSum^{\LowerShortcutCount+1}\LowerMeanWeight^{\LowerShortcutCount}
 =\frac{\LowerRunSum}{1-\LowerRunSum\LowerMeanWeight}
 \le\frac{1/(8\LowerRankFraction)}{1-1/4}
 =\frac1{6\LowerRankFraction}.
\]
The same choice keeps the prefactor small: using
$\LowerPrice{i}=\LowerLogSize/(1024\LowerRankFraction)$ and
$\ln2\ge1/2$,
\[
 \e^{\LowerTilt\LowerPrice{i}}
 =\e^{\LowerLogSize/64}
 =\LowerVertices^{1/(64\ln2)}
 \le\LowerVertices^{1/32}.
\]
Substitution into \eqref{eq:lower-pattern-sum} yields
\[
 \Prb(\LowerDeletedEdge\text{ is bad}\mid\operatorname{rank}(\LowerDeletedEdge)=i)
 \le\frac{\LowerVertices^{-31/32}}{3\LowerRankFraction}
 \le\frac{\LowerVertices^{-31/32}}{2\LowerRankFraction}.
\]

This bound holds for every fixed $u$. Summing over all vertices the
probability that their incident edge has rank $i$ and is bad counts
each bad rank-$i$ edge twice. The rank-conditioning event has probability
$1/\LowerMatchingSize$ at each vertex, and
$\LowerVertices/\LowerMatchingSize=2$, so the same bound holds for the
random edge of rank $i$.
\end{proof}

\subsection{Retaining enough price and choosing a fixed matching}

Having bounded the probability that each rank is bad, we now compare
the total price lost to bad ranks with the total available positive
price. Define
\begin{equation}
\label{eq:lower-total-price}
 \LowerTotalPrice=\sum_{i=\LowerFirstRank}^{\LowerMatchingSize}\LowerPrice{i}.
\end{equation}

The elementary bounds $\LowerFirstRank\le2\LowerPrefixFraction\LowerMatchingSize$ and
$\LowerFirstRank-1\ge\LowerPrefixFraction\LowerMatchingSize/2$ hold for the parameters in
\eqref{eq:lower-parameters}. The harmonic sum satisfies
\begin{align}
 \sum_{i=\LowerFirstRank}^{\LowerMatchingSize}\frac1i
 &\ge\int_{\LowerFirstRank}^{\LowerMatchingSize+1}\frac{\dd t}{t}
  =\ln\frac{\LowerMatchingSize+1}{\LowerFirstRank}\notag\\
 &\ge\ln\frac{\LowerLogSize}{512\LowerLevel}\notag\\
 &\ge\frac{\LowerLevel\ln2}{2}.
 \label{eq:lower-harmonic}
\end{align}
The final inequality follows from $2^{\LowerLevel/2}\ge512\LowerLevel$, which holds
at $\LowerLevel=32$ and remains true as $2^{\LowerLevel/2}/\LowerLevel$ increases for $\LowerLevel\ge32$. Thus
\begin{equation}
\label{eq:lower-price-lower}
 \LowerTotalPrice=\frac{\LowerLogSize\LowerMatchingSize}{1024}\sum_{i=\LowerFirstRank}^{\LowerMatchingSize}\frac1i
 \ge\frac{\LowerVertices\LowerLogSize\LowerLevel\ln2}{4096}.
\end{equation}

Let $\LowerBadPrice$ be the sum of the prices of all bad positive edges of the
random ordered matching. By \FullRef{Lemma~\ref*{lem:lower-bad-probability}}{lem:lower-bad-probability},
linearity of expectation, and the decreasing-integrand bound for
$\sum i^{-2}$,
\begin{align}
 \E \LowerBadPrice
 &\le\frac{\LowerLogSize \LowerVertices^{-31/32}\LowerMatchingSize^2}{2048}
       \sum_{i=\LowerFirstRank}^{\LowerMatchingSize}\frac1{i^2}\notag\\
 &\le\frac{\LowerLogSize \LowerVertices^{-31/32}\LowerMatchingSize^2}{2048(\LowerFirstRank-1)}\notag\\
 &\le\frac{\LowerLogSize \LowerVertices^{1/32}}{2048\LowerPrefixFraction}.
 \label{eq:lower-bad-price}
\end{align}
Combining \eqref{eq:lower-price-lower} and
\eqref{eq:lower-bad-price} gives
\begin{align*}
 \frac{\E \LowerBadPrice}{\LowerTotalPrice}
 &\le\frac{2\LowerVertices^{-31/32}}{\LowerPrefixFraction\LowerLevel\ln2}\\
 &\le\frac{\LowerLogSize \LowerVertices^{-31/32}}{64\LowerLevel^2}\\
 &\le\frac{1}{64\LowerLevel^2}<\frac12.
\end{align*}
For the penultimate inequality, $\LowerVertices^{31/32}\ge \LowerLogSize$ follows from
$(31/32)2^{\LowerLevel}\ge \LowerLevel$. Hence some ordered matching has
$\LowerBadPrice\le\LowerTotalPrice/2$, or equivalently
\begin{equation}
\label{eq:lower-retention-rule}
 \sum_{\substack{i\ge\LowerFirstRank\\i\text{ good}}}\LowerPrice{i}
 \ge\LowerTotalPrice/2.
\end{equation}

The random choice has served only to prove existence. To specify one
fixed instance for each $\LowerLevel$, represent an ordered matching by
the sequence
\[
 (\LowerEndpointA{1},\LowerEndpointB{1},\ldots,
   \LowerEndpointA{\LowerMatchingSize},\LowerEndpointB{\LowerMatchingSize}),
\]
where $\LowerEndpointA{i}<\LowerEndpointB{i}$ are the endpoints of
$\LowerMatchingEdge{i}$. Enumerate these sequences lexicographically
and select the first matching satisfying
\eqref{eq:lower-retention-rule}. There are finitely many ordered
matchings, and all the tests are shortest-path computations with
rational costs. The existence argument therefore proves that this
finite deterministic search terminates. Its role is to specify the
example unambiguously; no efficiency claim is needed.

\subsection{The actual instance and greedy's behavior}

Fix the ordered matching just selected. The actual graph consists of
$\LowerPath$ and only the good matching edges of rank
$i\ge\LowerFirstRank$. All prefix edges and all bad edges are excluded.
To obtain a simple graph even when a retained pair is adjacent on
$\LowerPath$, replace each retained edge $\LowerMatchingEdge{i}=\{u,v\}$
by a private degree-two vertex $\LowerSubdivision{i}$ and two edges
$\{u,\LowerSubdivision{i}\}$ and $\{\LowerSubdivision{i},v\}$, each of
cost $\LowerPrice{i}/2$. This replacement preserves the shortcut's
total cost; see \FullRef{Figure~\ref*{fig:lower-retained-instance}}{fig:lower-retained-instance}. All actual
edge costs are strictly positive and rational.

Present the retained matching pairs in increasing rank. Here $i$
continues to mean the edge's original matching rank, which need not be
its position in the filtered request sequence. The entire graph,
including every later retained shortcut, is available from the start.

\input{figures/lower-retained-instance}

Consider a retained edge $\LowerMatchingEdge{i}$. Inductively suppose that greedy has
bought exactly the previously retained matching paths and no edge of
$\LowerPath$. For the distance comparison, suppress the private
degree-two vertex of every retained shortcut, replacing its two edges
by one edge of the same total residual cost. Earlier retained shortcuts
then have cost zero, and the current and later ones keep their prices.
After deleting $\LowerMatchingEdge{i}$, the resulting residual graph is
a subgraph of $\LowerTestGraph{i}$. The latter also contains
all excluded prefix edges for free, all earlier bad edges for free,
and all later bad edges at their positive prices. On every shared
edge, its test cost agrees with its actual residual cost. Deleting
the extra edges cannot shorten an alternate
path. Since $i$ is good, every actual path between the endpoints of
$\LowerMatchingEdge{i}$ avoiding $\LowerMatchingEdge{i}$ has cost strictly greater than $\LowerPrice{i}$.

The private two-edge path representing $\LowerMatchingEdge{i}$ has cost $\LowerPrice{i}$. A simple
path between its endpoints using either of its two edges must use both,
because its private subdivision vertex has degree two; it is then
exactly that two-edge path. Thus this is the unique simple shortest
path. Greedy buys precisely its two edges, proving the induction.
Previously purchased matching paths are vertex-disjoint from the
current pair, so they cannot create a second simple path of the same
cost. If the algorithm permits shortest walks with gratuitous
zero-cost excursions, those excursions buy no additional edges and
leave the same paid behavior. The construction is therefore
independent of shortest-path tie-breaking.

\subsection{The competitive ratio}

It follows that
\[
 \ALG=\sum_{\substack{i\ge \LowerFirstRank\\i\text{ good}}}\LowerPrice{i}
 \ge \LowerTotalPrice/2.
\]
The path $\LowerPath$ connects every terminal pair, so $\OPT\le \LowerVertices-1$.
By \eqref{eq:lower-price-lower},
\begin{equation}
\label{eq:lower-ratio-h}
 \begin{aligned}
 \frac{\ALG}{\OPT}
 &\ge\frac{\LowerTotalPrice}{2(\LowerVertices-1)}\\
 &\ge\frac{\LowerVertices\LowerLogSize\LowerLevel\ln2}{8192(\LowerVertices-1)}\\
 &\ge\frac{\LowerLogSize\LowerLevel\ln2}{8192}.
 \end{aligned}
\end{equation}

Let $\LowerRequestCount{\LowerLevel}$ denote the number of retained terminal pairs. Each retained price
is at most
\[
 \LowerMaxPrice\le\frac{\LowerLogSize}{1024\LowerPrefixFraction}
 =\frac{\LowerLogSize^2}{262144\LowerLevel}.
\]
Therefore \eqref{eq:lower-price-lower} and the retention rule imply
\begin{equation}
\label{eq:lower-request-count}
 \begin{aligned}
 \LowerRequestCount{\LowerLevel}&\ge\frac{\LowerTotalPrice}{2\LowerMaxPrice}\\
 &\ge\frac{\LowerVertices\LowerLogSize\LowerLevel\ln2}{8192}\cdot\frac{262144\LowerLevel}{\LowerLogSize^2}
   =32\ln2\,\frac{\LowerVertices\LowerLevel^2}{\LowerLogSize},\\
 \LowerRequestCount{\LowerLevel}&\le \LowerMatchingSize=\LowerVertices/2.
 \end{aligned}
\end{equation}
In particular $\LowerRequestCount{\LowerLevel}\to\infty$. Since the lower bound is at least
$\LowerVertices/\LowerLogSize$, and $\LowerLevel\le \LowerLogSize/2$, we have
\begin{align*}
 \frac \LowerLogSize2&\le \LowerLogSize-\LowerLevel\le\log_2 \LowerRequestCount{\LowerLevel}\le \LowerLogSize,\\
 \LowerLevel-1&\le\log_2\log_2 \LowerRequestCount{\LowerLevel}\le \LowerLevel.
\end{align*}
Together with \eqref{eq:lower-ratio-h}, this proves
\[
 \frac{\ALG}{\OPT}
 \ge\frac{\ln2}{8192}
       (\log_2 \LowerRequestCount{\LowerLevel})(\log_2\log_2 \LowerRequestCount{\LowerLevel}),
\]
which proves \FullRef{Theorem~\ref*{thm:lower}}{thm:lower}.

To complete \FullRef{Corollary~\ref*{cor:tight}}{cor:tight}, we extend the bound to every
sufficiently large number $\NumPairs$ of terminal pairs. Since $\LowerVerticesAt{\LowerLevel+1}=\LowerVerticesAt{\LowerLevel}^2$ and
$\LowerLogSizeAt{\LowerLevel+1}=2\LowerLogSizeAt{\LowerLevel}$,
\eqref{eq:lower-request-count} gives
\[
 \LowerRequestCount{\LowerLevel+1}\ge32\ln2\,\frac{\LowerVerticesAt{\LowerLevel}^2(\LowerLevel+1)^2}{2\LowerLogSizeAt{\LowerLevel}}
 >\LowerVerticesAt{\LowerLevel}/2\ge \LowerRequestCount{\LowerLevel}.
\]
Choose $\LowerLevel$ with
$\LowerRequestCount{\LowerLevel}\le \NumPairs<\LowerRequestCount{\LowerLevel+1}$ and pad its instance by repeating an already served
pair. These repetitions change neither greedy's cost nor the optimum.
Moreover $\log_2 \NumPairs\le2\LowerLogSizeAt{\LowerLevel}$ and
$\log_2\log_2 \NumPairs\le \LowerLevel+1$, so \eqref{eq:lower-ratio-h} still yields
the same asymptotic lower bound, with an absolute constant smaller by
at most a factor of four.


%% file: figures/lower-test-graph.tex
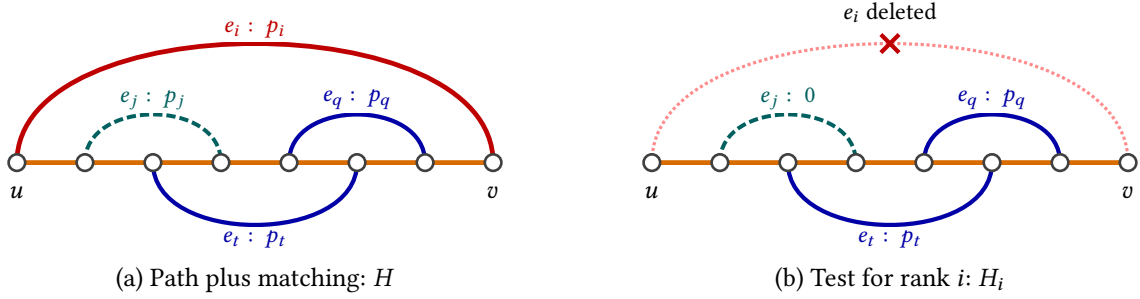
\begin{figure}[t]
\centering
\begin{tikzpicture}[
 font=\small,
 vertex/.style={circle,draw=black!75,fill=white,line width=1.0pt,inner sep=2.2pt},
 path edge/.style={orange!85!black,line width=1.8pt},
 earlier/.style={teal!75!black,line width=1.5pt,dash pattern=on 3.5pt off 1.5pt},
 later/.style={blue!65!black,line width=1.6pt},
 edge label/.style={font=\footnotesize,fill=white,inner sep=1.3pt},
 >=Latex]

 \begin{scope}
  \draw[path edge] (0,0)--(6.3,0);
  \draw[red!75!black,line width=1.8pt] (0,0)
   .. controls (0,2.1) and (6.3,2.1) ..
   node[edge label,above] {$\LowerMatchingEdge{i}:\ \LowerPrice{i}$} (6.3,0);
  \draw[earlier] (.9,0)
   .. controls (.9,.85) and (2.7,.85) ..
   node[edge label,above] {$\LowerMatchingEdge{j}:\ \LowerPrice{j}$} (2.7,0);
  \draw[later] (3.6,0)
   .. controls (3.6,.85) and (5.4,.85) ..
   node[edge label,above] {$\LowerMatchingEdge{q}:\ \LowerPrice{q}$} (5.4,0);
  \draw[later] (1.8,0)
   .. controls (1.8,-1.1) and (4.5,-1.1) ..
   node[edge label,below] {$\LowerMatchingEdge{t}:\ \LowerPrice{t}$} (4.5,0);
  \foreach \x in {0,.9,1.8,2.7,3.6,4.5,5.4,6.3}
   \node[vertex] at (\x,0) {};
  \node[below=5pt] at (0,0) {$u$};
  \node[below=5pt] at (6.3,0) {$v$};
  \node at (3.15,-1.55) {(a) Path plus matching: $\LowerAuxiliaryGraph$};
 \end{scope}

 \begin{scope}[xshift=8.4cm]
  \draw[path edge] (0,0)--(6.3,0);
  \draw[red!45,line width=1.2pt,densely dotted] (0,0)
   .. controls (0,2.1) and (6.3,2.1) .. (6.3,0);
  \draw[red!75!black,line width=1.6pt] (3.03,1.455)--(3.27,1.695);
  \draw[red!75!black,line width=1.6pt] (3.03,1.695)--(3.27,1.455);
  \node[edge label,above] at (3.15,1.8)
   {$\LowerMatchingEdge{i}$ deleted};
  \draw[earlier] (.9,0)
   .. controls (.9,.85) and (2.7,.85) ..
   node[edge label,above] {$\LowerMatchingEdge{j}:\ 0$} (2.7,0);
  \draw[later] (3.6,0)
   .. controls (3.6,.85) and (5.4,.85) ..
   node[edge label,above] {$\LowerMatchingEdge{q}:\ \LowerPrice{q}$} (5.4,0);
  \draw[later] (1.8,0)
   .. controls (1.8,-1.1) and (4.5,-1.1) ..
   node[edge label,below] {$\LowerMatchingEdge{t}:\ \LowerPrice{t}$} (4.5,0);
  \foreach \x in {0,.9,1.8,2.7,3.6,4.5,5.4,6.3}
   \node[vertex] at (\x,0) {};
  \node[below=5pt] at (0,0) {$u$};
  \node[below=5pt] at (6.3,0) {$v$};
  \node at (3.15,-1.55) {(b) Test for rank $i$: $\LowerTestGraph{i}$};
 \end{scope}
\end{tikzpicture}
\caption{The matching pairs arbitrary vertices of the comparison path;
its edges are additional shortcuts, not edges selected from the path.
The eight displayed vertices give a schematic of the construction with ranks $j<i<q<t$; orange edges belong to $\LowerPath$ and cost $1$ in both panels, earlier matching edges become free ($0$), and every later matching edge stays available at its original price.
In the test graph $\LowerTestGraph{i}$, the crossed-out guide marks the deleted edge $\LowerMatchingEdge{i}$ and is
not available for travel. Rank $i$ is good precisely when every remaining
route from $u$ to $v$ costs more than
$\LowerPrice{i}$.}
\label{fig:lower-test-graph}
\end{figure}

%% file: figures/lower-path-pattern.tex
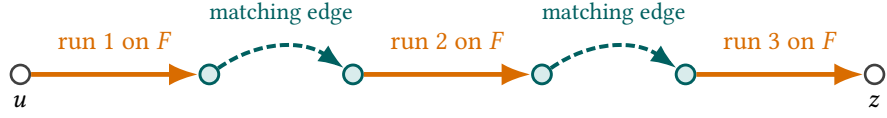
\begin{figure}[t]
\centering
\begin{tikzpicture}[
 font=\small,
 vertex/.style={circle,draw=black!75,fill=white,line width=1.0pt,inner sep=2.5pt},
 exposed/.style={vertex,draw=teal!75!black,fill=teal!15,line width=1.1pt},
 run/.style={orange!85!black,line width=1.8pt,->},
 jump/.style={teal!75!black,line width=1.5pt,dash pattern=on 3.5pt off 1.5pt,->},
 >=Latex]
 \node[vertex,label=below:$u$] (u) at (0,0) {};
 \node[exposed] (a) at (2.5,0) {};
 \node[exposed] (b) at (4.4,0) {};
 \node[exposed] (c) at (6.9,0) {};
 \node[exposed] (d) at (8.8,0) {};
 \node[vertex,label=below:$z$] (z) at (11.3,0) {};

 \draw[run] (u)--node[above=5pt] {run 1 on $\LowerPath$}(a);
 \draw[jump] (a) to[bend left=35]
  node[above=4pt,font=\footnotesize] {matching edge}(b);
 \draw[run] (b)--node[above=5pt] {run 2 on $\LowerPath$}(c);
 \draw[jump] (c) to[bend left=35]
  node[above=4pt,font=\footnotesize] {matching edge}(d);
 \draw[run] (d)--node[above=5pt] {run 3 on $\LowerPath$}(z);
\end{tikzpicture}
\caption{A pattern with $\LowerShortcutCount=2$ revealed matching pairs and $\LowerShortcutCount+1=3$ path runs. The picture lays out
the walk in traversal order, not in the vertex order of $\LowerPath$;
each orange run is nonempty and monotone (moving left or right) on $\LowerPath$. Only the teal matching
pairs are exposed, and passing through a vertex along $\LowerPath$ does not
expose its matching partner. The partners of $u$ and $z$ remain unexposed until the final test checks whether $z=\LowerPartner{u}$ (the
other endpoint of the deleted edge incident to $u$).}
\label{fig:lower-path-pattern}
\end{figure}

%% file: figures/lower-retained-instance.tex
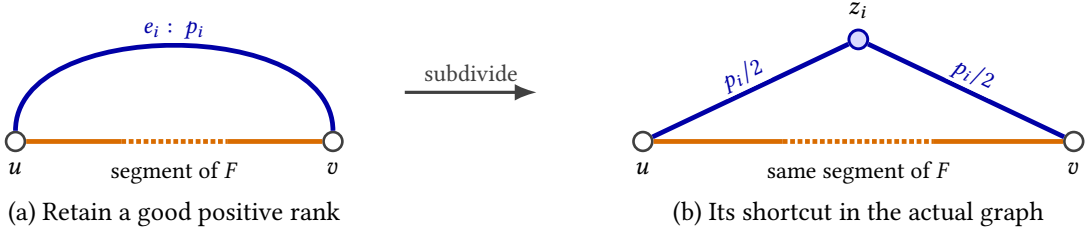
\begin{figure}[t]
\centering
\begin{tikzpicture}[
 font=\small,
 vertex/.style={circle,draw=black!75,fill=white,line width=1.0pt,inner sep=2.5pt},
 private vertex/.style={vertex,draw=blue!65!black,fill=blue!15,line width=1.1pt},
 path edge/.style={orange!85!black,line width=1.8pt},
 shortcut/.style={blue!65!black,line width=1.8pt},
 price/.style={font=\footnotesize,fill=white,inner sep=1.5pt},
 >=Latex]

 \node[vertex,label=below:$u$] (a) at (0,0) {};
 \node[vertex,label=below:$v$] (b) at (4.2,0) {};
 \draw[path edge] (a)--(1.4,0);
 \draw[path edge,densely dotted] (1.4,0)--(2.8,0);
 \draw[path edge] (2.8,0)--(b);
 \node[price,below=6pt] at (2.1,0) {segment of $\LowerPath$};
 \draw[shortcut] (a) .. controls (0,1.65) and (4.2,1.65) ..
  node[price,above] {$\LowerMatchingEdge{i}:\ \LowerPrice{i}$} (b);
 \node at (2.1,-0.95) {(a) Retain a good positive rank};

 \draw[->,line width=1.2pt,black!75] (5.15,.65)--node[above,font=\footnotesize] {subdivide}(6.9,.65);

 \node[vertex,label=below:$u$] (ac) at (8.3,0) {};
 \node[vertex,label=below:$v$] (bc) at (14,0) {};
 \node[private vertex,label=above:$\LowerSubdivision{i}$] (private) at (11.15,1.35) {};
 \draw[path edge] (ac)--(10.15,0);
 \draw[path edge,densely dotted] (10.15,0)--(12.15,0);
 \draw[path edge] (12.15,0)--(bc);
 \node[price,below=6pt] at (11.15,0) {same segment of $\LowerPath$};
 \draw[shortcut] (ac)--node[price,above,sloped] {$\LowerPrice{i}/2$}(private);
 \draw[shortcut] (private)--node[price,above,sloped] {$\LowerPrice{i}/2$}(bc);
 \node at (11.15,-0.95) {(b) Its shortcut in the actual graph};
\end{tikzpicture}
\caption{Only good ranks $i\ge\LowerFirstRank$ become terminal pairs,
presented in increasing original rank. Each retained shortcut receives
its own degree-two subdivision vertex $\LowerSubdivision{i}$, so the actual graph is simple
even if its endpoints are consecutive on $\LowerPath$; prefix and bad
matching edges are absent from this graph. The drawing shows one retained
shortcut (with the dotted orange segment standing for the intervening path edges): greedy buys the blue path of total cost $\LowerPrice{i}$ for this request, whereas the orange comparison path $\LowerPath$ connects every retained pair for total cost $\LowerVertices-1$.}
\label{fig:lower-retained-instance}
\end{figure}

%% file: methodology.tex
\section*{Acknowledgments}
Some of the proofs in this paper were developed with the help of OpenAI's GPT Astra, which also contributed
to the writing and exposition, with Gemini Argon assisting in some parts. The authors are responsible for the correctness of the results and the final
content of the paper. Lean files formalizing the main
theorem are available on GitHub at \url{https://github.com/igholami/online-steiner-forest-lean}.